\documentclass[a4paper,12pt]{scrartcl}
\usepackage[a4paper,bindingoffset=0mm]{geometry}
\usepackage[utf8]{inputenc}
\usepackage{amsmath}
  \numberwithin{equation}{section}
\usepackage{amssymb}
\usepackage{graphicx}
\usepackage{hyperref}
\usepackage{enumitem}

\hypersetup{
    colorlinks=true,
    linkcolor=blue,
    filecolor=magenta,      
    urlcolor=cyan,
}

\title{Numerical Relativity, Holography\\
    and the Quantum Null Energy Condition}
\author{Philipp Stanzer}
\makeindex
\date{\today}    

\begin{document}

\begin{titlepage}
\newgeometry{tmargin=3cm,bmargin=3cm,lmargin=3.5cm,rmargin=2.5cm,headsep=1.5cm,footskip=1.5cm}
\begin{figure}[ht]
\begin{center}
\vspace{-2cm}
\includegraphics[scale=0.3]{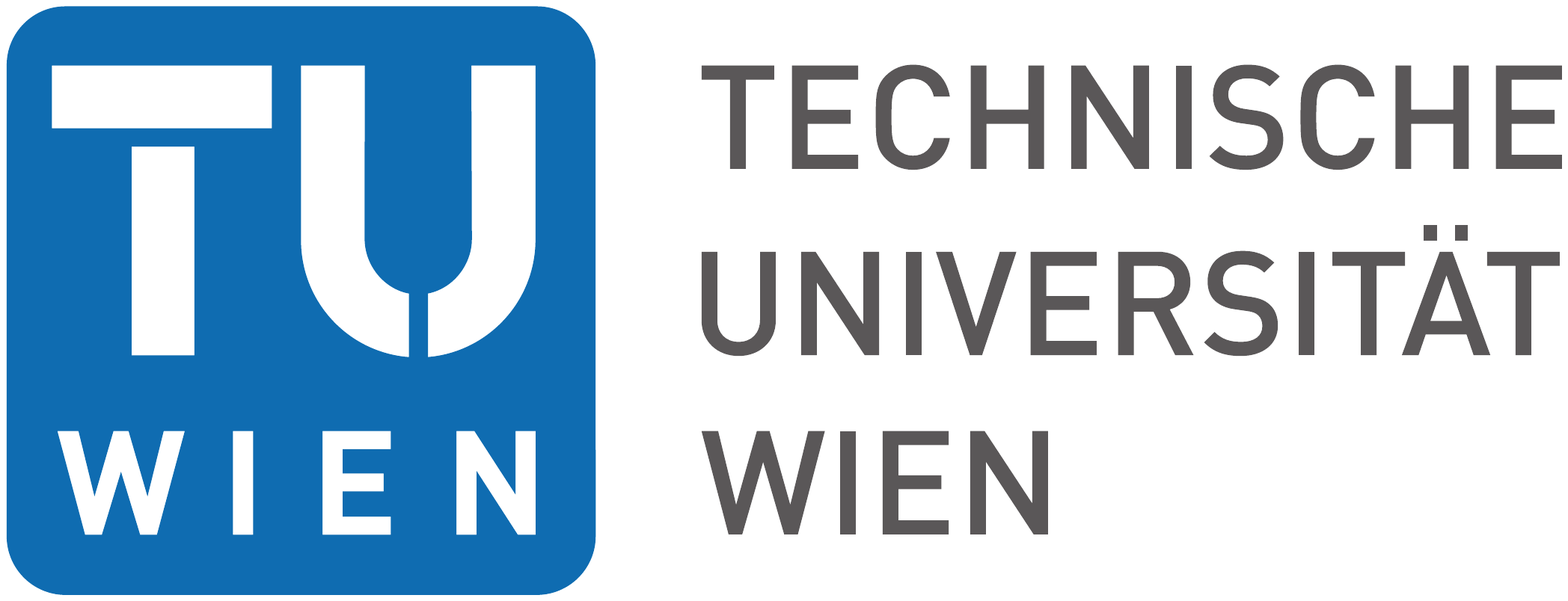}\\[1.0cm]
\end{center}
\end{figure}
\vspace{-3cm}
\begin{center}
{\LARGE Dissertation\\[1.0cm]}
{\LARGE\textbf{Numerical Relativity, Holography\\
    and the Quantum Null Energy Condition}\\[1.0cm]}
{\normalsize Ausgeführt zum Zwecke der Erlangung des Akademischen Grades \\
 eines Doktors der technischen Wissenschaften}\\[1cm]
{\normalsize unter der Leitung von\\
Assoc.-Prof. Priv.-Doz. Dr. Daniel Grumiller\\
Institut für theoretische Physik\\
Technische Universität Wien, AT\\[1cm]}

{\normalsize eingereicht an der Technischen Universität Wien\\
Fakultät für Physik\\[1cm]}
{\normalsize von\\

Dipl.-Ing. Philipp Stanzer\\

Mollgasse 13/2\\

1180 Wien}\\[2cm]

\noindent\begin{tabular}{ll}

\makebox[2cm]{Wien, am 31.07.2020}\hspace*{1cm} & \makebox[4cm]{\hrulefill}\\

& \\[1ex]

\end{tabular}
\vspace*{2.5cm}

\begin{tabular}{p{5cm}@{}p{5cm}@{}p{5cm}@{}}
 \makebox[4cm]{\hrulefill} & \makebox[4cm]{\hrulefill} & \makebox[4cm]{\hrulefill} \\
 Prof. Daniel Grumiller    & Prof. Umut G\"ursoy       & Prof. Paul Romatschke     \\
 (Betreuer)                & (Gutachter)               & (Gutachter)                 \\
\end{tabular}
\end{center}
\end{titlepage}

\cleardoublepage

\section*{Kurzfassung}
    
    Die Quanten-Null-Energiebedingung (QNEC) ist die einzige bekannte, lokale E\-ner\-gie\-bedingung f\"ur Quantentheorien.
    Im Gegensatz zu den klassischen Energiebedingungen wurde QNEC nicht postuliert, sondern hat ihren Ursprung in der Quan\-ten-Fokussierungsvermutung.
    Au{\ss}erdem wurde sie be\-reits f\"ur mehrere Spezial\-f\"al\-le und allgemein in mehr als drei Raumzeitdimensionen bewiesen.
    Des Weiteren ist ihr zentraler Bestandteil eine intrinsisch quantenmechanische Observable, die Verschr\"ankungsentropie.\\
    Die direkte Berechnung der Verschr\"ankungsentropie in einer Quantenfeldtheorie ist extrem schwierig, w\"ahrend sie unter Verwendung des holographischen Prinzips durch eine einfache geometrische Gr\"o{\ss}e bestimmt werden kann.
    Das ho\-lo\-gra\-phi\-sche Prinzip stellt eine Beziehung zwischen Eichtheorien ohne Gravitation und Quantengravitationstheorien mit einer zus\"atzlichen Dimension her.
    Das be\-kanntes\-te Beispiel f\"ur diese Dualit\"at ist die AdS/CFT Korrespondenz.
    Holographie bie\-tet die M\"oglichkeit sowohl etwas \"uber stark gekoppelte Feldtheorien als auch \"uber Quantengravitation zu lernen.
    Das Studium von QNEC wird in diesem Zusammenhang zweifellos zu neuen Erkenntnissen f\"uhren.\\
    \\
    Der Fokus dieser Arbeit liegt auf $2$- und $4$-dimensionalen Feldtheorien.
    Wir untersuchen unterschiedlich komplexe Systeme mit numerischen und (sofern m\"oglich) analytischen Methoden.
    Im Vakuum, in thermischen Zust\"anden, Ungleich\-ge\-wichts\-zust\"anden und einem Modell f\"ur Schwerionen-Kollisionen ist QNEC immer erf\"ullt und manchmal auch ges\"attigt.
    Gleichzeitig kann QNEC st\"arker oder schw\"a\-che\-r als die klassische Null-Energiebedingung sein.\\
    Interessant ist, dass QNEC$_2$ in zwei Dimensionen bei Vorhandensein von Materie in der Gravitationstheorie nicht ges\"attigt sein kann.
    Die R\"uckwirkung eines massiven skalaren Teilchens auf die Geometrie bietet ein gutes Beispiel bei dem sogar die Differenz zur S\"attigung bekannt ist.
    Betrachtet man hingegen ein massives selbst-wechselwirkendes Skalarfeld, f\"uhrt das Potential zu Phasen\"uberg\"angen von kleinen zu gro{\ss}en schwarzen L\"ochern.
    Wir verwenden QNEC$_2$ in der dualen Feldtheorie als Werkzeug, um stark gekoppelte, dynamische Systeme besser zu verstehen.
    Aus den Eigenschaften von QNEC$_2$ im Grundzustand kann man bereits Aussagen \"uber Phasen\"uberg\"ange in den thermischen Zust\"anden treffen.

\cleardoublepage

\section*{Abstract}

    The quantum null energy condition (QNEC) is the only known consistent local energy condition in quantum theories.
    Contrary to the classical energy conditions which are simply postulated and known to be violated in some classical systems and quantum field theory, QNEC is a consequence of the more general quantum focussing conjecture. It has been proven for several special cases and in general for quantum field theories in three or more spacetime dimensions.
    QNEC involves an intrinsically quantum property of the theory under consideration, the entanglement entropy.\\
    While entanglement entropy is notoriously hard to calculate in quantum field theory, the holographic principle provides a simple geometric description.
    In general the holographic principle relates a gauge theory without gravity to a theory of quantum gravity in one dimension higher.
    The most famous example of this gauge/gravity duality is the AdS/CFT correspondence.
    Holography provides a way to learn about strongly coupled field theories as well as quantum gravity and investigating QNEC in this context will undoubtedly lead to new insights.\\
    \\
    In this thesis the focus is put on $2$- and $4$-dimensional field theories, where we study systems of increasing complexity with numerical and (whenever possible) analytical methods.
    In vacuum, thermal states, globally quenched states and a toy model for heavy ion collisions we find that QNEC is always satisfied and sometimes saturated, while it can be a stronger or weaker condition than the classical null energy condition.\\
    Interestingly in two dimensions QNEC$_2$ cannot be saturated in the presence of bulk matter.
    The backreaction of a massive scalar particle provides an example where the finite gap to saturation is precisely known.
    Considering a massive self-interacting scalar field coupled to Einstein gravity leads to phase transitions from small to large black holes, determined by its potential.
    The dual field theory provides a rich example to use QNEC$_2$ as a tool to learn about strongly coupled dynamical systems.
    In particular knowing QNEC$_2$ in the ground state allows us to make statements about the phase structure of the thermal states.

\cleardoublepage

\section*{Acknowledgements}

    First of all, I want to thank everyone who is reading my thesis, because this justifies all the effort put into it.
    A special thanks goes to my supervisor Daniel Grumiller who supported me since I joined his research group for my master thesis.
    Most importantly he gave me the opportunity to continue my studies as a PhD student at TU Wien.
    Another important person is Christian Ecker who is a former member of Daniel's research group in Vienna.
    In the beginning he was co-supervisor and mentor, which turned into colleague, friend and climbing partner over time.
    The discussions with Daniel and Christian as well as many current and former members of the institute for theoretical physics at TU Wien have been crucial for my research.
    Similarly I want to thank my collaborators who provided a lot of knowledge, interesting and fun discussions leading to four papers: Stefan Stricker, Wilke van der Schee, Shahin Sheikh-Jabbari and Hesam Soltanpanahi.\\
    In the last year of my PhD studies the doctoral school DKPI provided me the opportunity to visit Paul Romatschke in Boulder, Colorado and Umut G\"ursoy in Utrecht, Netherlands for three month each, which I am very grateful for.
    In Boulder Paul Romatschke and Marcus Benghi Pinto taught me a lot about quantum field theory in some kind of private lectures.
    In Utrecht Umut G\"ursoy and many other members of the institute for theoretical physics contributed to my understanding of various topics around holography and string theory.
    I really enjoyed both visits and do not want to miss these experiences!\\
    I am very thankful that Paul and Umut agreed to read my thesis and prepare their reports for my defense.\\
    \\
    Completing a PhD is a very strenuous endeavor and requires a good balance between research and other activities.
    I want to thank all my friends and family outside of academia who very successfully helped me to distract my thoughts (such that I could return to my research with a fresh mind).
    Especially I want to thank my parents and grandparents who supported me in every way possible.
    They encouraged my curiosity for nature and science from childhood on.
    Last but not least I want to thank my wife for her love, trust and support and for believing in me more than I did myself.\\
    \\
    This paragraph is not only a way to say thank you to all the people mentioned above, but also to remind myself, that nothing I have achieved would have been possible without anyone of them.
    
\cleardoublepage

\section*{Publications}
    The first publication [1] was a direct result of my master thesis and serves as a foundation for all the work during my PhD studies.
    Publications [2] and [3] provide the research results discussed in sections \ref{sec:QNEC_D=4} and \ref{sec:QNEC_D=2}.
    Since the most recent study [4] was carried out while writing this thesis, some but not all of the results are included in section \ref{sec:QNEC_D=2}.
    
    \paragraph{Peer Reviewed Journals}
    \begin{enumerate}[label={[\arabic*]}]
        \item C.~Ecker, D.~Grumiller, P.~Stanzer, S.~A. Stricker and W.~van~der Schee, ``Exploring nonlocal observables in shock wave collisions,'' \emph{JHEP} \textbf{11} (2016) 054, \href{http://arXiv.org/abs/1609.03676}{\texttt{ 1609.03676}}.
        
        \item C.~Ecker, D.~Grumiller, W.~van~der Schee and P.~Stanzer, ``Saturation of the Quantum Null Energy Condition in Far-From-Equilibrium Systems,'' \emph{Phys.~Rev.~D} \textbf{97} (2018), no.~12, 126016, \href{http://arXiv.org/abs/1710.09837}{\texttt{1710.09837}}.
        
        \item C.~Ecker, D.~Grumiller, W.~van~der Schee, M.~M.~Sheikh-Jabbari and P.~Stanzer, ``Quantum Null Energy Condition and its (non)saturation in 2dCFTs,'' \emph{\mbox{SciPost} Phys.} \textbf{6} (2019), no.~3, 036, \href{http://arXiv.org/abs/1901.04499}{\texttt{1901.04499}}.
    \end{enumerate}
    
    \paragraph{Submitted}
    \begin{enumerate}[label={[4]}]
        \item C.~Ecker, D.~Grumiller, H.~Soltanpanahi and P.~Stanzer, ``QNEC2 in deformed holographic CFTs,'' \emph{submitted to JHEP}, \href{http://arXiv.org/abs/2007.10367}{\texttt{ 2007.10367}}.
    \end{enumerate}

\cleardoublepage

\tableofcontents

\cleardoublepage

\section{Introduction}
 \label{sec:Intro}

Bringing together two apparently disjoint fields of research often resulted in great success and progress for both fields individually and science in general.
Historical and recent examples exist in abundance, within physics and even across disciplines.
Focussing on theoretical physics, the unification of \textit{electricity} and \textit{magnetism} by Maxwell in the 19th century is one of the most compelling examples, such that nowadays it is hard to imagine them as different fields.
\textit{Electromagnetism} became known as one of the four fundamental interactions.
In the 1960s it was realized, that electromagnetism and the \textit{weak interaction} can be described as two aspects of the same force.
Despite being very different at low energies, above the \textit{electroweak scale} of about $250$ GeV those two fundamental interactions unify into only one, described as \textit{electroweak theory} (EWT).
For this example of combining two fields of physics, in 1979 the Nobel prize was awarded to Glashow, Salam and Weinberg.
Similarly the third fundamental force, the \textit{strong interaction}, describing the interaction of quarks and gluons in \textit{quantum chromodynamics} (QCD) may serve as an example for the amazing results of combining different ideas in science.
By incorporating the \textit{Higgs mechanism} into EWT and combining with QCD, the \textit{standard model of particle physics} (SM) was established by Salam and Weinberg.
The fourth fundamental interaction, \textit{gravity} has not yet been successfully combined with the SM into a \textit{theory of everything} (TOE).
A promising candidate for such a theory is \textit{string theory}, originally developed to describe the strong interaction of subatomic particles in the 1960s.
After QCD proved to be the better model for that, string theory was repurposed and turned into a framework for a fundamental TOE.\\
\\
More recently, with the \textit{holographic principle} 't Hooft and Susskind \cite{Hooft:1993gx,Susskind:1995vu} created a way to relate gravitational theories to gauge theories of one dimension lower.
The most concrete realization of this duality was found by Maldacena in 1997 \cite{Maldacena:1997re}.
In the context of type IIB superstring theory he established a relation between \textit{anti-de Sitter} (AdS) space and \textit{conformal field theories} (CFT), known as the  \textit{AdS/CFT correspondence}.
This opened a whole new direction for research in theoretical physics and continues to provide new topics for investigation.
Impacts of this development reach into high energy physics \cite{Gubser:2009fc,CasalderreySolana:2011us}, quantum information \cite{Ryu:2006ef,Hubeny:2007xt}, condensed matter physics \cite{Hartnoll:2009sz,McGreevy:2009xe,Sachdev:2011wg,Iqbal:2011ae} and even neutron star physics \cite{Hoyos:2016zke} with a connection to gravitational wave research \cite{Ecker:2019xrw}.
Another important field of physics affected by this development is \textit{general relativity} (GR) and the search for a theory of \textit{quantum gravity}.\\
\\
The gauge/gravity duality can be used in two ways.
Firstly, the gauge theory can be used to study the properties of \textit{black holes} (BH) to gain insight into (quantum) gravity and string theory \cite{Harlow:2014yka,Anous:2016kss}.
The second approach, relevant in this work, is to use GR as a tool to study the dual field theory (in a certain limit) \cite{Erdmenger:2007cm,Gursoy:2010fj}.
When using a sophisticated theory like GR as a "tool", one will undoubtedly encounter a number of problems and limitations.
A prime example for this is the \textit{two body problem}, solved trivially in Newtonian gravity.
In GR on the other hand, even with supercomputers it was practically impossible to compute more than a single rotation until Pretorious' groundbreaking work in 2005 \cite{Pretorius2005}.
The governing equations of this theory, Einstein's field equations, are a set of non-linear coupled \textit{partial differential equations} (PDE) and can be solved analytically only in very few special cases (mostly in vacuum) with a lot of symmetries.
For even more complex problems it is no longer possible to do all the calculations by hand or even analytically using the aid of computer algebra systems.
At this point the connection to the field of computer science comes to help.
With improvements in hardware and software over the past decades coming in very handy, more and more problems in GR can be tackled \textit{numerically}.
The progress can be tracked from solving Einstein's equations numerically over half a century ago \cite{Hahn-Lindquist1964} to simulating the collision of two black holes in 2005 \cite{Pretorius2005}.
The latest progress led to the detection of gravitational waves via providing gravitational wave forms as templates for the large scale experiments at LIGO and VIRGO \cite{Abbott2016}.\\
\\
With holography and GR as tools to investigate quantum field theories (QFT) physicists made huge progress, especially with problems that can not be solved with known QFT methods, like perturbation theory or lattice calculations.
An excellent example is \textit{entanglement entropy} (EE) which is notoriously hard to calculate in QFT, but has a very simple description in the dual gravity theory \cite{Ryu:2006ef}.
EE is an intrinsically quantum observable, as it is a measure for the amount of entanglement between different parts of Hilbert space.
In this work, the partitioning of Hilbert space is a result of the spatial separation of the so-called \textit{entangling region} and its surrounding.
Using holography, EE has been studied in a huge number of settings and theories analytically as well as numerically, including toy models for \textit{heavy ion collisions} (HIC) \cite{Ecker:2016thn}.
Based on the success of applying holography to non-local observables like EE, we investigate a newly developed concept of QFT in this work, that we briefly outline in the next few paragraphs.\\
\\
The \textit{quantum null energy condition} (QNEC) was proposed by Bousso et al.~in 2015 \cite{Bousso:2015mna} and as the name suggests, is a condition on the \textit{energy momentum tensor} (EMT) of some QFT.
Similar to the \textit{classical energy conditions} (EC) which are fundamental for many applications of GR, it is supposed to be true in any reasonable, physically relevant QFT.\\
The classical ECs provide `physically reasonable' assumptions based on things like the positivity of energy or the speed of light as the upper limit for energy flow.
When they were introduced in the 1960's, one of the main motivations was to prove the existence of singularities \cite{Penrose:1964wq,Hawking:1965cc}, but the sheer amount of theorems in classical GR based on the ECs (e.g.~area theorems in BH mechanics) \cite{Hawking:1973,wald2010general,carroll2004spacetime,poisson2004relativist} shows how important they are.
Although it was known from the beginning, that some of them are violated in the presence of quantum fields \cite{Epstein:1965zza}, especially the \textit{null energy condition} (NEC) is still used quite frequently.
The NEC restricts the null projection of the EMT $T_{\mu\nu}$ to be positive
\begin{equation}
    T_{\mu\nu} k^\mu k^\nu \geq 0\,, \qquad \qquad k_\mu k^\mu = 0\,,
\end{equation}
where $k^\mu$ is a null vector.
Several attempts have been made to find consistent ECs for quantum theories \cite{Ford:1994bj,Martin-Moruno:2013wfa}, with QNEC not only the most recent amongst them, but also the only condition where the EMT is constrained locally.
QNEC relates the projection of the EMT $T_{\mu\nu}$ of some QFT along the null vector $k^\mu$ to the second variation of EE $S''$ along the same null vector (explained in detail in section \ref{sec:QNEC})
\begin{equation}
    \langle T_{\mu\nu} k^\mu k^\nu \rangle \geq  \frac{1}{2 \pi} S''\,.
\end{equation}
Soon after the conjecture was made, proofs for certain theories were found \cite{Bousso:2015mna,Bousso:2015wca,Koeller:2015qmn}.
One of them applies to field theories with holographic dual, opening the door for investigating QNEC in more complicated settings with the holographic tools mentioned above.
Of special interest is the highly non-trivial system modeling HICs, where energy densities can get negative, violating the classical ECs while QNEC holds.
We studied the properties of QNEC in this and many other QFTs in two and four dimensions \cite{Ecker:2017jdw,Ecker:2019ocp}.\\
\\
All these efforts described above led towards the goal of combining a theory of gravity (e.g.~Einstein's GR) with the standard model of particle physics, the most accurate theory ever developed.
This closes the loop from GR as a tool to understand QFT, back to solving questions that arise in the context of GR in order to make progress towards a theory of quantum gravity.
This search for a TOE, goes back to Einstein himself and other great physicists of this and the previous century.
In our work we use holography as a tool in order to investigate applications, consequences and properties of QNEC.\\
\\
This thesis is organized as follows.
After the introduction, a short review of the theoretical background is given in section \ref{sec:TheoBG}.
This includes GR and black holes, the ECs, EE and Holography.
In section \ref{sec:NumBG} the numerical background is presented, first numerical relativity and then the methods applied to calculate EE and QNEC.
Then QNEC is introduced in section \ref{sec:QNEC} and the available proofs presented.
Results in several systems in $4$-dimensional QFTs are presented in section \ref{sec:QNEC_D=4}.
Results for $2$-dimensional field theories are shown in section \ref{sec:QNEC_D=2}.
In section \ref{sec:End} there are some future directions besides a summary and a conclusion.
Appendix \ref{app:Application} shows how Einstein's equations are solved for the HIC toy model and appendix \ref{app:PnP} explains how EE and QNEC can be tackled by paper and pencil.

\cleardoublepage

\section{Theoretical Background}
 \label{sec:TheoBG}

In this section I will review the four theoretical ingredients of this thesis, general relativity, energy conditions, entanglement entropy and holography.
Since QNEC is the core of this work, a separate chapter is dedicated to it later on.
The aim is to keep the following sections self-contained, such that they can be read or skipped individually, depending on the readers choice.

\subsection{General Relativity and Black Holes}
 \label{sec:GR_BH}

More than a hundred years ago Einstein published his theory of gravity and introduced a completely new concept of space and time.
Treating space and time as dynamical entities led to a geometric description of gravity.
He postulated the governing equations, a set of coupled non-linear second order PDEs for the spacetime metric to describe the movement of matter due to curvature of spacetime and vice versa
\begin{align}
 G_{\mu\nu} = 8 \pi \, T_{\mu\nu}\,. \label{eq:Einsteinequations}
\end{align}
The Einstein tensor $G_{\mu\nu}$ describes the geometry of spacetime and the EMT $T_{\mu\nu}$ contains all information about the matter content (see section \ref{sec:EinsteinEqs_Vacuum} below for details).
Only a few weeks after this groundbreaking publication, Schwarzschild found the first analytic solution under the assumptions of spherical symmetry and vacuum, i.e.~the absence of matter ($T_{\mu\nu}\!=\!0$).
This solution describes the geometry around a central mass and already contains two interesting features,
\begin{itemize}
 \item a coordinate singularity,
 \item a spacetime singularity.
\end{itemize}
As the name suggest, the coordinate singularity can be overcome by choosing a different set of coordinates and nothing special happens to observers reaching this location.
Looking at the causal structure of the Schwarzschild solution, one realizes that there exists a surface, coinciding with the coordinate singularity, that can only be crossed in one direction by objects moving at the speed of light or slower.
This surface is called \textit{event horizon} of a black hole, as not even light manages to escape from its inside.
The spacetime singularity is located at the center of the Schwarzschild solution.
In this case it is not possible to circumvent the problems arising, because in contrast to the coordinate singularity, the curvature of spacetime becomes infinitely large.
Later on, further BH-solutions to \eqref{eq:Einsteinequations} were found, including charge and angular momentum as well as multiple black holes.
Especially in the second half of the past century, black holes were investigated by a lot of well known physicists like
 Bekenstein, Hawking and Penrose who found several properties, theorems and conjectures like:
\begin{itemize}
 \item black holes do have entropy (BH-thermodynamics) \cite{Bekenstein:1973ur},
 \item black holes do emit radiation and evaporate (information paradox) \cite{Hawking:2005kf},
 \item there are no naked singularities (cosmic censorship) \cite{Penrose:1979}.
\end{itemize}
Black holes are of central interest in this thesis for two reasons.
On one hand, they are part of the reason why ECs were introduced. See section \ref{sec:EC} for a detailed discussion of the ECs.
Learning about quantum versions of the ECs is necessary for the description of black holes in a theory of quantum gravity.
On the other hand in holography (see section \ref{sec:Holo}) black holes are dual to a thermal state in the corresponding field theory.
Since we are interested in such states (among others), we have to deal with black holes in our holographic calculations.

\subsubsection{Einstein's Equations and Vacuum Solutions}
 \label{sec:EinsteinEqs_Vacuum}

The Einstein tensor used in equation \eqref{eq:Einsteinequations} contains the spacetime metric $g_{\mu\nu}$ and combinations of its derivatives, called the Riemann tensor $R^\tau_{\,\,\mu\sigma\nu}$, the Ricci tensor $R_{\mu\nu}\!=\!R^\tau_{\,\,\mu\tau\nu}$ and the Ricci scalar $R\!=\!R^\mu_{\,\,\mu}$
\begin{equation}
    G_{\mu\nu} = R_{\mu\nu} - \frac{1}{2} \, g_{\mu\nu} \, R\,,
\end{equation}
where the Riemann tensor is defined via the Christoffel symbols $\Gamma^\rho_{\mu\nu}$
\begin{align} \label{eq:RGamma}
    R^\rho_{\,\,\sigma\mu\nu} &= \partial_\mu \Gamma^\rho_{\,\,\nu\sigma} - \partial_\nu \Gamma^\rho_{\,\,\mu\sigma} + \Gamma^\rho_{\,\,\mu\lambda}\Gamma^\lambda_{\,\,\nu\sigma} - \Gamma^\rho_{\,\,\nu\lambda}\Gamma^\lambda_{\,\,\mu\sigma}\,,\\ 
    \Gamma^\rho_{\,\,\mu\nu} &= \frac{1}{2} g^{\rho\sigma} (\partial_\nu g_{\sigma\mu} + \partial_\mu g_{\sigma\nu}-\partial_\sigma g_{\mu\nu})\,.
\end{align}
For the signature of the metric we use $(-,+,...,+)$ throughout this work.\\
\\
A possible addition to Einstein's equation \eqref{eq:Einsteinequations} is the cosmological constant $\Lambda$, which is sometimes referred to as part of the geometry, but other times seen as a constant energy density and therefore part of the matter content.\\
Given some EMT, solving Einstein's equations will give a spacetime metric, describing the suitable geometry for the matter content encoded by the EMT.
In this section we will keep the cosmological constant and consider only vacuum solutions where $T_{\mu\nu}\!=\!0$
\begin{align}
    G_{\mu\nu} + \Lambda \, g_{\mu\nu} = 0\,. \label{eq:EEQ_vac}
\end{align}
In general there are three different cases for the value of the cosmological constant.
Firstly it can be zero ($\Lambda\!=\!0$).
In this scenario the obvious solution to Einstein's equations is a constant metric with vanishing Riemann tensor (and therefore the Ricci tensor and scalar are also vanishing).
The second case is a positive cosmological constant, $\Lambda\!>\!0$, which has the so-called \textit{de Sitter} space as solution.
It can be viewed as a hyperboloid satisfying
\begin{equation}
    -x_0^2 + \sum x_i^2 = L^2\,,
\end{equation}
where $L$ is a positive constant with dimension of length, the so-called de Sitter radius.
The curvature scalar $R$ of spacetime is a positive constant and related to the de Sitter radius and the cosmological constant via
\begin{equation}
    R = \frac{2 d}{d-2} \Lambda, \qquad \Lambda = \frac{(d-1)(d-2)}{2 L^2}\,,
\end{equation}
where $d$ is the dimension of spacetime.\\
\\
Of greater interest for this thesis is the third option, a negative cosmological constant ($\Lambda\!<\!0$), leading to the so-called anti-de Sitter solution.
This spacetime can be viewed as a pseudo-sphere satisfying
\begin{equation}
    -x_0^2 + \sum x_i^2 = -L^2\,,
\end{equation}
where $L$ is a positive constant with dimension of length, the so-called AdS radius.
The curvature scalar $R$ of spacetime is a negative constant and related to the AdS radius and the cosmological constant via
\begin{equation}
    R = \frac{2 d}{d-2} \Lambda, \qquad \Lambda = \frac{-(d-1)(d-2)}{2 L^2}\,,
\end{equation}
where $d$ is the dimension of spacetime again.
This geometry can be expressed via different coordinate patches.
The most important in this work are the Poincar\'e patch
\begin{equation}
     \mathrm{d}s^2 = \frac{L^2}{r^2} \, \mathrm{d}r^2 + \frac{r^2}{L^2} \, \eta_{\mu\nu} \, \mathrm{d}x^\mu \, \mathrm{d}x^\nu\,,
\end{equation}
where $\eta_{\mu\nu}$ is the Minkowski metric, and global AdS
\begin{equation} \label{eq:globalAdS}
    \mathrm{d}s^2 = L^2 \left( -\cosh^2 \! \rho \, \mathrm{d}\tau^2 + \mathrm{d}\rho^2 + \sinh^2 \! \rho \, \mathrm{d}\Omega^2_{d-2}\right)\,,
\end{equation}
where $\mathrm{d}\Omega^2_{d-2}$ is the metric of the round unit sphere in $d\!-\!2$ dimensions.\\
\\
Often we are interested in geometries that approach the AdS solution asymptotically close to the boundary at $r\!=\!\infty$ ($\rho\!=\!\infty$), but differ in the bulk.
For example the AdS-Schwarzschild solution has a black hole in the center and approaches AdS at the boundary, while still being a vacuum solution to Einstein's equations.

\subsubsection{Bulk Action}
 \label{sec:HamiltonianApproach}
 
If one considers systems with matter content, it is often more convenient to start from the action principle to obtain the equations of motion.
The Einstein-Hilbert action covers the geometrical part
\begin{equation}
    S_{EH} = \frac{1}{2 \kappa} \int \mathrm{d}^dx \sqrt{-g} \, \left( R + 2 \Lambda \right)\,, \label{eq:EHAction}
\end{equation}
the matter part is covered by an appropriate Lagrangian, e.g.~a scalar field with potential $V(\phi)$
\begin{equation}
    S_M = \int \mathrm{d}^dx \sqrt{-g } \, \, \mathcal{L} = - \int \mathrm{d}^dx \sqrt{-g } \left( \frac{1}{2}(\partial \phi)^2 + V(\phi) \right)\,, \label{eq:matterAction}
\end{equation}
and the Gibbons-Hawking-York boundary term $S_{GHY}$ and possibly counter-terms for renormalizability $S_{ct}$ need to be considered as well.
Variation of this action with respect to the metric leads to Einstein's equations \eqref{eq:Einsteinequations}, while variation with respect to the scalar field gives the equation of motion for the scalar field, e.g.~the Klein-Gordon equation.
In general, solving this set of coupled differential equations is not possible with analytical methods and hence \textit{numerical relativity} is needed (see section \ref{sec:NumRel}).
Amongst others this has been done for massless and massive scalar and/or vector fields with various initial distributions.\\
\\
Of course solutions of this set of equations, given by \eqref{eq:Einsteinequations} combined with the equation(s) of motion for the matter field(s), can still be asymptotically AdS and therefore relevant and interesting for this work.

\subsubsection{General Relativity in Lower Dimensions}

Since GR is a very complicated and computation intense field, it is useful to find simplifications to make calculations feasible/possible.
One approach is to study systems with a high degree of symmetry, like spherical-, rotational- or translation symmetry, to study homogeneous and isotropic systems.
A different approach is to study GR in a different number of dimensions.
Since perturbative calculations have a long tradition in physics, one possibility is to take the limit of an infinite number of dimensions and then use $\frac{1}{D}$ as perturbative parameter \cite{Emparan:2020inr}.\\
\\
More relevant for the work at hand is the approach of going to lower dimensions.
The lowest spacetime dimension where Einstein gravity exists is three, so we shall focus on three spacetime dimensions when discussing lower-dimensional examples.
An added bonus of this is that the dual QFT (if there exists one) is two-dimensional, and a lot more is known about two-dimensional QFTs as compared to their higher-dimensional counterparts.
This gives lower-dimensional gravity and holography a high degree of analytic control, allowing to tackle problems that are hard to address in higher dimensions.
The analogue of the Kerr solution was found by Ba\~nados, Teitelboim and Zanelli (BTZ) \cite{Banados:1992wn,Banados:1992gq},
\begin{equation}
    \mathrm{d}s^2 = - \frac{\left( r^2 - r_+^2 \right)\left( r^2 - r_-^2 \right)}{L^2r^2} \mathrm{d}t^2 + \frac{L^2r^2\,\mathrm{d}r^2}{\left( r^2 - r_+^2 \right)\left( r^2 - r_-^2 \right)} + r^2 \left( \mathrm{d}\phi - \frac{r_+r_-}{Lr^2} \mathrm{d}t \right)^2\,.
\end{equation}
The BTZ black hole is locally equivalent to AdS, but still is a BH since it has an event horizon.

\subsection{Energy Conditions}
 \label{sec:EC}

Since QNEC is the central topic of this thesis, it is crucial to understand the whole story of ECs, from a classical and a quantum point of view.
We will see that there is a need for a general quantum EC, which provides the main motivation behind this work, as mentioned in the introduction.
For these two reasons, this part of the theoretical background will be a bit more elaborate than the other ones.
While writing this thesis, an excellent review article was published \cite{Kontou:2020bta} with many references for diving even deeper into the matter.\\
\\
In general relativity the field equations allow for arbitrary forms of matter, even if this results in situations considered unphysical by common sense, like closed timelike curves or wormholes.
ECs not only pose restrictions on these cases, but also provide mathematical convexity conditions that are turned into further convexity conditions by various theorems to capture physical properties like positivity of energy and attractiveness of gravity.

\subsubsection{What are Energy Conditions?}

The ECs were introduced as coordinate independent restrictions on the EMT in the 1960s \cite{Hawking:1973,wald2010general,carroll2004spacetime,poisson2004relativist}.
Since restrictions on the components of an arbitrary EMT are not intuitively accessible, Hawking and Ellis introduced a classification into four different types of EMTs.
Two of them being physical, namely ordinary matter as \textit{type I} (containing the perfect fluid) and \textit{type II} which also allows for the vacuum energy, leaving the other \textit{types III} and \textit{IV} to systems that are not realized in nature.
We will briefly introduce the four most common classical ECs and their interpretation using the perfect fluid type of EMT as example
\begin{equation}
    T_{\mu\nu} = \mathrm{diag}(\rho,p_1,...,p_d)\,. \label{eq:EMTperfectfluid}
\end{equation}
\paragraph{The Null Energy Condition (NEC)}
uses a null vector to project the EMT, achieving a coordinate invariant expression and demands this projection to be positive
\begin{equation} \label{eq:NEC}
    T_{\mu\nu} k^\mu k^\nu \geq 0\,, \qquad \qquad k_\mu k^\mu = 0\,.
\end{equation}
Interpreting this condition, using the EMT for a perfect fluid \eqref{eq:EMTperfectfluid}, this amounts to the sum of energy density $\rho$ and a single pressure component $p_i$ to be positive, allowing them individually to be negative though
\begin{equation}
    \rho + p_i \geq 0\,.
\end{equation}
\paragraph{The Weak Energy Condition (WEC)}
considers timelike vectors for the projection, demanding its positivity again
\begin{equation}
    T_{\mu\nu} t^\mu t^\nu \geq 0\,, \qquad \qquad t_\mu t^\mu < 0\,.
\end{equation}
For the perfect fluid this demands the positivity of the energy density in addition to the NEC
\begin{equation}
\rho \geq 0\,, \qquad \qquad \rho + p_i \geq 0\,.
\end{equation}
\paragraph{The Dominant Energy Condition (DEC)}
uses timelike vectors again, but demands not only the projection to be positive but also the energy flow to be non-spacelike
\begin{equation}
    T_{\mu\nu} t^\mu t^\nu \geq 0\,, \qquad \qquad (T_{\mu\nu} t^\mu)^2 \leq 0, \qquad \qquad t_\mu t^\mu < 0\,.
\end{equation}
For a perfect fluid this amounts to
\begin{equation}
    \rho \geq |p_i|\,,
\end{equation}
while implying the WEC (and therefore also the NEC).

\paragraph{The Strong Energy Condition (SEC)}
uses timelike vectors again, but employs the trace of the EMT as bound for the projection
\begin{equation}
    T_{\mu\nu} t^\mu t^\nu \geq T_{\,\,\mu}^\mu t_\nu t^\nu\,, \qquad \qquad t_\mu t^\mu < 0\,.
\end{equation}
For a perfect fluid this amounts to the pressure not dominating the energy density, implying only the NEC but not the WEC
\begin{equation}
    \rho + p_i \geq 0\,, \qquad \qquad \rho + \sum p_i \geq 0\,.
\end{equation}
All these conditions are `pointwise', meaning they are local conditions on the EMT.
Giving up this property, it is possible to write down averaged ECs, described later in section \ref{sec:ECsandQT}.

\subsubsection{What are Energy Conditions Good For?}

After having introduced all these possible conditions, some used more often than others, the natural question of their applications arises.
Mainly they provide an assumption for general proofs and theorems in classical GR.
This gives the advantage that it is not necessary to know the EMT explicitly, having to repeat the proof for every possible matter content of the theory.
Using the field equations of your gravitational theory of choice allows to translate the ECs from restrictions on the EMT to restrictions on the geometry.
We restrict ourselves to classical Einstein gravity and therefore the ECs can be translated to restrictions on the Ricci tensor, using Einstein's equations \eqref{eq:Einsteinequations}.
A simple example is the so-called \textit{null curvature condition} (NCC), given by combining \eqref{eq:NEC} and \eqref{eq:Einsteinequations}
\begin{equation} \label{eq:NCC}
    0 \leq 8\pi \, T_{\mu\nu} k^\mu k^\nu = \left( R_{\mu\nu} - \frac{1}{2} \, g_{\mu\nu} \, R \right) k^\mu k^\nu = R_{\mu\nu} k^\mu k^\nu\,.
\end{equation}
These conditions are then used in the Raychaudhuri equation for the focussing/ex\-pan\-sion $\theta$ of a congruence of null geodesic with tangent vector $k^\mu$
\begin{equation} \label{eq:Raychaudhuri}
    \dot \theta = \omega^2 - \sigma^2 - \frac{\theta^2}{D-2} - R_{\mu\nu} k^\mu k^\nu\,,
\end{equation}
where the dot denotes derivatives w.r.t.~an affine parameter $\tau$, $\omega$ is the twist, $\sigma$ the shear of the geodesic congruence and $D$ is the dimension of spacetime (for the precise definitions see e.g.~\cite{wald2010general}).
Similarly this works for the other ECs as well.
The prime example for their application are the singularity theorems by Hawking and Penrose \cite{Penrose:1964wq,Hawking:1971vc}, but also the laws of BH thermodynamics \cite{Hawking:1971tu}.
Further applications are no-go theorems concerning spacetimes with wormholes and things like warp drives or time travel \cite{Hawking:1991nk,Morris:1988tu}.

\paragraph{Example: Singularity Theorems.} \label{subsec:SingTheo}
The big bang or the center of the Schwarz\-schild solution are prime examples of singularities.
These singularities are based on perfect symmetry, isotropy and homogeneity.
That this might be a problem can be seen from Newtonian gravity.
In this theory of gravity, a singularity can only be formed from collapsing matter if there is no rotation or perturbation present.
The singularity theorems of GR show that singularities are not an artifact, but a generic, unavoidable feature of certain spacetimes.
We will outline briefly how these singularity theorems work, but refer the reader to the literature (e.g.~\cite{wald2010general}) for a full proof.\\
\\
In general, singularity theorems rely on some convexity condition, some trapping condition and certain properties of the geometry.
Convexity conditions are provided by the ECs like in \eqref{eq:NCC} and combined with the geometric assumption of a twist-free geodesic congruence, Rauchaudhuri's equation \eqref{eq:Raychaudhuri} can be reduced to the inequality
\begin{equation}
    \dot \theta \geq -\frac{\theta^2}{D-2}\,.
\end{equation}
The trapping condition ensures that the expansion $\theta$ is negative at some point $\tau_0$ along the congruence.
Integration from $\tau_0$ to some $\tau_1$ yields
\begin{equation}
    -\frac{1}{\theta_1} \geq -\frac{1}{\theta_0} - \frac{\tau_1 -\tau_0}{D-2}\,,
\end{equation}
where we can see that the right hand side has a zero at finite $\tau_1$ and $\theta_0$ being the negative expansion at $\tau_0$.
This means that $\theta_1$ tends to $-\infty$, since the zero on the r.h.s.~is approached from above.
Infinitely negative expansion within proper time means that the geodesic congruence collapses to a point and indicates a singularity.

\subsubsection{Problems with Energy Conditions}

There are several things about the ECs being criticized for a good reason.
One of them is their arbitrary character, being invented by someone as physically `reasonable assumption', to provide mathematical assumptions required in the proofs of the attractiveness of gravity, a lower bound to energy and the existence of singularities.
Indeed the first appearance happened to be exactly that way.
Further criticism concerns the fact that they are solely classical physics.
In fact all of the ECs mentioned so far are violated easily, taking semi-classical or quantum effects into account.
Among those violations are the Casimir effect, quantum vacuum states near black hole horizons, squeezed vacuum states and also Hawking radiation to name only a few.
Even worse, most of them can be violated by classical matter fields, that are not even exotic.
In the following we want to give some examples of how to violate the ECs most commonly used in physics.

\paragraph{SEC violation:}
The SEC has been used in many proofs similar to the one described in section \ref{subsec:SingTheo}.
In spite of its popularity, the SEC is also the easiest to violate among all relevant ECs, even without the need of quantum effects.
One of the examples is the expanding universe.
The SEC does not allow for a positive cosmological constant and also excludes inflationary theories.
The second example to violate the SEC is simply a free (non-) minimally coupled massive scalar field like the Higgs.
In addition to those violations, there are the semi-classical and quantum effects mentioned above.

\paragraph{DEC violation:}
Assuming a Freedman-Lema\^itre-Robertson-Walker metric, the dominant EC does not allow for a static universe.
Further, if the universe is expanding (as we know it is), the DEC predicts that the rate of expansion slows down \cite{ludvigsen1999general}.
This contradicts observations.
Similarly a negative cosmological constant is excluded as well by the DEC.
From the matter point of view, non-minimally coupled scalar fields violate this condition as well.
In addition to those violations, there are the semi-classical and quantum effects mentioned above.

\paragraph{WEC violation:}
The WEC is more useful since the only classical violations are a negative cosmological constant and again the non-minimally coupled scalar field.
Many of the singularity theorems have been modified such that the assumption of the WEC replaces the SEC.
But once more, there are the semi-classical and quantum effects mentioned above, violating the WEC.

\paragraph{NEC violation:}
Within the framework of classical GR, the most universal of the local ECs is the NEC.
It is only violated by non-minimally coupled scalar fields.
The NEC is used in most modern versions of all the important theorems and proofs of GR, replacing the other ECs.
But also NEC is prone to violations by the semi-classical and quantum effects mentioned above.

\paragraph*{Personal note:}
In my opinion these issues with the violation of ECs did not receive enough attention over the past decades. Almost all textbooks on GR ignore the fact that there are violations known that affect all ECs and subsequently all proofs recited in the books.
Good coverage of this topic can be found in Matt Vissers book from 1994 \cite{visser1996lorentzian} and his contribution together with Martin-Moruno to a new book by Lobo \cite{Martin-Moruno:2017exc,Lobo:2017oab} as well as a recent review article by Kontou \cite{Kontou:2020bta}.

\subsubsection{Energy Conditions in the Quantum World}
 \label{sec:ECsandQT}

In view of all these problems with the classical pointwise ECs, there have been several attempts to improve or rescue the concept of ECs.

\paragraph{Averaged Energy Conditions:}
With negative energy densities and the Casimir effect in mind, one option is to allow these negative values locally, but demand that on average the energy density is still positive.
This leads us to the idea of integrating the pointwise ECs along some curve in spacetime, demanding the result of this integral to be positive.
This can be done easily for the NEC and WEC, but is more complicated (while still possible) for the DEC and SEC.
We will use only the \textit{averaged null energy condition} (ANEC), since it can be obtained as limit of the other conditions and provides the weakest assumption
\begin{equation}
    \int_\gamma T_{\mu\nu} k^\mu k^\nu \mathrm{d}\lambda \geq 0\,.
\end{equation}
The integral is taken along some null geodesic $\gamma$ with $k^\mu$ being its tangent vector.
Further there exist proofs for the ANEC to be true in any unitary quantum field theory on Minkowski space \cite{Faulkner:2016mzt,Hartman:2016lgu}.
Given that, the ANEC is a very powerful tool to be used for further generalizing the singularity theorems and other proofs in GR.
Actually most theorems can be modified so that only ANEC has to be assumed as convexity condition.
The drawback of this kind of averaging is that energy densities can become arbitrarily large and negative in some region.
A related criticism is that ANEC is a non-local condition, while intuitively one may expect the stress tensor to be bounded locally.
We shall see in section \ref{sec:QNEC} that this intuition is justified.

\paragraph{Quantum Inequalities:}
A similar approach was taken by Roman and Ford when they introduced quantum (energy) inequalities (QI) in the 90s \cite{Ford:1994bj}.
In spite of taking the average of some EC over a certain region in spacetime, they wanted to restrict the amount of violation to the classical EC.
This is realized by limiting the average to a small region (contrary to the whole geodesic of ANEC) by adding some test function $\phi(\lambda)$ with compact support to the integral
\begin{equation}
    \int_\gamma T_{\mu\nu} k^\mu k^\nu \phi(\lambda) \mathrm{d}\lambda \geq 0\,.
\end{equation}
Now the negative energy density must be compensated for by positive energy density nearby (and not at infinite future for instance).
A similar concept is their \textit{quantum interest conjecture} \cite{Ford:1999qv}.

\paragraph{Semiclassical Energy Conditions:}
A very different approach to making the ECs compatible with quantum physics was taken by Martin-Moruno and Visser in 2013 \cite{Martin-Moruno:2017exc}.
They keep the pointwise character of the ECs and introduce quantum corrections to the right-hand side of the inequalities.
As an example we look at the \textit{semiclassical WEC}
\begin{equation}
    \langle T_{\mu\nu} t^\mu t^\nu \rangle \geq - \zeta \frac{\hbar N}{L^4} (u^\lambda t_\lambda)^2\,,
\end{equation}
where $\zeta$ is a constant of order unity, $N$ is the number of fields in the theory in question, $L$ is a system dependent constant and $u^\lambda$ is the systems four-velocity.
Considering the Casimir effect, $L$ would be related to the distance of the parallel plates.
While this approach works very well in terms of quantifying or limiting the amount of violation of the classical ECs, it contains a number of arbitrary system-dependent constants.
For this reason there is no way to find a general version or proof for such semiclassical ECs.\\
\\
All of these approaches have their advantages and drawbacks, but none of them is fully satisfying.
The only known possibility overcoming all problematic issues mentioned above is the QNEC, proposed by Bousso et.~al in 2015 \cite{Bousso:2015mna}, described in detail in section \ref{sec:QNEC}.

\subsection{Entanglement Entropy}
 \label{sec:EE}

In classical systems the thermodynamic entropy approaches zero as the temperature goes to zero as well (although this can never be reached in finite amount of time).
In quantum theory systems at zero temperature can still possess entropy originating from their entanglement.
In some sense this EE gives a measure of `how much quantum a system is'.
Originating in the field of quantum information theory, there exist a number of other quantities or observables like mutual information or Renyi entropy as well, which measure how strong the quantum nature of a system is \cite{Plenio:2007zz}.\\
\\
EE contributes to the understanding of quantum physical phenomena in many fields of physics.
For example in condensed matter physics quantum phase transitions cannot be described with classical quantities, as they do not take entanglement into account.
Here EE is a candidate for an order parameter, distinguishing different phases.
Some of these systems at quantum critical points can be described with conformal field theories.\\
\\
Of special interest for this thesis is the holographic description of EE.
In the context of the AdS/CFT correspondence, an especially simple method of calculating EE in situations where it is not accessible any other way, was found by Ryu and Takayanagi \cite{Ryu:2006ef}.
Coming back to this after introducing holography in general in the next section, we will now introduce the concept and definition of EE alongside some examples and properties.

\subsubsection{Definition}

\begin{figure}
	\begin{center}
	\includegraphics[height=.23\textheight]{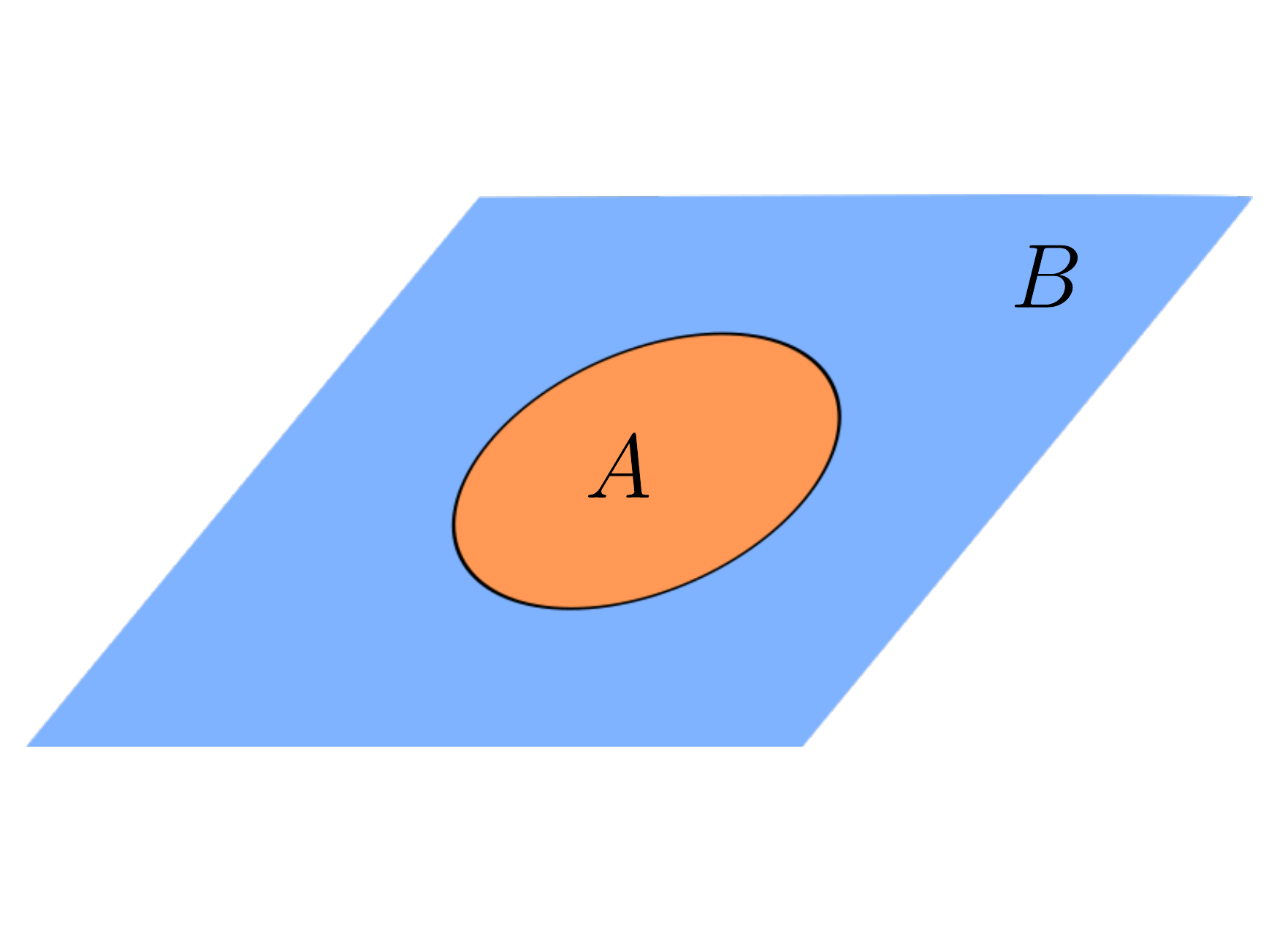}\includegraphics[height=.23\textheight]{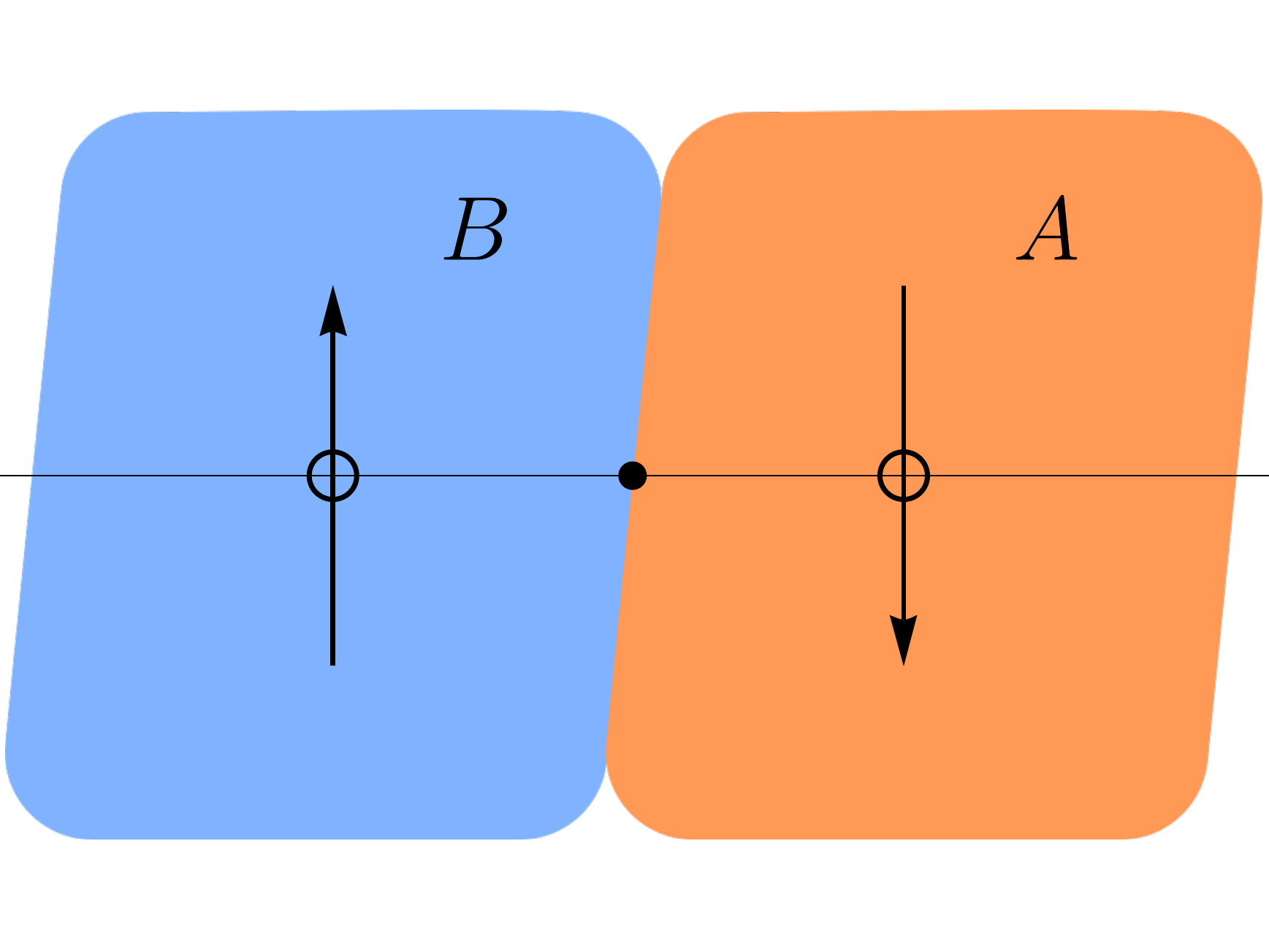}
	\caption{Left: Dividing a system into the entangling region $A$ and its surrounding $B$. The boundary of the entangling region $\partial A$ (called \textit{entangling surface}) is shown in black.
	Right: Applying the division to a spin chain of length two. The entangling region $A$ contains one spin and the surrounding $B$ the other one, while the entangling surface is highlighted by the black dot.}
	\label{fig:EE_def}
	\end{center}
\end{figure}

Consider a quantum mechanical system or a QFT at zero temperature, which is described by the pure ground state $\lvert \Psi \rangle$.
For a non-degenerate wave function the density matrix is given by
\begin{align}
 \rho_{tot} = \lvert \Psi \rangle \langle \Psi \rvert\,,
\end{align}
and the von Neumann entropy is zero
\begin{align}
 S_{tot} = - \mathrm{tr}\left( \rho_{tot} \, \, \log \rho_{tot} \right) = 0\,.
\end{align}
If we divide this system into two subsystems $A$ and $B$ (see figure \ref{fig:EE_def}), we can write the total Hilbert space as direct product of the two Hilbert spaces of the subsystems $A$ and $B$
\begin{align}
 \mathcal{H}_{tot} = \mathcal{H}_A \otimes \mathcal{H}_B\,.
\end{align}
An observer in \textit{entangling region} $A$ without access to its surrounding $B$ computes observables with the reduced density matrix $\rho_A$
\begin{align}
 \rho_A = \mathrm{tr}_B \left( \rho_{tot} \right)\,,
\end{align}
where the trace is taken over the subspace $\mathcal{H}_B$ only.
The EE of the subsystem $A$ is defined as the von Neumann entropy of the reduced density matrix $\rho_A$
\begin{align} \label{eq:EEdef}
 S_A = - \mathrm{tr}_A \left( \rho_A \, \, \log \rho_A \right)\,.
\end{align}
This quantity provides a measure for how strongly entangled the state $\lvert \Psi \rangle$ is with respect to partitioning into subsystems $A$ and $B$.

\paragraph{Example: two spin system.}
A very simple case to demonstrate this is a spin chain with only two sites for simplicity.
As shown in figure \ref{fig:EE_def} the chain can be split into two subsystems consisting of one site with two spin values each $\mathcal{H}_A\!=\!\{|\uparrow\rangle_A, |\downarrow\rangle_A\}, \mathcal{H}_B\!=\!\{|\uparrow\rangle_B, |\downarrow\rangle_B\}$.
A possible state is then given by
\begin{equation}
    |\psi\rangle = \frac{1}{2} \left( |\uparrow\rangle_A + |\downarrow\rangle_A \right) \otimes \left( |\uparrow\rangle_B +|\downarrow\rangle_B \right)\,,
\end{equation}
with density matrix
\begin{equation}
    \rho = |\psi\rangle \langle\psi| = \frac{1}{4} \left( |\uparrow\rangle_A + |\downarrow\rangle_A \right)\left( \langle\uparrow|_A + \langle\downarrow|_A \right) \otimes \left( |\uparrow\rangle_B +|\downarrow\rangle_B \right) \left( \langle\uparrow|_B +\langle\downarrow|_B \right)\,.
\end{equation}
To calculate the entanglement entropy of subsystem $A$ we need the reduced density matrix
\begin{equation}
    \rho_A = \mathrm{Tr}_B |\psi\rangle \langle\psi| = \frac{1}{2} \left( |\uparrow\rangle_A + |\downarrow\rangle_A \right)\left( \langle\uparrow|_A + \langle\downarrow|_A \right)\,.
\end{equation}
From this we can see that $\rho_A\!=\!\rho_A^2$ and know that we are dealing with a pure product state with vanishing entanglement entropy
\begin{equation}
    S_A = - \mathrm{Tr}\left[\rho_A \ln(\rho_A)\right] = - \mathrm{Tr}\left[\rho_A^2 \ln(\rho_A^2)\right] = - \mathrm{Tr}\left[2 \rho_A \ln(\rho_A)\right] = 0\,.
\end{equation}
Repeating this calculation for a Bell state, which is a maximally mixed one
\begin{equation}
    |\psi\rangle = \frac{1}{\sqrt{2}} \left( |\uparrow\rangle_A \otimes |\uparrow\rangle_B + |\downarrow\rangle_A \otimes |\downarrow\rangle_B \right)\,,
\end{equation}
leads to the following density matrix
\begin{equation}
    \rho = |\psi\rangle \langle\psi| = \frac{1}{2} \left( |\uparrow\rangle_A \otimes |\uparrow\rangle_B + |\downarrow\rangle_A \otimes |\downarrow\rangle_B \right) \left( \langle\uparrow|_A \otimes \langle\uparrow|_B + \langle\downarrow|_A \otimes \langle\downarrow|_B \right)\,.
\end{equation}
Taking the trace over subregion $B$ once more leaves us with the reduced density matrix
\begin{equation}
    \rho_A = \mathrm{Tr}_B |\psi\rangle \langle\psi| = \frac{1}{2} \left( |\uparrow\rangle_A \langle\uparrow|_A + |\downarrow\rangle_A \langle\downarrow|_A\right)\,.
\end{equation}
Calculating the EE using the matrix representation of $\rho_A\!=\!\mathrm{diag}(\tfrac{1}{2},\tfrac{1}{2})$ leads to one e-bit of EE
\begin{equation}
    S_A = - \mathrm{Tr}\left[\rho_A \ln(\rho_A)\right] = \ln(2)\,.
\end{equation}

\subsubsection{Properties of Entanglement Entropy}

As shown by the two examples in the previous section, EE has the following properties with respect to the states considered:
\begin{itemize}
    \item $S_A\!=\!0$\\
    for any pure state (i.e.~if the state can be written as product of components of the two Hilbert spaces),
    \item $\mathrm{max}(S_A)\!=\!\ln [\mathrm{dim}(\mathcal{H}_A)]$\\
    is the upper bound to entanglement entropy in a given system. This bound is only reached for maximally entangled states like the Bell state.
\end{itemize}
In addition to that, EE is invariant under any unitary change of basis $S_A(\rho_A)\!=\!S_A(U^\dagger \rho_A U)$ and satisfies the following inequality for systems with two entangling regions $A$ and $B$, called \textit{strong subadditivity}
\begin{equation}
    S_A + S_B \geq S_{A \cup B} + S_{A \cap B}\,.
\end{equation}

\subsubsection{Entanglement Entropy in Quantum Field Theory}

Calculating EE directly from the density matrix in a field theory is an almost impossible task.
A more feasible method is called the replica trick and can be applied in many different CFTs.
The replica trick employs the \textit{Renyi entropy}
\begin{equation} \label{eq:RenyiEntropy}
    S^{Renyi}_n = \frac{\ln[\mathrm{Tr}_A \rho^n_A]}{1-n}\,,
\end{equation}
involving powers of the density matrix.
This simplifies the calculation of EE a bit, since powers are much easier to calculate than the logarithm.\\
\\
The $n$-th power of the density matrix can be expressed via the path integral over a manifold obtained by glueing together $n$ copies of the original spacetime, generating a $n$-sheeted Riemann surface.
This way one can express the \textit{Renyi entropies} via the path integral and after analytic continuation and taking the limit $n\!\to\!1$, it results in the EE
\begin{align}
    &S_A = \lim_{n \to 1} S^{Renyi}_n = \lim_{n \to 1} \frac{\ln[\mathrm{Tr}_A \rho^n_A]}{1-n} = -\partial_n \ln[\mathrm{Tr}_A \rho^n_A] |_{n=1} = \ln[\rho_A \mathrm{Tr}_A \rho_A]\,.
\end{align}
Using this method, it is possible to calculate the EE in a $2$-dimensional CFT on a plane \cite{Holzhey:1994we,Calabrese:2004eu}
\begin{align}
 S_A = \frac{c}{3} \, \log \! \left( \frac{\ell}{\epsilon} \right)\,, \label{eq:EE_CFT2}
\end{align}
where $c$ is the central charge of the CFT and $\ell$ is the length of the entangling region.
To regulate the result an ultraviolet (UV) cutoff $\epsilon$ is introduced, providing a minimal length scale that is taken into account.
In the limit $\epsilon\!\to\!0$, all short range correlations across the boundary of the entangling region contribute and the EE diverges $S_A\!\to\!\infty$.
Since they used the enhanced symmetries in $2$-dimensional CFTs, it is still notoriously hard to calculate EE in higher dimensions directly from field theory.
An extension of the relativistic CFT result to Galilean CFT is presented in \cite{Bagchi:2014iea}.

\subsection{Holography}
 \label{sec:Holo}

I will not attempt to give an introduction to string theory at this point, but rather quote the necessary arguments and formulas and refer the reader to excellent books, explaining the connection between holography and string theory \cite{CasalderreySolana:2011us,Ammon:2015wua}.

\subsubsection{Holographic Principle and AdS/CFT Correspondence}

Inspired by the formula for the entropy of black holes found by Bekenstein and Hawking \cite{Bekenstein:1973ur}, which does not scale with its volume but with the area of the horizon
\begin{equation}
    S_{BH} = \frac{A}{4 G_N \hbar},
\end{equation}
’t Hooft \cite{Hooft:1993gx} and Susskind \cite{Susskind:1995vu} introduced the holographic principle.
Similar to a hologram all information contained in the black hole can be stored on its horizon, a surface in one dimension less.
This insight led to the relation of a field theory without gravity in $d$ dimensions to a gravitational theory in $d\!+\!1$ dimensions, both having the same information content.
Due to this property of describing the same information in two different ways, it is often referred to as gauge/gravity duality.
With this dual approach, depending on the system in question, it is possible to switch between the related theories depending on which is more convenient for the task at hand and then translate the results to the other description if needed.
On one hand this allows to learn more about the gauge/gravity duality itself, if both theories are well known.
On the other hand this gives a very powerful tool to study theories that are not accessible in an easy way by using its dual theory that is under more control.\\
\\
The most famous realization of this principle appears in string theory and was conjectured by Maldacena \cite{Maldacena:1997re} in 1997.
Today this duality is known as AdS/CFT correspondence, because it relates type IIB superstring theory on $AdS_5\!\times\!S^5$ and $\mathrm{SU}(N_c) \, \mathcal{N}\!=\!4$ supersymmetric Yang-Mills theory (SYM), a CFT on the $4$-dimen\-sio\-nal boundary of AdS space.
The Yang-Mills coupling $g_{YM}$ and the rank of the gauge group $N_c$ of the field theory are related to the parameters of the superstring theory via
\begin{equation}
    g_{YM}^2 = 2\pi g_s\,, \qquad 2 g_{YM}^2 N_c = \frac{L^4}{l_s^4}\,,
\end{equation}
where $g_s$ is the string coupling, $l_s$ is the string length and $L$ is the curvature radius of both $AdS_5$ and the $S^5$.\\
In the limit of large $N_c$ and strong 't Hooft coupling $\lambda\!=\!g_{YM}^2 N_c$ this duality allows to make calculations in classical GR and translate them to a strongly coupled CFT.
This strong/weak character is one of the most powerful properties of gauge/gravity dualities.
In this case the weakly coupled gravitational theory allows to obtain results from comparatively easy calculations, which can then be translated to the strongly coupled field theory.
In the field theory itself the calculation would have been impossible even with the most recent methods.
Certainly there are other examples where it is the other way around and weakly coupled field theory can be used to study strong gravitational fields and learn more about black holes and maybe even quantum gravity.\\
There is a lot of evidence but no proof for the AdS/CFT correspondence or the holographic principle.

\subsubsection{Holographic Renormalization}
 \label{sec:HoloRen}
 
In the method of holographic renormalization \cite{deHaro:2000vlm}, the so-called Fefferman-Graham (FG) expansion of the metric plays a central role
\begin{equation} \label{eq:FGmetric}
    \mathrm{d}s^2_{FG} = g_{\mu\nu} \mathrm{d}x^\mu \mathrm{d}x^\nu = L^2 \left( \frac{\mathrm{d}\rho^2}{4 \rho^2} + \frac{1}{\rho} g_{ab}^{FG} \mathrm{d}x^a \mathrm{d}x^b \right)\,,
\end{equation}
where $g_{ab}^{FG}$ is the induced metric at the spacetime boundary.
If $g_{\mu\nu}$ satisfies Einstein's equations \eqref{eq:Einsteinequations}, it follows from the \textit{Fefferman-Graham-theorem} that $g_{ab}^{FG}$ can be expanded near the AdS boundary
\begin{equation} \label{eq:FGexpansion}
    g_{ab}^{FG} = g_{ab}^{(0)} + \rho g_{ab}^{(1)} + ... + \rho^{\frac{d}{2}} \left( \ln(\rho) \, h_{ab}^{(d)} + g_{ab}^{(d)} \right) + ... \,\,.
\end{equation}
The EMT $T_{\mu\nu}$ of the dual field theory is given by the following relation
\begin{equation} \label{eq:holoEMT}
    \langle T_{\mu\nu}(x) \rangle = -\frac{2}{\sqrt{g_{(0)}}} \frac{\delta S_{ren}}{\delta g_{(0)}(x)}\,, \qquad S_{ren} = S_{EH} + S_M + S_{GH} + S_{ct}\,.
\end{equation}
The renormalized gravitational action $S_{ren}$ is varied with respect to the boundary metric, the zeroth coefficient of the FG expansion $g_{(0)}$, while in the action the full expansion is used.
The renormalized action contains the gravitational \eqref{eq:EHAction} and matter \eqref{eq:matterAction} parts as well as suitable boundary terms $S_{GH}$ and counter-terms $S_{ct}$.
It turns out that only expansion coefficients up to order $d$ are relevant for the EMT.
In some systems considered in this work, the geometry is not known analytically and therefore the coefficients must be extracted from numerical simulations, as described in appendix \ref{app:Application}.\\
\\
Local observables like the EMT cannot capture the non-local features of quantum mechanics and therefore other observables are of interest as well.
Using state of the art methods, two-point functions and EE are some of the very few non-local observables, which can be computed using their dual geometric probes.

\subsubsection{Holographic Entanglement Entropy}
 \label{sec:HEE}

\begin{figure}
  \centering
  \includegraphics[width=0.59\textwidth]{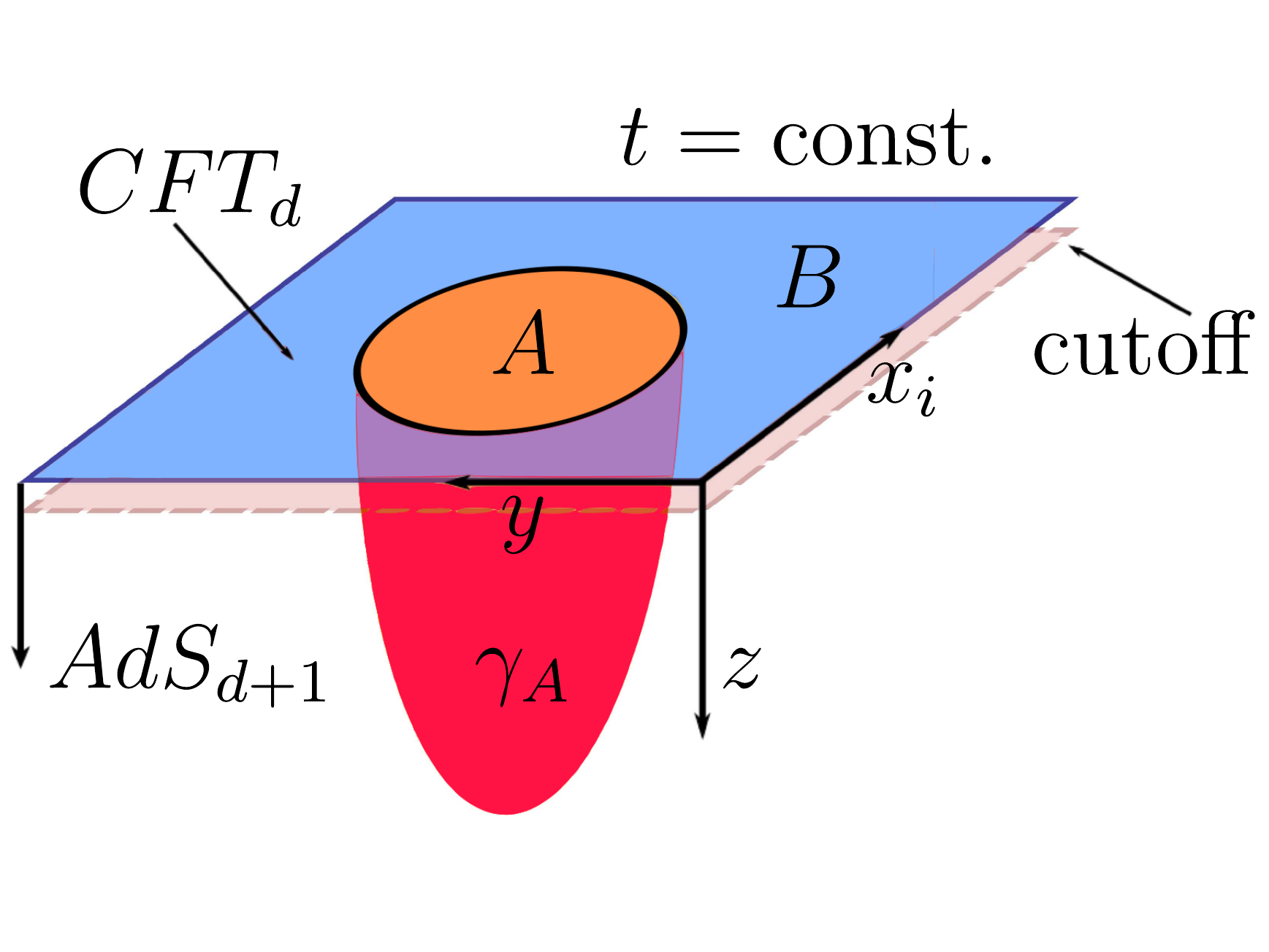}
  \caption{Illustration of the area formula for the holographic EE. The minimal surface $\gamma_A$ shares the boundary with the entangling region $A$. The holographic direction is denoted by $z$.}
  \label{fig:HEE}
\end{figure}

As introduced previously in section \ref{sec:EE}, EE is a measure for the entanglement of two quantum systems $A$ and $B$ (see figure \ref{fig:EE_def}).
It is defined for a subsystem $A$ as the von Neumann entropy of the reduced density matrix $\rho_A$ where all degrees of freedom of the (complementary) subsystem $B$ are traced out \eqref{eq:EEdef}.
Apart from $2$-dimensional CFT \eqref{eq:EE_CFT2}, it is almost impossible to find analytic results for the EE.
Performing similar calculations in higher dimensions with field theoretical methods is a very hard task.
Using the methods of the AdS/CFT correspondence, an \textit{area law} to calculate the EE in gauge theories dual to $d\!+\!1$-dimensional static geometries was proposed by Ryu and Takayanagi in 2006 \cite{Ryu:2006ef}
\begin{align}
 S_A = \frac{\mathrm{Area}(\gamma_A)}{4 \, G_N}\,, \label{area_formula}
\end{align}
where $G_N$ is Newton's constant in $d\!+\!1$ dimensions and $\gamma_A$ denotes the \textit{RT-surface}.
This surface is defined by having the smallest surface area among all possible surfaces in AdS space, sharing their boundary with the subsystem $A$ of the field theory located on the AdS boundary (see figure \ref{fig:HEE}).
Another requirement is that the RT-surface is homologous to the entangling region.
Hubeny et al.~\cite{Hubeny:2007xt} generalized this approach to time dependent backgrounds by generalizing the minimal surface to an extremal one.
The HRT prescription requires to choose the time slice with the RT-surface with the largest surface area, while the RT-surface has still minimal area on the selected time slice.
Looking at the $2$-dimensional result \cite{Holzhey:1994we,Calabrese:2004eu}, one can see that the results match and the central charge of the CFT is related to AdS spacetime via
\begin{align}
 c = \frac{3 \, L}{2 \, G_N}\,,
\end{align}
where $L$ is the AdS radius.

\cleardoublepage

\section{Numerical Background}
 \label{sec:NumBG}

In this section the other big component of this work, numerical calculations, will be reviewed.
The first topic, \textit{numerical relativity}, deals with the solution of Einstein's equations \eqref{eq:Einsteinequations}.
Studying only systems where the geometry is known analytically is limited to very symmetric cases.
Including also numerical solutions provides a larger sample of different systems to investigate, even highly dynamic or less symmetric ones.
This includes vacuum solutions as well as systems that involve a non-zero EMT.
In this work the following two numerical solutions were considered:
\begin{itemize}
    \item Gravitational shock wave collisions in $AdS_5$ dual to a toy model for HICs,
    \item Phase transition from small to large BTZ black brane caused by a massive scalar field in $AdS_3$.
\end{itemize}
The second topic, \textit{solving for extremal surfaces}, focuses on techniques needed for calculating EE holographically.
These methods will be applied to analytic and numeric solutions of Einstein's equations and are extended for calculating QNEC in section \ref{sec:NumQNEC}.

\subsection{Numerical Relativity}
 \label{sec:NumRel}

Solving Einstein's equations \eqref{eq:Einsteinequations} in full generality, which are a set of coupled non-linear partial differential equations, is a highly involved task.
In the rare examples where analytic solutions are available, simplifying assumptions such as spatial homogeneity and/or isotropy are imposed.
Investigating systems with less (or no) symmetries usually requires the help of numerical methods.
Particularly systems without time-translation invariance, which are relevant for this work, usually need to be solved numerically.\\
\\
Since the start in the 1960s \cite{Hahn-Lindquist1964}, numerical relativity made a huge progress.
In the 1970s the first simulations of colliding black holes were successfully performed \cite{Smarr-Cadez-deWitt-Eppley1976}.
Since then the development of computer hardware and the improvement of algorithms allowed to tackle more involved problems.
Nowadays it is possible to simulate (full) $4$-dimensional systems of binary black holes with strong and dynamical gravitational fields \cite{Pretorius2005}.
Of special interest is the computation of gravitational wave forms which provide important templates for large detection experiments such as LIGO or VIRGO amongst others
 and made the first direct detection of gravitational waves \cite{Abbott2016} possible.
 
\subsubsection{Approaches}
 \label{sec:NumRelApproaches}

Considering dynamical spacetimes, the problem can be divided into the search of initial conditions on a hypersurface and the evolution to neighboring hypersurfaces.
Even finding initial conditions can be a complicated task, as they must satisfy constraint equations which are part of Einstein’s equations.\\
\\
The most common approach is called \textit{ADM formalism}, introduced by Arnowitt, Deser and Misner in the late 1950s \cite{Arnowitt:1962hi}.
They used a \textit{3+1 decomposition} where spacetime is split into spacelike hypersurfaces and the evolution is performed in the time direction.
This corresponds to a Hamiltonian formulation of general relativity, briefly discussed in section \ref{sec:HamiltonianApproach}.
The Lagrangian can be written as
\begin{equation}
    \mathcal{L} = -\gamma_{ij} \partial_t \pi^{ij} - N H - N_i P^i - 2 \partial_i \left( \pi^{ij} N_j - \frac{1}{2} \pi N^i + \nabla^i N \sqrt{\gamma} \right)\,,
\end{equation}
where the $d\!-\!1$-dimensional metric $\gamma_{ij}$ are the generalized coordinates, $\pi^{ij}$ the conjugate momenta, $H$ the \textit{Hamiltonian constraint}, $P^i$ the \textit{momentum constraint} and their Lagrange multipliers $N$ and $N_i$ called \textit{lapse function} and \textit{shift vector}.
Variation of this Lagrangian leads to evolution equations
\begin{align}
    \partial_t \gamma_{ij} &= \frac{2N}{\sqrt{\gamma}} \left(\pi_{ij}-\frac{1}{2}\pi \gamma_{ij} \right) + \nabla_j N_i + \nabla_i N_j\,, \\
    \partial_t \pi^{ij} &= -N\sqrt{\gamma} \left( R^{ij} - \frac{1}{2}R \gamma^{ij} \right) + \frac{2N}{\sqrt{\gamma}} \gamma^{ij} \left(\pi^{mn}\pi_{mn}-\frac{1}{2}\pi \pi^ij \right) + \\
    + &\sqrt{\gamma} \left( \nabla^i \nabla^j N - \gamma^{ij} \nabla^n \nabla_n N \right) + \nabla_n (\pi^{ij} N^n) - (\nabla_n N^i) \pi^{nj} - (\nabla_n N^j) \pi^{ni}\,, \nonumber
\end{align}
as well as constraint equations
\begin{equation}
    H=0\,, \qquad \qquad P^i = 0\,,
\end{equation}\enlargethispage{\baselineskip}%
while the lapse and shift remain unconstrained.
This represents the freedom to choose a gauge in GR.
These equations need to be solved numerically, using standard techniques (some of which are presented in the following section \ref{sec:NumRelMethods}).
More details can be found in review articles and textbooks like \cite{Baumgarte:2010ndz}.
The ADM formalism is suited very well for astrophysical applications where Minkowski space is considered as boundary condition for the calculational domain.
ADM in its original form leads to an initial value problem which is only weakly hyperbolic which usually spoils the numerical stability of the time evolution.
This issue was resolved by Baumgarte, Shapiro, Shibata and Nakamura \cite{Baumgarte:1998te,Shibata:1995} in the 1990s who developed the \textit{BSSN formalism}, which realizes a strongly hyperbolic formulation and nowadays forms the basis of state of the art numerical relativity codes for merger simulations.\\
\\
A different approach is the \textit{characteristic formulation}, where lightlike slices (called \textit{characteristics}) are used for the foliation of spacetime instead of spacelike slices.
In this approach Einstein's equations \eqref{eq:Einsteinequations} can be cast into the schematic form
\begin{equation}
    \partial_\lambda F = H_F[F,G]\,, \qquad \qquad \partial_u \partial_\lambda G = H_G[F,G,\partial_u G]\,,
\end{equation}
where $\lambda$ denotes the radial null coordinate along the characteristics defined by $u\!=\!const.$
$F$ represent \textit{hypersurface variables} that are determined by functions on a single characteristic, denoted by $H_F$.
The \textit{evolution variables} $G$ and their $u$-derivatives can be determined similarly by integrating the functions $H_G$ along the characteristic.
It is possible to solve for all variables on a single null-hypersurface and then use the evolution variables $G$ to propagate the solution to the next characteristic.
Similarly to the ADM formalism, the techniques for solving the differential equations varies widely.
Some methods are introduced in the following section \ref{sec:NumRelMethods}.\\
\\
Opposed to \textit{3+1 formalisms} which are restricted to a bounded domain, the characteristic formulation was tailored to study radiation at null infinity.
The biggest advantage of the characteristic formulation is that Einstein's equations form a nested set of ordinary differential equations (ODE) on the characteristics.
The computational implementation of an evolution scheme depends on the version of the formalism and the initial value problem, but most of them have additional advantages in common \cite{Winicour2012}.
For example there are no elliptic constraints on the initial conditions and therefore no iterative constraint solvers are needed.
Further the characteristics extend to null infinity, so the behavior at the boundary can be described without extrapolations.
It turns out that the characteristic formulation is particularly well suited for solving the initial value problems in the context of AdS/CFT.

\subsubsection{Methods}
 \label{sec:NumRelMethods}
 
Before specific methods and algorithms for solving differential equations numerically can be discussed, the first step is to convert a set of differential equations into \textit{finite difference equations} (FDE) on a grid, covering the computational domain of interest.
A simple example may look like
\begin{align} \label{eq:ODEtoFDE1}
 \frac{\mathrm{d}y}{\mathrm{d}x} &= f(x,y)\,, \\
 \frac{y_{k} - y_{k-1}}{x_k - x_{k-1}} &= f \! \left( \frac{1}{2}(x_k - x_{k-1}), \frac{1}{2}(y_k - y_{k-1}) \right)\,, \label{eq:ODEtoFDE2}
\end{align}
where $k$ denotes the grid points.
For a system of $N$ first order ODEs converted to FDEs on a grid of size $M$, a solution consists of values for $N$ functions on $M$ grid points, i.e.~of $N\!\times\!M$ variables.
Using a \textit{multidimensional Newton method}, the equations are written in matrix form.
The matrix has a special block diagonal form, which reduces the resources needed for inverting it to solve the equations.\\
Independent of the approaches discussed in the previous section \ref{sec:NumRelApproaches}, the following methods (among many others) can be used to solve the resulting set of differential equations.

\paragraph{(Pseudo-) Spectral Methods} \cite{Boyd:2001cfsm,Grandclement:2007sb,Press2007} make use of a set of complete basis functions $\phi_k(x)$ to represent the solution $u(x)$ of some differential equation as well as its derivatives as series
\begin{equation} \label{eq:spectralSeries}
    u(x) \approx \sum_{k=0}^N u_k \phi_k(x)\,, \qquad \partial_x u(x) \approx \sum_{k=0}^N u_k \partial_x  \phi_k(x)\,,
\end{equation}
where the $\tilde u_k$ are called \textit{spectral coefficients}.
In the case $N\!=\!\infty$ the representation is exact and therefore the approximation error depends on the number of basis functions $N$ considered.
The following differential equation and basis functions serve as a simple example
\begin{equation}
    \partial^2_x u(x) = s(x)\,, \qquad \phi(x) = e^{i k x}\,.
\end{equation}
Replacing $u(x)$ and $s(x)$ by their expansions
\begin{equation}
    -\sum_{k=0}^N u_k k^2 e^{i k x} = \sum_{k=0}^N s_k e^{i k x}\,,
\end{equation}
yielding a relation between the expansion coefficients
\begin{equation}
    u_k = -\frac{s_k}{k^2}\,.
\end{equation}
To approximate the solution $u(x)$ via \eqref{eq:spectralSeries}, one needs to compute the coefficients of the known source $s(x)$ and insert into the relation above.\\
While this minimal example illustrates the idea, more difficult problems profit from choosing more suitable basis functions and other subtleties \cite{Baumgarte:2010ndz}.
For applications in numerical relativity the \textit{pseudo-spectral method} with \textit{Chebychev polynomials} as basis functions turned out to be suited especially well.
The Chebychev polynomials are defined on the interval $[-1,1]$ as
\begin{equation}
    T_n(\cos\theta) = \cos(n\theta)\,.
\end{equation}
The biggest advantage of spectral methods is the exponential convergence with the number $N$ of basis functions considered (finite difference methods converge with some power of the number of gridpoints for comparison).
On the downside, spectral methods are not as straight forward to implement and don't work, if the solution can't be represented by the basis functions very well (e.g.~if discontinuities appear).

\paragraph{Runge-Kutta methods} \cite{Gear:1971,Dormand:1980,Press2007} provide a technique for solving initial value problems approximately using numerical calculations.
They are not only easy to understand and implement (compared to other methods), they also bring a very good mix of accuracy, stability and computational intensity to the table.\\
\\
The simplest way to integrate a differential equation $\frac{\mathrm{d} y}{\mathrm{d}x}\!=\!f(x,y)$ numerically is the \textit{Euler method}, given by the formula
\begin{equation}
    y_{n+1} = y_n + h f(x_n, y_n)\,,
\end{equation}
which propagates the solution from one gridpoint $x_n$ to the adjacent $x_{n+1}$ with step size $h$.
While this formula is asymmetric and takes into account the information about the derivative only at one point (resulting in an error of order $\mathcal{O}(h^2)$), Runge-Kutta methods improve on that by taking into account the derivative information also between the gridpoints.
To obtain the results presented in section \ref{sec:HIC}, the \textit{classic fourth order Runge-Kutta} method (also known as RK4) was employed.
This requires the evaluation of $f(x,y)$ at four points
\begin{align}
    k_1&=h f(x_n, y_n)\,, \nonumber \\
    k_2&=h f(x_n+\frac{h}{2 }, y_n+\frac{k_1}{2})\,, \nonumber \\
    k_3&=h f(x_n+\frac{h}{2 }, y_n+\frac{k_2}{2})\,, \\
    k_4&=h f(x_n+h, y_n+k_3)\,, \nonumber \\
    y_{n+1}&=y_n + \frac{k_1}{6} + \frac{k_2}{3} + \frac{k_3}{3} + \frac{k_4}{6} + \mathcal{O}(h^5)\,. \nonumber
\end{align}
As indicated, a fourth order algorithm has an error term of order $\mathcal{O}(h^5)$, while still being simple to write down and implement.\\
\\
If combined with an adaptive step size algorithm, RK4 is a very powerful tool.
On the other hand, if the requirements on accuracy and/or efficiency are very high, there exist superior methods (described in \cite{Press2007} among others).

\subsection{Solving for Extremal Surfaces}
 \label{sec:GeoEQ}

\begin{figure}
	\begin{center}
	\includegraphics[height=.33\textheight]{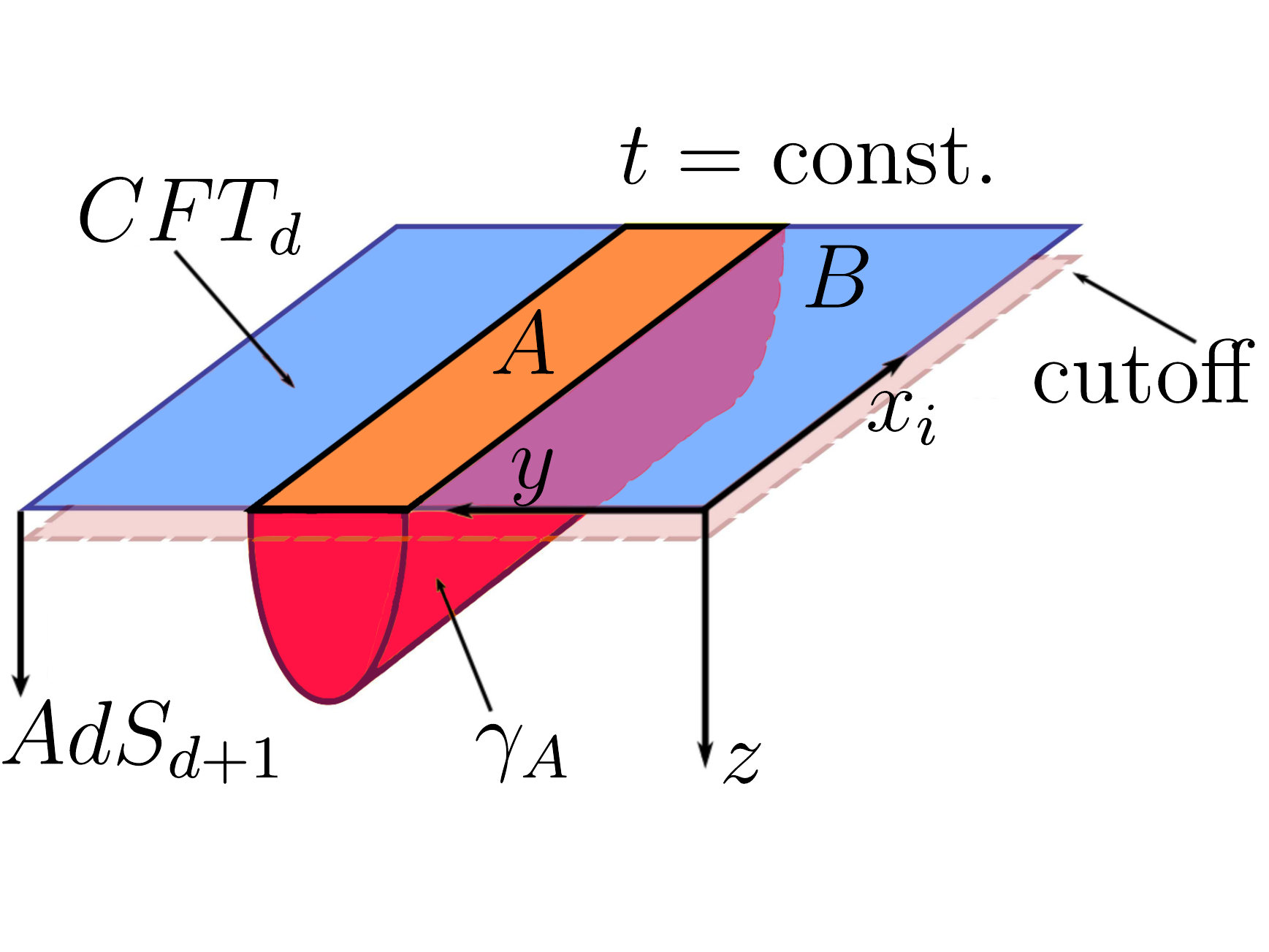}
	\caption{Choosing a strip as entangling region $A$ reduces the problem of finding \mbox{$d\!-\!2$-dimensional} extremal surfaces $\gamma_A$ to calculating geodesics, connecting the boundary points of $A$ on the $y$-axis.}
	\label{fig:stripRegion}
	\end{center}
\end{figure}

The main challenge of calculating QNEC, is to find the extremal surfaces related to EE via holography (discussed in section \ref{sec:HEE}) and its variation (defined in section \ref{sec:QNEC}).
Similar to Einstein's equations we are dealing with PDEs and it may not be possible to solve them analytically.
Especially in geometries given only by numerical solutions to Einstein's equations \eqref{eq:Einsteinequations}, it is necessary to approach the calculation of EE and QNEC numerically as well.\\
\\
In this work we focus on $3$- and $5$-dimensional AdS space, such that extremal surfaces are found by solving the geodesic equation.
In the $3$-dimensional case it is obvious that extremal `surfaces' are geodesics defined by the area functional
\begin{equation} \label{eq:area_functional_3D}
    \mathcal{A} = \int \mathrm{d}\tau \sqrt{g_{\mu\nu} \dot x^\mu \dot x^\nu}\,,
\end{equation}
where the dot denotes derivatives w.r.t.~the affine parameter $\tau$.
In $5$ dimensions the restriction to a special kind of entangling regions effectively reduces the problem to calculating geodesics as well.
This means choosing so-called \textit{strip regions} with finite extent in one spatial direction $y$, but covering the whole of the other (in our case two) spatial directions $x_i$ (shown in figure \ref{fig:stripRegion}).
The $5$-dimensional line element can then be written as
\begin{equation}
    \mathrm{d}s^2 = h_{\alpha\beta} \, \mathrm{d}x^\alpha \, \mathrm{d}x^\beta + \phi_1^2 \, \mathrm{d}x_1^2 + \phi_2^2 \, \mathrm{d}x_2^2\,, \label{eq:auxiliaryspacetime}
\end{equation}
introducing the auxiliary fields $\phi_i(x^\alpha)$ and the metric $h_{\alpha\beta}$ reduced to the coordinates $x^\alpha = (z,t,y)$.
The corresponding area functional is given by
\begin{equation} \label{eq:area_functional_5D}
    \mathcal{A} = \int \mathrm{d}x_1 \int \mathrm{d}x_2 \int \mathrm{d}\tau \sqrt{ \phi_1^2 \, \phi_2^2 \, h_{\alpha\beta} \, \dot x^\alpha \, \dot x^\beta}\,,
\end{equation}
where the $x_i$ integration yields a constant volume factor.
We divide by this factor and in turn consider \textit{entanglement entropy densities} per Killing volume (i.e.~the volume in the homogenous directions).
Such a simplification is only valid for geometries, homogeneous and isotropic in $x_i$.\\
\\
The geodesic equation can be found by variation of the action \eqref{eq:area_functional_3D} or \eqref{eq:area_functional_5D} with respect to the embedding functions $x^\mu$, which leads to
\begin{equation} \label{eq:geodesicEQ}
    \ddot x^\mu + \Gamma^\mu_{\alpha\beta} \, \dot x^\alpha \, \dot x^\beta = 0\,,
\end{equation}
where the Christoffel symbols are associated with the metric $g_{\mu\nu}$ or $g^{aux}_{\mu\nu} = \phi_1^2 \phi_2^2 h_{\alpha\beta}$ respectively.\\
\\
Finding curves, connecting the boundary points $y_1$ and $y_2$ of the entangling region, that solve the geodesic equation \eqref{eq:geodesicEQ} is a prime example for a so-called \textit{two point boundary value problem}.
These kind of problems are defined by a differential equation that is required to satisfy boundary conditions at two (or more) parameter values $\tau_i$.
There are two standard methods to solve such a two point boundary value problem, \textit{shooting} and \textit{relaxation} \cite{Press2007}, which are both iterative procedures.

\subsubsection{Methods}

\paragraph{Shooting methods} consist of two parts: integration and root finding.
As a simple example two coupled first order differential equations, subject to one boundary condition at the starting point $y_1$ and one boundary condition at the endpoint $y_2$ can be used.
The first part, integration, treats the problem as initial value problem that can be solved with a variety of methods (see e.g.~\cite{Press2007} or Runge-Kutta in section \ref{sec:NumRelMethods}).
All these methods require two initial conditions at $y_1$.
Since the problem provides only one of them, the other parameter can be chosen freely.
The integration with a random value for the free parameter will not lead to a solution satisfying the boundary condition at the endpoint $y_2$.
The second part is used to minimize the discrepancy between the solution of the integration and the desired result at $y_2$.
Root finding is an iterative process, fine-tuning the free parameter for the initial value problem.
Again there are a number of algorithms available \cite{Press2007}.\\
\\
Another variant of the shooting method is to start integrating from both boundaries and try to match the results at some point in the interior.
This can be especially helpful if the boundary conditions are specified at singular points or the numeric integration crashes even before the endpoint is reached.

\paragraph{Relaxation methods} employ a very different approach.
First the equations are written as FDEs [see equations \eqref{eq:ODEtoFDE1} and \eqref{eq:ODEtoFDE2}], discretized on the whole parameter range.
Then an initial guess for the solution, that neither has to satisfy the differential equation nor the boundary conditions, is made for all gridpoints.
The \textit{relaxation} is then the iterative process of improving the guess to gradually bringing it closer to the real solution.
This approach needs more computational power than shooting, but has other advantages.
Especially when the boundary conditions are very delicate or the solution shows chaotic behavior, which makes the fine-tuning nearly impossible.
The key to efficient relaxation methods is the quality of the ansatz.\\
\\
Although the authors of \cite{Press2007} encourage to "shoot first, and only then relax", for the problems treated in this thesis, the relaxation method is suited very well.
I will give a detailed summary of a relaxation algorithm that is used to solve the geodesic equation in the remainder.
Consider an arbitrary set of FDEs
\begin{align}
 0 = E_{j,k} = (y_{j,k} - y_{j,k-1}) - (x_k - x_{k-1}) \, \, \, f_{j,k} \! \! \left( \frac{1}{2}(x_k - x_{k-1}), \frac{1}{2}( y_{j,k} - y_{j,k-1} ) \right)\,,
\end{align}
where $j\!=\!0, ...,\!N\!-\!1$ labels the equation and $k\!=\!1, ...,\!M\!-\!1$ specifies the position $x_k$.
This are $N$ equations at $M\!-\!1$ points for $M\!\times\!N$ variables.
The missing $N$ equations to solve the system are provided by the boundary conditions
\begin{align}
 0 = E_{j,0} = B_j(x_0,y_{j,0})\,, \qquad
 0 = E_{j,M} = C_j(x_M,y_{j,M})\,,
\end{align}
where $B_j$ contains $n_1$ boundary conditions at the starting point and $C_j$ provides $n_2\!=\!N\!-\!n_1$ boundary conditions at the endpoint (while the remaining entries of $B$, $C$ and $E$ are zero).\\
The relaxation method needs an ansatz $y_{j,k}$ for the values of the $N$ variables at the $M$ points.
These values are corrected by small increments $\Delta y_{j,k}$, such that $y_{j,k}\!+\!\Delta y_{j,k}$ is an improved solution to the FDEs.
The new values are then used as ansatz to start all over again until some error criterion is fulfilled.
The error may be calculated like
\begin{align}
 err = \frac{1}{N M} \sum_{j,k} \Delta y_{j,k}\,. \label{FDEerror}
\end{align}
To find the increments $\Delta y_{j,k}$ a multidimensional Taylor expansion of the FDEs to first order is used
\begin{align}
 E_{j,k}(y_{j,k} &+ \Delta y_{j,k}, y_{j,k-1} + \Delta y_{j,k-1}) = \nonumber \\
  &= E_{j,k}(y_{j,k},y_{j,k-1}) + \sum_{n=0}^{N-1} \frac{\partial E_{j,k}}{\partial y_{n,k-1}} \, \Delta y_{n,k-1} + \sum_{n=0}^{N-1} \frac{\partial E_{j,k}}{\partial y_{n,k}} \, \Delta y_{n,k}\,.
\end{align}
For a solution, the updated equations $E_{j,k}$ must be zero, therefore
\begin{align}
 -E_{j,k} = \sum_{n=0}^{N-1} \hat S_{j,n} \, \Delta y_{n,k-1} + \sum_{n=0}^{N-1} \tilde S_{j,n} \, \Delta y_{n,k}\,, \label{updatedFDE}
\end{align}
where
\begin{align}
 \hat S_{j,n} = \frac{\partial E_{j,k}}{\partial y_{n,k-1}}, \qquad \tilde S_{j,n} = \frac{\partial E_{j,k}}{\partial y_{n,k}}\,.
\end{align}
A similar expansion of the equations at the first boundary leads to
\begin{align}
 -E_{j,0} = \sum_{n=0}^{N-1} \tilde S_{j,n} \, \Delta y_{n,0}\,, \qquad j = n_2, n_2+1, ..., N-1\,,
\end{align}
where
\begin{align}
 \tilde S_{j,n} = \frac{\partial E_{j,0}}{\partial y_{n,0}}\,.
\end{align}
At the second boundary this gives
\begin{align}
 -E_{j,M} = \sum_{n=0}^{N-1} \hat S_{j,n} \, \Delta y_{n,M-1}\,, \qquad j = 0, ..., n_2-1 \,,
\end{align}
with
\begin{align}
 \hat S_{j,n} = \frac{\partial E_{j,M}}{\partial y_{n,M-1}}\,.
\end{align}
Combining $\hat S_{j,k}$ and $\tilde S_{j,k}$ into one matrix
\begin{align}
 S_{j,n} &= \hat S_{j,n}, \, \, \mathrm{for} \, \, \, n = 0,...,N-1  \,, \\
 S_{j,n} &= \tilde S_{j,n}, \, \, \mathrm{for} \, \, \, n = N,...2N-1 \,,
\end{align}
one gets a $N\!\times\!2N$ matrix at every point $k$.
To solve the equations \eqref{updatedFDE} above for $\Delta y_{j,k}$, all these matrices are combined like in figure \ref{fig:s_matrix} and the linear equation
\begin{align}
 S.v = b\,,
\end{align}
is solved, where $v$ is the solution vector related to the corrections $\Delta y_{j,k}$ and $b$ contains the FDEs $E_{j,k}$.
The matrix $S$ has now a block diagonal form which can be dealt with efficiently, using a form of the Gaussian elimination for sparse matrices \cite{Press2007}, exploiting the form of $S$.
This makes it possible to tackle problems with large grid sizes and many equations.

\begin{figure}
  \centering
    \begin{minipage}[htb]{\textwidth}
      \begin{align}
      \left(
      \begin{array}{*{20}c}
	X & X & X & . & . & . & . & . & . & . & . & . & . & . & . & . & . & . \\
	X & X & X & . & . & . & . & . & . & . & . & . & . & . & . & . & . & . \\
	I & I & I & X & X & X & . & . & . & . & . & . & . & . & . & . & . & . \\
	I & I & I & X & X & X & . & . & . & . & . & . & . & . & . & . & . & . \\
	I & I & I & X & X & X & . & . & . & . & . & . & . & . & . & . & . & . \\
	. & . & . & I & I & I & X & X & X & . & . & . & . & . & . & . & . & . \\
	. & . & . & I & I & I & X & X & X & . & . & . & . & . & . & . & . & . \\
	. & . & . & I & I & I & X & X & X & . & . & . & . & . & . & . & . & . \\
	. & . & . & . & . & . & I & I & I & X & X & X & . & . & . & . & . & . \\
	. & . & . & . & . & . & I & I & I & X & X & X & . & . & . & . & . & . \\
	. & . & . & . & . & . & I & I & I & X & X & X & . & . & . & . & . & . \\
	. & . & . & . & . & . & . & . & . & I & I & I & X & X & X & . & . & . \\
	. & . & . & . & . & . & . & . & . & I & I & I & X & X & X & . & . & . \\
	. & . & . & . & . & . & . & . & . & I & I & I & X & X & X & . & . & . \\
	. & . & . & . & . & . & . & . & . & . & . & . & I & I & I & X & X & X \\
	. & . & . & . & . & . & . & . & . & . & . & . & I & I & I & X & X & X \\
	. & . & . & . & . & . & . & . & . & . & . & . & I & I & I & X & X & X \\
	. & . & . & . & . & . & . & . & . & . & . & . & . & . & . & I & I & I \\
      \end{array} \nonumber
      \right)
      \end{align}
    \end{minipage}
  \caption{In case of three equations on 6 grid points with two initial conditions at the first grid point and one on the last grid point, the $S$-matrix looks like this. For better visualization, $I$ stands for $\hat S$, $X$ stands for $\tilde S$ and the dots represent zeros.}
  \label{fig:s_matrix}
\end{figure}

\subsubsection{We do Not Shoot but Relax}

Since we are working in (asymptotically) AdS spacetime where the distances start diverging once we get close to the boundary, fine-tuning parameters for shooting turns out to be rather tedious and shows almost chaotic behavior.
The fixed boundary points required from the holographic description of EE clearly favour the relaxation method (although there are situations and applications for shooting as will be discussed at the end of this section).
This becomes especially clear when full control over the boundary points is needed to extend this method for calculating QNEC in section \ref{sec:NumQNEC}.\\
\\
In order to apply the relaxation method to the geodesic equation \eqref{eq:geodesicEQ}, we introduce a non-affine parameter $\sigma$ which is often useful for numerical purposes
\begin{equation} \label{eq:geodesicEQ_jacobian}
    \ddot x^\mu + \Gamma^\mu_{\alpha\beta} \, \dot x^\alpha \, \dot x^\beta = -J \, \dot x^\mu\,,
\end{equation}
where dot denotes the derivative with respect to $\sigma$ and
\begin{equation}
    J(\sigma) = \left. \frac{\mathrm{d}^2 \tau}{\mathrm{d}\sigma^2} \middle/ \frac{\mathrm{d} \tau}{\mathrm{d}\sigma} \right.\,.
\end{equation}
This counters the effect of AdS that gridpoints, equidistant in the affine parameter $\tau$, cluster near the boundary.
As \eqref{eq:geodesicEQ_jacobian} are three second order differential equations, we need to rewrite them into six first order ODEs by introducing the derivatives of $x^\mu$ as separate variables $p^\mu$
\begin{align}
    p^\mu &= \dot x^\mu\,, \\
    \dot p^\mu + \Gamma^\mu_{\alpha\beta} \, p^\alpha \, p^\beta &= -J \, p^\mu\,.
\end{align}
Following \eqref{eq:ODEtoFDE1} and \eqref{eq:ODEtoFDE2} this can be brought into a suitable form for the relaxation algorithm.\\
\\
The last ingredient is a suitable ansatz.
In general it is possible to use the analytically known solution for the vacuum AdS geometry.
A big strength of the relaxation method is that previous results can be used as ansatz for future calculations.
A prime example is a time evolution, where we can take the solution at some instant of time as ansatz for the next computation and reduce the number of iterations (and therefore the duration) significantly.\\
\\
If the generic ansatz of a vacuum AdS geodesic is not good enough and relaxation takes too long or runs into troubles like leaving the computational domain, shooting can be used to solve the geodesic equation and find an ansatz.
The result from shooting won't hit the boundary at the desired points, but sufficiently well to serve as improved ansatz for the relaxation.
This procedure is a bit more involved but leads to good results.\\
\\
The really big advantage of shooting is that most of the time it is sufficient to ask \textit{Mathematica's NDSolve} for the result.
To compute EE, where exact control over the boundary points is not required, shooting can be superior, especially if the system is static and/or isotropic and/or homogeneous.
The important subtlety is to shoot from the turning point of the geodesic in the bulk to the boundary.
If the entangling region is of importance, the initial values need to be fine-tuned.
This reduces the advantage of shooting notably.

\cleardoublepage

\section{Quantum Null Energy Condition}
 \label{sec:QNEC}

\begin{figure}
	\begin{center}
	\includegraphics[height=.23\textheight]{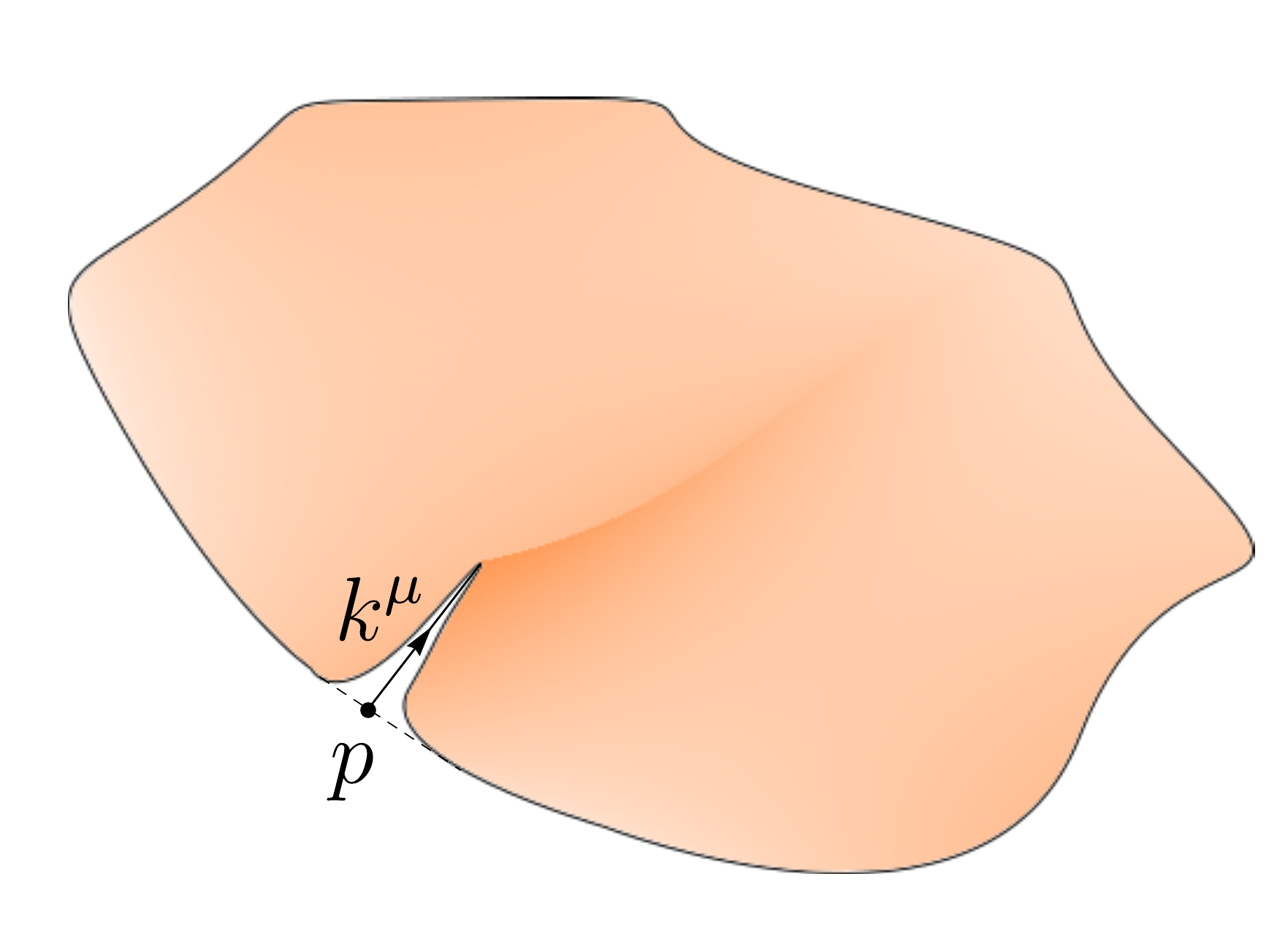}
	\caption{The orange surface is the entangling region, while the black line depicts its boundary, the entangling surface. $p$ denotes the point at which the EMT is evaluated and the entangling surface is deformed along the null vector $k^\mu$. The dashed line indicates the shape of the entangling surface without the deformation.}
	\label{fig:nulldeformation}
	\end{center}
\end{figure}

The QNEC emerged from a greater concept, the \textit{quantum focussing conjecture} (QFC) \cite{Bousso:2015mna}, rather than being the attempt to create a new EC.
Despite that, QNEC is currently the only known local quantum EC and does not require system-dependent constants like some of the approaches discussed in section \ref{sec:ECsandQT}.\\
As special case of the QFC, QNEC states that the null projection of the energy momentum tensor at the point $p$ is bounded from below by the second functional derivative of the EE $S''$ w.r.t.~a deformation of the entangling region along the null vector $k^\mu$, as shown in figure \ref{fig:nulldeformation}
\begin{equation} \label{eq:QNEC}
  \langle T_{\mu\nu} k^\mu k^\nu \rangle = T_{kk} \geq  \frac{1}{2 \pi} S''\,.
\end{equation}
To simplify the equations in the remainder of the thesis we introduce the abbreviation $T_{kk}$ for the expectation value of the null projection of the EMT, defined above.
Its precise definition and how this inequality is obtained from the QFC is discussed in the next section.
Following the proofs for free field theories and QFTs with holographic duals \cite{Bousso:2015wca,Koeller:2015qmn}, very recently a general proof was published, using methods of field theory as well as differential geometry \cite{Balakrishnan:2017bjg} (see section \ref{sec:QNECproofs}).
A very relevant feature of QNEC presented in section \ref{sec:stronger2D}, is the altered form in $2$-dimensional CFTs, we refer to as QNEC$_2$, which puts an even stronger condition on the EMT.
Although the aforementioned proofs exist, there is a lot to be learned about QNEC by studying it explicitly in physically relevant systems.
Investigating the properties and relations of QNEC is an exciting open problem.
The question of NEC violation and positivity of energy, the behavior of QNEC in highly dynamical situations and saturation of the inequality are tackled in sections \ref{sec:QNEC_D=4} and \ref{sec:QNEC_D=2} as well.

\subsection{Quantum Focussing Conjecture}
 \label{sec:QFC}

The intriguing thing about QNEC is that it arises as part of the QFC and therefore is not postulated like the classical ECs, but might be derived from fundamental principles.
In order to properly introduce QNEC, it is useful to review this derivation by following the original publication by Bousso et.~al. \cite{Bousso:2015mna}.
Therefore we will start with the \textbf{generalized second law (GSL)} of thermodynamics \cite{Bekenstein:1974ax}
\begin{equation}
    \mathrm{d}S_{gen} \geq 0\,.
\end{equation}
In this generalization of the second law of thermodynamics, Bekenstein replaced the entropy with the generalized entropy
\begin{equation} \label{eq:genEntropy}
    S_{gen} = S_{BH} + S_{out}\,,
\end{equation}
where $S_{BH}\!=\!A/(4G\hbar)$ is the entropy of all black holes in the system and $S_{out}$ is the matter entropy outside these black holes, given by the von Neumann entropy.
That this generalization was needed, can be understood using the gedankenexperiment of throwing a box of some mass into a black hole.
The entropy associated with the box is then lost, violating the classical second law of thermodynamics.
The GSL accounts for that via the increased horizon area.
A further generalization allows the description of the entropy of arbitrary surface on light sheets (not only black hole horizons) in terms of their area \cite{Bousso:1999xy}.\\
\\
Another integral part is the \textit{classical focussing theorem}.
The expansion of a congruence of geodesics is defined as
\begin{equation} \label{eq:classicExpansion}
    \theta = \lim_{\mathcal{A} \to 0} \frac{1}{\mathcal{A}}\frac{\mathrm{d}\mathcal{A}}{\mathrm{d}\lambda}\,,
\end{equation}
where $\mathcal{A}$ is the infinitesimal area element spanned by neighboring geodesics and $\lambda$ is the affine parameter along those geodesics.
The change of the expansion scalar is governed by the Raychaudhuri equation \eqref{eq:Raychaudhuri} mentioned already in section \ref{sec:EC} about the classical energy conditions.\\
\\
The focussing theorem states that \textit{along hypersurface-orthogonal geodesic congruences the expansion is non-positive, if the null curvature condition \eqref{eq:NCC} is satisfied}.
\begin{equation}
    \dot\theta\leq0
\end{equation}
This can be seen directly from the Raychaudhuri equation \eqref{eq:Raychaudhuri}.
Hypersurface orthogonality guarantees vanishing twist $\omega\!=\!0$, while the NCC ensures the last expression to be positive.
$\theta^2$ and $\sigma^2$ being manifestly positive as well concludes the proof.\\
\\
In order to find a quantum version of that statement, the idea was to use the generalized entropy for defining the \textit{quantum expansion} $\Theta$ as follows
\begin{equation}
    \Theta[V(y);y_1] = \frac{4G\hbar}{\sqrt{^V g(y_1)}} \frac{\delta S_{gen}}{\delta V(y_1)} = \lim_{\mathcal{A} \to 0} \frac{4G\hbar}{\mathcal{A}}S'_{gen}\,.
\end{equation}
The generalized entropy takes the role of the area-element of the congruence and $V(y)$ specifies one slice $N$ of the hypersurface spanned by the geodesic congruence, orthogonal to the entangling surface.
The affine parameter $\lambda$, together with the coordinate $y$ along the entangling surface forms a coordinate system on the hypersurface.
The area element $\sqrt{^V g(y)}$ ensures that the functional derivative is taken by unit area of the slice selected by $V(y)$.
This \textit{quantum focussing conjecture} now states that \textit{the quantum expansion cannot increase at $y_1$, if the slice of $N$ defined by $V(y)$ is infinitesimally deformed along the same direction at $y_2$}
\begin{equation}
    \frac{\delta}{\delta V(y_2)} \Theta[V(y);y_1] \leq 0\,.
\end{equation}
This definition of quantum focussing contains two different cases.
Either $y_1$ and $y_2$ are different (\textit{off-diagonal part}) or they coincide (\textit{diagonal part}).
In the former case the conjecture can be proven using strong sub-additivity \cite{Bousso:2015mna}, since the area part (localized at $y_1$) in the generalized entropy is annihilated by the derivative with respect to the distant generator $y_2$.
The latter case leads to QNEC
\begin{equation}
    0 \geq \frac{\delta}{\delta V(y_1)} \frac{4G\hbar}{\sqrt{^V g(y_1)}} \frac{\delta S_{gen}}{\delta V(y_1)}\,.
\end{equation}
In this case we can relate to figure \ref{fig:nulldeformation} by identifying $y_1$ with the point $p$ and the slice of $N$ is generated by the null vector $k^\mu$ and the affine parameter $\lambda$.
Inserting for the generalized entropy \eqref{eq:genEntropy} and using the definition of the classical expansion \eqref{eq:classicExpansion}, this can be written as
\begin{equation}
    0 \geq \lim_{\mathcal{A}\to 0} \frac{\partial}{\partial \lambda} \left( \theta + \frac{4G\hbar}{\mathcal{A}} S'_{out} \right) = \lim_{\mathcal{A} \to 0} \left[ \theta' + \left( \frac{4G\hbar}{\mathcal{A}} S''_{out} - S'_{out} \theta \right) \right] \,,
\end{equation}
where prime denotes the derivative w.r.t.~$\lambda$.
Making use of the Raychaudhuri equation \eqref{eq:Raychaudhuri} and assuming Einstein gravity, we find
\begin{equation}
    0 \geq \lim_{\mathcal{A} \to 0} \left[ \omega^2 - \sigma^2 - \frac{\theta^2}{3} - 8\pi G T_{\mu\nu} k^\mu k^\nu + \left( \frac{4G\hbar}{\mathcal{A}} S''_{out} - S'_{out} \theta \right) \right]\,,
\end{equation}
which, in the case of arbitrary hypersurface orthogonal congruences with shear and expansion vanishing at some point, leads us to QNEC
\begin{equation} \label{eq:QNECfromQFC}
    T_{kk} \geq \lim_{\mathcal{A} \to 0} \frac{\hbar}{2 \pi \mathcal{A}} S''_{out}\,.
\end{equation}
In the classical limit $\hbar \to 0$ the classical NEC \eqref{eq:NEC} is recovered.

\subsection{Proofs of QNEC}
 \label{sec:QNECproofs}

Shortly after the publication of the QFC the first proof for QNEC was found.
In \cite{Bousso:2015wca} it was shown that QNEC holds for free or superrenormalizable bosonic field theories on flat background (or on bifurcate killing horizons with null-tangent $k^\mu$ in curved backgrounds).
Their strategy was to discretize the null hypersurface $N$ along the transverse direction into small pencils, such that EE can be calculated on every pencil via the replica trick.
This allowed the authors to calculate its second variation and complete the proof by taking the continuum limit of the discrete pencils.
Special attention was needed in the $2$-dimensional case where the discretization is not possible due to the lack of a transverse direction.
Here the analytic continuation of the higher dimensional case to $d\!=\!2$ does the trick.
This proof is entirely within QFT and does not rely on the gravitational origin of the QFC.
A generalization to fermionic field theories was published in 2019 \cite{Malik:2019dpg}.
Only a couple months after the first proof, QNEC was proven for different types of theories \cite{Koeller:2015qmn}.
This time the focus was put on holographic theories with well behaved Einstein gravity dual, as discussed in the next section.
In 2017 Balakrishnan et al.~\cite{Balakrishnan:2017bjg} found a general proof for QNEC in relativistic QFTs in dimension $d\geq3$, sketched in section \ref{sec:QNECgeneralProof}.
It was also shown that the diagonal part of the QFC (i.e.~QNEC) is saturated in QFT of dimension three or higher \cite{Leichenauer:2018obf}.
A discussion of QNEC in curved spacetimes can be found in \cite{Fu:2017evt}.

\subsubsection{Holographic Proof}
 \label{sec:QNECholoProof}

\begin{figure}
	\begin{center}
	\includegraphics[height=.33\textheight]{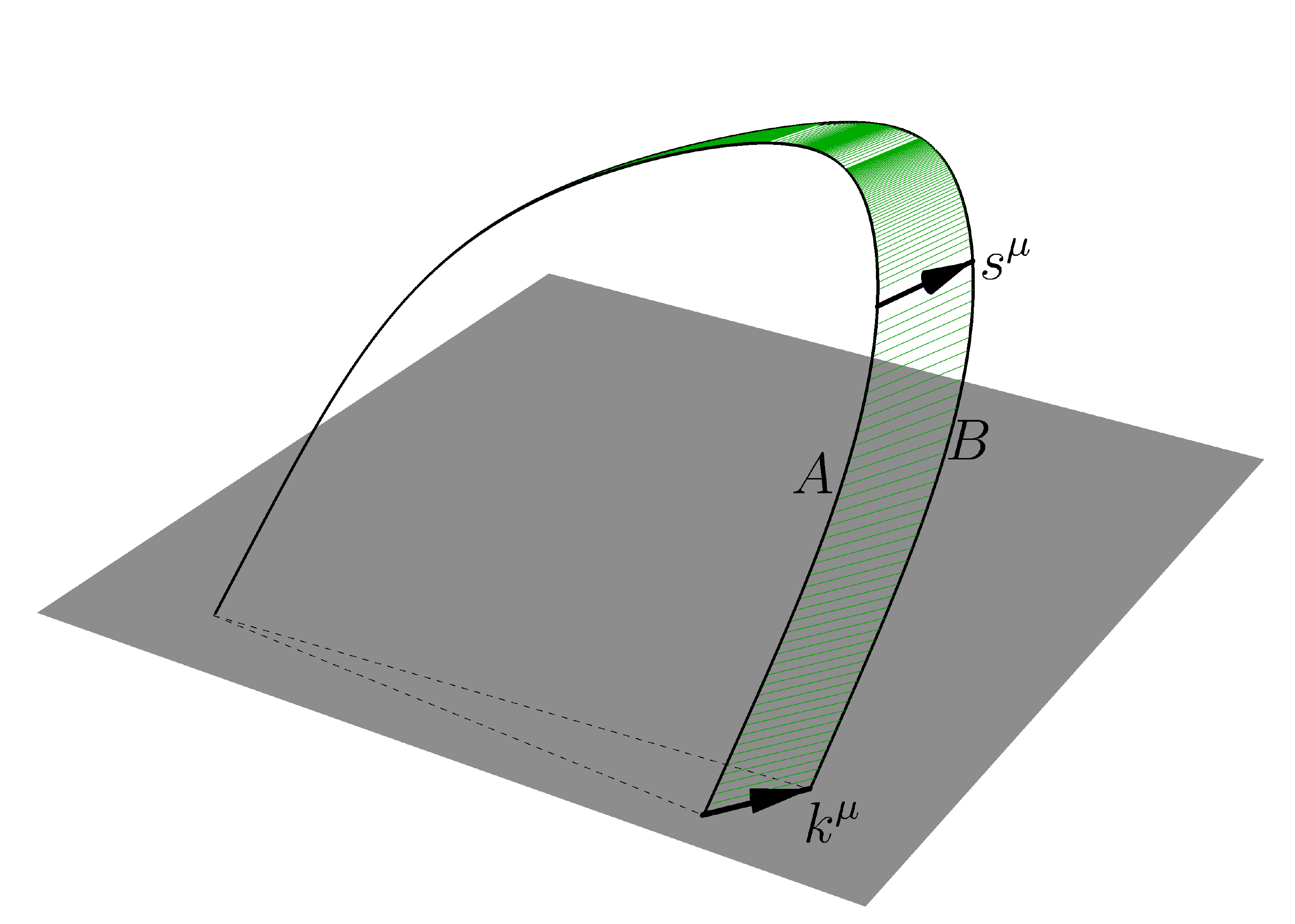}
	\caption{Application of Entanglement wedge nesting in the holographic proof of QNEC. The two extremal surfaces $A$ and $B$ are separated by the vector $k^\mu$ at the boundary (gray plane). If $k^\mu$ is null or spacelike, EWN leads to the separation vector $s^\mu$ in the bulk being null or spacelike.}
	\label{fig:sVector}
	\end{center}
\end{figure}

In order to proof QNEC, there are always two ingredients needed.
First the EMT of the field theory, which can be obtained from the holographic renormalization outlined in section \ref{sec:HoloRen} in this case.
The near boundary expansion of the bulk metric $G_{\mu\nu}$ up to order $z^d$ for a Poincar\'e invariant system is given by
\begin{equation} \label{eq:MetricExpansionHoloProof}
    \mathrm{d}s^2 = G_{\mu\nu} \mathrm{d}x^\mu \mathrm{d}x^\nu = \frac{L^2}{z^2} \left( \mathrm{d}z^2 + \left[ f(z) \eta_{ij} + \frac{16 \pi G_N}{d L^{d-1}} z^d t_{ij} \right]  \mathrm{d}x^i \mathrm{d}x^j + o(z^d) \right)\,,
\end{equation}
where $\eta_{ij}$ is the boundary metric and $f(z)$ contains only powers of $z$ less than $d$, but its explicit form depends on the theory [for a CFT $f(z)\!=\!1$].
The coefficient of $z^d$ denoted by $t_{ij}$ is not necessarily equal to the EMT, but for a CFT on Minkowski space, they are the same.
Especially the null components are the same $t_{kk}\!=\!T_{kk}$, since the difference between the tensors is only proportional to the boundary metric $\eta_{ij}$.\\
\\
The other part contains the derivatives of EE, which one can express following the holographic description presented in section \ref{sec:HEE}.
The area of the extremal surface related to EE is given by the area functional
\begin{equation} \label{eq:Areafunctional}
    A = \int \mathrm{d}z \, \mathrm{d}^{d-2}y \sqrt{H[X]}\,,
\end{equation}
where $H_{\alpha\beta}$ is the induced metric on the extremal surface given by the embedding functions $X^\mu(z,y^a)$, depending on the radial coordinate $z$ and the intrinsic coordinates on the surface $y^a$.
Since we are interested in the change of EE with respect to a null deformation of the entangling region, we need the variation of the area with respect to a change in the embedding functions
\begin{equation} \label{eq:areavariation}
    \delta A = \left. -\frac{L^{d-1}}{z^{d-1}} \int \mathrm{d}^{d-2} y \sqrt{h} \frac{g_{ij} \partial_z X^i}{\sqrt{1+g_{lm} \partial_z X^l \partial_z X^m}} \delta X^j \right|_{z=\epsilon}\,,
\end{equation}
where we rescaled the metrics $g_{ij}\!=\!\frac{z^2}{L^2} G_{ij}$ and $h_{ij}\!=\!\frac{z^2}{L^2} H_{ij}$.
The potentially infinite expression will be regulated by a cutoff and $\delta A$ is evaluated at $z\!=\!\epsilon$.
The next step is to find the embedding functions and insert them into the above expression.
This is done by solving the equations of motion obtained by extremizing the area functional \eqref{eq:Areafunctional}
\begin{equation}
    \frac{1}{\sqrt{H}} \partial_\alpha \left( \sqrt{H} H^{\alpha\beta} \partial_\beta X^\mu \right) + H^{\alpha\beta} \Gamma^\mu_{\nu\sigma} \partial_\alpha X^\nu \partial_\beta X^\sigma = 0\,,
\end{equation}
where $\Gamma^\mu_{\nu\sigma}$ is the bulk Christoffel symbol \eqref{eq:RGamma}.
With some clever gauge choice and considering only powers of $z$ lower than $d$ in the expansion \eqref{eq:MetricExpansionHoloProof}, this reduces to
\begin{equation}
    z^{d-1} \partial_z \left( z^{1-d} \sqrt{h} h^{zz} f \partial_z X^i \right) + \partial_a \left(f \sqrt{h} h^{ab} \partial_b X^i   \right) = 0\,,
\end{equation}
which can be solved order by order, yielding
\begin{equation} \label{eq:embeddingfunction}
    X^i(z,y^a) = X^i_{bdry}(y^a) + \frac{1}{2(d-2)} z^2 K^i(y^a) + ... + \frac{1}{d} z^d \left( V^i(y^a) + W^i(y^a) \log z\right) + o(z^d)\,.
\end{equation}
The omitted terms with powers between $2$ and $d$, denoted by the ellipsis as well as the logarithmic term $W^i$ are all state independent and fixed by geometric invariants of the entangling region and vanish for surfaces with flat boundary.
$K^i$ is the trace of the extrinsic curvature of the boundary of the entangling region, which is given for flat backgrounds by
\begin{equation}
    K^i = \frac{1}{\sqrt{h}} \partial_a \left( \sqrt{h} h^{ab} \partial_b X^i_{bdry} \right)\,.
\end{equation}
Putting together the pieces and inserting \eqref{eq:embeddingfunction} into \eqref{eq:areavariation} we find
\begin{equation}
    \frac{1}{L^{d-1}\sqrt{h}}\frac{\delta A}{\delta X^i} = -\frac{1}{(d-2)\epsilon^{d-2}}K_i + ... - W_i \log \epsilon - V_i + ... \,.
\end{equation}
Here both ellipsis refer to the terms obtained from the geometric invariants, showing power law divergence behavior or are finite and state dependent.
Taking into account the properties of the extrinsic curvature, the entangling region and the null vector $k^i$, it can be shown that most terms vanish for null deformations \cite{Leichenauer:2018obf} and we are left with
\begin{equation} \label{eq:nullvariation}
    k^i \frac{\delta A}{\delta X^i(y)} = - L^{d-1} \sqrt{h(y)} k^i V_i(y)\,.
\end{equation}
The crucial point in the proof that links the null variation of the extremal surface \eqref{eq:nullvariation} to the metric expansion \eqref{eq:MetricExpansionHoloProof} is the geometric property that \textbf{extremal surfaces are not causally connected}.
This property is also known as \textit{entanglement wedge nesting} (EWN), sketched in figure \ref{fig:sVector}.
More precisely, any vector $s^\mu$ pointing from one extremal surface $A$ to another extremal surface $B$ must be spacelike or null, if the boundary of $A$ and $B$ are spacelike or null separated.
This property was proven by Wall \cite{Wall:2012uf}, relying on the classical NEC in the bulk.
Constructing $s^\mu$ explicitly as sum of the two orthogonal null vectors $l^\mu$ and $k^\mu$
\begin{equation}
    s^\mu = \alpha l^\mu + \beta k^\mu\,,
\end{equation}
allows us to express the non-causality as $0\!\leq\!s^2\!=\!\alpha \beta$.
The coefficients are chosen as
\begin{equation}
    \alpha = g_{\mu\nu} l^\mu \partial_\lambda X^\nu\,, \qquad \beta = g_{\mu\nu} k^\mu \partial_\lambda X^\nu\,,
\end{equation}
where $\lambda$ is the parameter for the null deformation at the boundary.
In the near boundary limit this means $\alpha\!\to\!1$ and $\beta\!\to\!0$, which requires the leading contribution to $\beta$ to be positive.
To conclude the proof, we need to construct $\beta$, using all the ingredients prepared above
\begin{equation}
    g_{\mu\nu} k^\mu \partial_\lambda X^\nu = \left( k_i \Delta k^i + \frac{1}{d} k_i \partial_\lambda V^i + \frac{16 \pi G_N}{d L^{d-a}} T_{kk} \right) z^d + o(z^d)\,.
\end{equation}
Demanding $k^\mu$ to be orthogonal to $\partial_a X^\mu$ and $\partial_z X^\mu$ gives additional constraints on $k^i \Delta k_i$, such that we find
\begin{align}
    &\textrm{for}\,\,d\!>\!2: \quad \frac{8 \pi G_N}{L^{d-1}} T_{kk} \geq -k_i \partial_\lambda V^i\,, \\
    &\textrm{for}\,\,d\!=\!2: \quad \frac{8 \pi G_N}{L} T_{kk} \geq -k_i \partial_\lambda V^i + (k_i V^i)^2\,.
\end{align}
Inserting the expression for the variation of the area \eqref{eq:areavariation} and considering only infinitesimal deformations at the point $y\!=\!p$, the variation and derivative are given by
\begin{align}
    &\textrm{for}\,\,d\!>\!2: \quad 2 \pi T_{kk} \geq \frac{S''}{\sqrt{h}}\,, \\
    &\textrm{for}\,\,d\!=\!2: \quad 2 \pi T_{kk} \geq S'' + \frac{4 G_N}{L} \left(S' \right)^2\,. \label{eq:holoproofQNEC_2}
\end{align}
The result in $d\!>\!2$ is precisely the one obtained from the QFC in \eqref{eq:QNECfromQFC} for a local deformation.
This proof works regardless of the holographic interpretation, since the vector $s^\mu$ being not timelike is a purely geometric statement.
While QNEC itself is a field theoretic inequality, holography relates it to the geometrical proof presented above.

\subsubsection{General Proof}
 \label{sec:QNECgeneralProof}

The proof by Balakrishnan et al.~\cite{Balakrishnan:2017bjg} was inspired by the holographic proof and is built on the same ideas as the ANEC proofs by Faulkner et al.~\cite{Faulkner:2016mzt} and Hartman et al.~\cite{Hartman:2016lgu}, namely \textit{properties of modular Hamiltonians} and \textit{causality considerations}.\\
\\
First they establish an integrated version of QNEC
\begin{equation} \label{eq:intQNECGenProof}
    Q_-(A,B;y) \equiv \int_{\partial A}^{\partial B} \mathrm{d}x^- T_{kk} - \frac{1}{2\pi} \frac{\delta S(B)}{\delta x^-} + \frac{1}{2\pi} \frac{\delta S(A)}{\delta x^-} \geq 0\,,
\end{equation}
where $A$ and $B$ are two spatial regions, related by the deformation of the entangling region.
The EWN property explained above,  guarantees the necessary causal relation between the two regions.\\
\\
The next important idea is to study a correlation function of two probe operators $\mathcal{O}_B$ and $\mathcal{O}_{\bar A}$
\begin{equation} \label{eq:corrFunGenProof}
    f(s) = \frac{\langle \psi | \mathcal{O}_B e^{-i s K_B} e^{i s K_A} \mathcal{O}_{\bar A} | \psi \rangle}{\langle \Omega | \mathcal{O}_B e^{-i s K^0_B} e^{i s K^0_A} \mathcal{O}_{\bar A} | \Omega \rangle}\,,
\end{equation}
where $\psi$ is the considered state, $\Omega$ represents the vacuum and $K$ and $K^0$ are their respective modular Hamiltonians, while the action $e^{-i s K}$ on the operators is called modular flow.
The modular Hamiltonian of some state is defined via its density matrix
\begin{equation}
     K = - \log \rho\,,
\end{equation}
which can be applied to reduced density matrices as well, resulting in modular Hamiltonians of subsystems.\\
\\
While the modular Hamiltonians can be calculated using the replica trick, the correlation function \eqref{eq:corrFunGenProof} is related to the integrated form of QNEC \eqref{eq:intQNECGenProof}.
A number of further considerations and calculations (for which we refer the interested reader to the original paper \cite{Balakrishnan:2017bjg}) finish the proof of QNEC in relativistic QFTs in dimension $d\!\geq\!3$.

\subsection{Stronger Inequality for 2D CFTs}
 \label{sec:stronger2D}

Applying the holographic proof mentioned above to $2$-dimensional boundary theories, led to a stronger QNEC inequality \eqref{eq:holoproofQNEC_2}.
An additional term proportional to Newton's constant and the AdS radius appears.
Using the holographic relation $c\!=\!3L/2G_N$, the prefactor can be written in terms of the central charge of the CFT and therefore restore the purely field theoretic character of the inequality
\begin{equation} \label{eq:QNEC_2}
    \langle T_{\mu\nu} k^\mu k^\nu \rangle \geq \frac{1}{2\pi} \left[ S'' + \frac{6}{c} (S')^2 \right]\,,
\end{equation}
where $S'$ is the first variation of EE w.r.t.~a null deformation of the entangling region.
We will refer to this stronger form valid only for $2$-dimensional CFTs and derformations thereof as QNEC$_2$.\\
\\
One intriguing observation regarding QNEC$_2$ is that the additional term ensures that both sides transform the same way under diffeomorphisms.
The transformation behavior of the EMT under infinitesimal diffeomorphisms is given by
\begin{equation}
    \delta_\xi T = 2 T \xi' + \xi T' - \frac{c}{12} \xi'''\,.
\end{equation}
By inserting the transformation behavior of EE \cite{Wall:2011kb}
\begin{equation} \label{eq:EEtrafo}
    \delta_\xi S = \xi S' + \frac{c}{12} \xi'\,,
\end{equation}
directly into \eqref{eq:QNEC_2} we find after rearranging the terms
\begin{equation}
    \delta_\xi \left[ S'' + \frac{6}{c} (S')^2 \right] = 2 \left[ S'' + \frac{6}{c} (S')^2 \right] \xi' + \xi \left[ S'' + \frac{6}{c} (S')^2 \right]' - \frac{c}{12} \xi'''\,,
\end{equation}
that they transform with an \textit{infinitesimal Schwarzian derivative} as well.
This feature plays an important role in the proof of QNEC$_2$ saturation for states dual to Ba\~nados geometries presented in section \ref{sec:Banados}.

\subsection{Numerical Implementation}
 \label{sec:NumQNEC}

\begin{figure}
	\begin{center}
	\includegraphics[height=.33\textheight]{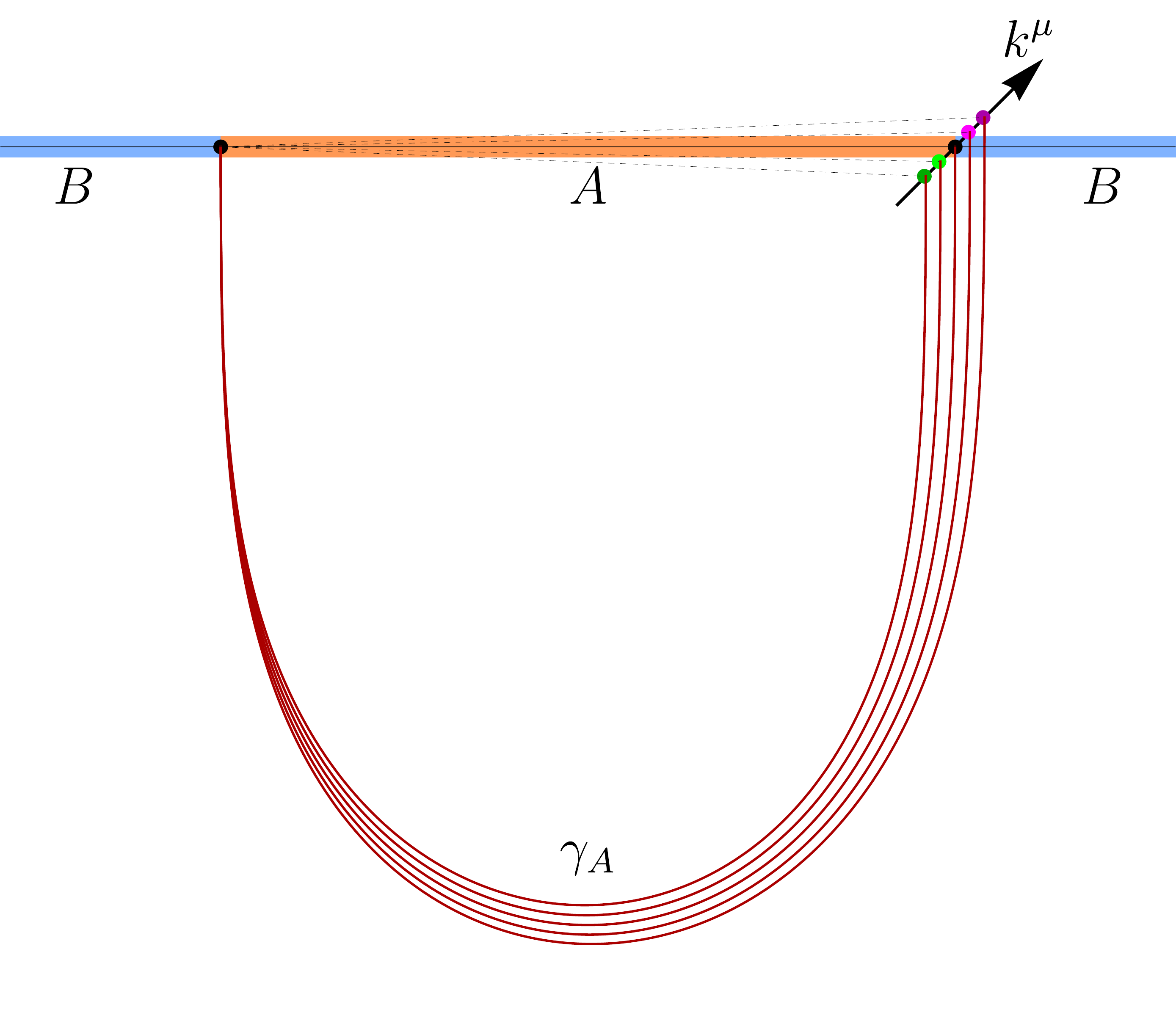}
	\caption{Example for the definition of QNEC in a $2$-dimensional field theory. The entangling region (orange) and its surrounding (blue) are separated by the entangling `surface' (black dots). To obtain QNEC, one boundary point is shifted along the null vector $k^\mu$, generating deformed extremal surfaces $\gamma_A$ (red).}
	\label{fig:2Ddeformation}
	\end{center}
\end{figure}

\begin{figure}
	\begin{center}
	\includegraphics[height=.23\textheight]{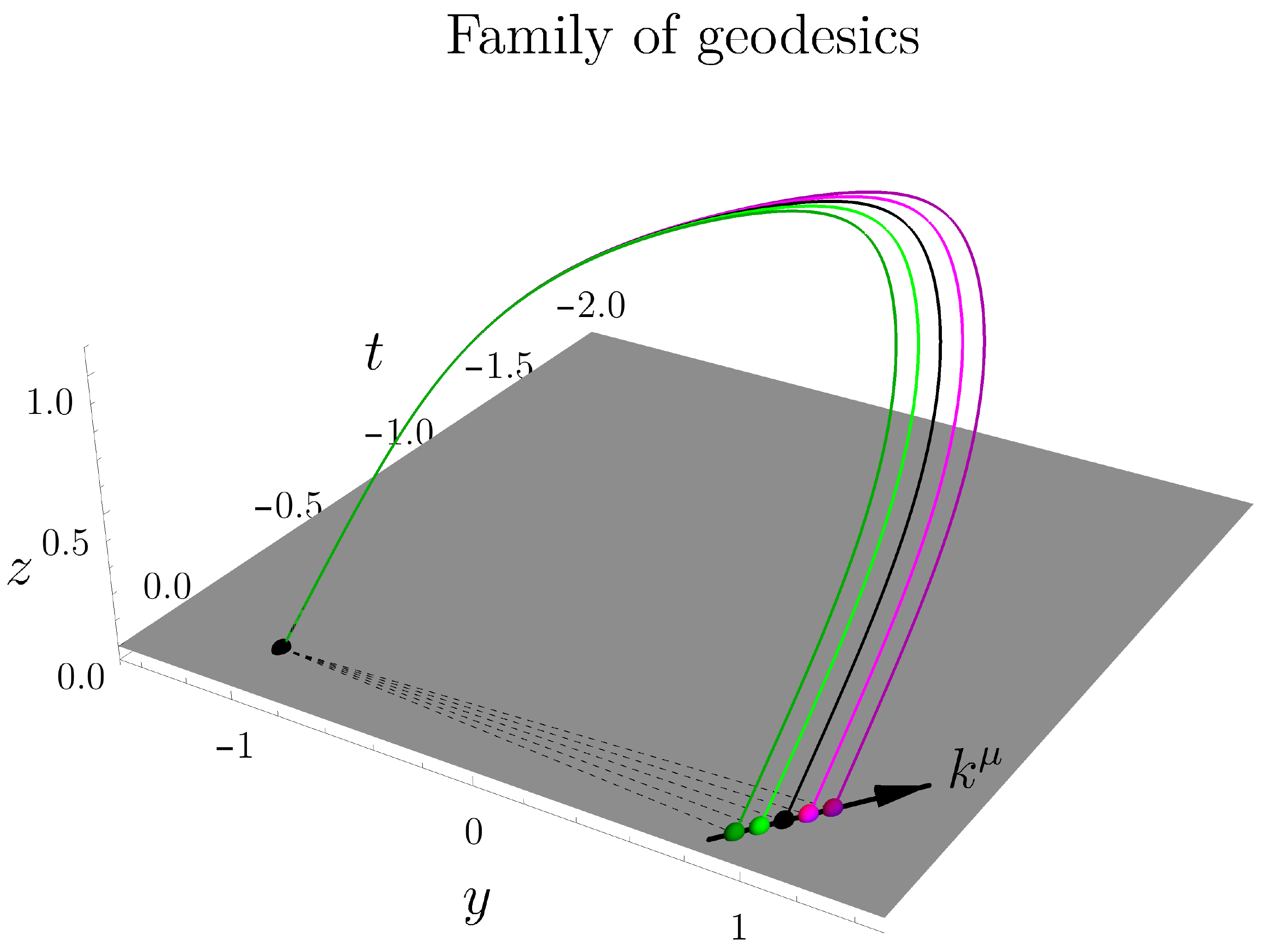}
	\quad
	\includegraphics[height=.23\textheight]{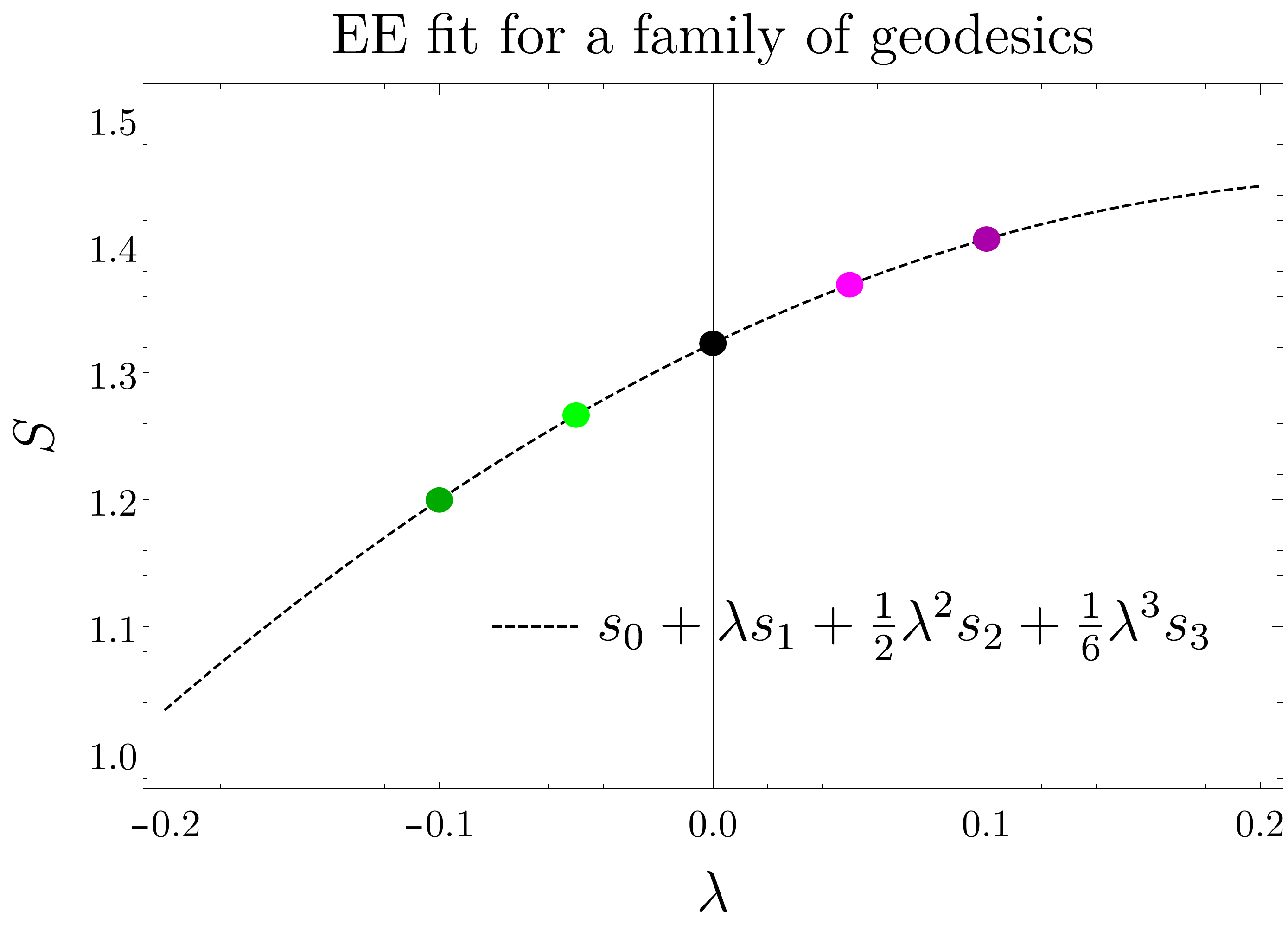}
	\caption{Left panel: A family of geodesics generated by the null deformation of the entangling surface (black dot) along $k^\mu$ at the boundary (gray).
	Right panel: EE and cubic fit as function of the affine parameter $\lambda$ along $k^\mu$. Every value for EE (colored dots) is obtained from the corresponding geodesic with the same color in the left panel.}
	\label{fig:EEinterpolation}
	\end{center}
\end{figure}

In order to calculate QNEC numerically we need several ingredients, introduced in previous chapters.
In section \ref{sec:HEE} the holographic description of EE was introduced, using extremal surfaces (see figure \ref{fig:HEE}).
In section \ref{sec:GeoEQ} it was shown that it is sufficient to consider geodesics and described the methods to solve the geodesic equation (see figure \ref{fig:stripRegion}).
How to obtain QNEC from EE was introduced in \ref{sec:QFC}.
Figure \ref{fig:2Ddeformation} shows a $2$-dimensional boundary field theory (spatial direction along the horizontal line and the time direction going up), which is split into the orange entangling region $A$ and the surrounding $B$.
The entangling surface reduces to two points in this case and is shown as black dots between $A$ and $B$.
Often we refer to the distance between these two point as \textit{separation}, especially when we study how QNEC depends on the size of the entangling region.
The null deformation along $k^\mu$ moves the right boundary point to one of the colored points and changes the entangling region to the corresponding dashed line.
This causes different extremal surfaces $\gamma_A$, sketched in red.\\
To put this description into practice we numerically calculate $n\!=\!5$ geodesics with the left endpoint fixed and the right endpoint shifted a multible of $\lambda\!=\!0.05$ along the deformation vector $k^\mu$, shown by the dark and light green, dark and light pink as well as the central black point in figure \ref{fig:2Ddeformation}.
In the left plot of figure \ref{fig:EEinterpolation} the geodesics are shown in the colors according to their shift at the boundary (gray surface).
The EEs corresponding to the length of the calculated geodesics are shown in the right panel of figure \ref{fig:EEinterpolation}, using the same color code as before.
From this data a cubic fit (dashed black curve) is generated, which is then used to find the first and second derivatives of EE w.r.t.~the affine parameter $\lambda$ along $k^\mu$ at $\lambda\!=\!0$.

\cleardoublepage

\section{QNEC in D=4}
 \label{sec:QNEC_D=4}

When studying QNEC in $4$ dimensions, we use some symmetries of spacetime to simplify the numerical calculations.
In this spirit we consider systems where all but one spatial directions are homogeneous and isotropic, such that the problem can be reduced to effectively $3$-dimensional AdS space.
This allows us to calculate geodesics (instead of higher dimensional surfaces), but also restricts our insight to strip like entangling regions as described in section \ref{sec:GeoEQ} and figure \ref{fig:stripRegion}.
In order to calculate QNEC, the boundary of the entangling region must be deformed, which amounts to shifting not only one point, but the whole boundary line/surface in the homogeneous directions.
Similar to our previous work on EE \cite{Ecker:2016thn}, we start with basic examples and work through to more complicated and more interesting systems, starting from vacuum and thermal states in the CFT and a global quench all the way to the system of colliding shock waves which mimics a heavy ion collision.\\
\\
All our examples use $5$-dimensional metrics of the form
\begin{equation}
    \label{eq:genMetric5d}
    \mathrm{d}s^2=2\mathrm{d}t\,(F\mathrm{d}y-\mathrm{d}z/z^2)-A\mathrm{d}t^2+R^2\left(e^B\mathrm{d}\mathbf{x}_{\perp}^{2}+e^{-2B}\mathrm{d}y^2\right)\,,
\end{equation}
where the functions $A$, $B$, $F$ and $R$ can depend on boundary coordinates $t$, $y$ and the AdS radial coordinate $z$.\\
\\
In these examples we have two different approaches to study QNEC.
The first one is to investigate its dependence on the size of the entangling region.
The second one applies only to time dependent systems and focuses on the change of QNEC during the evolution of a given system.

\subsection{Vacuum State -- Pure Anti-de Sitter Space}

\begin{figure}
	\begin{center}
	\includegraphics[height=.22\textheight]{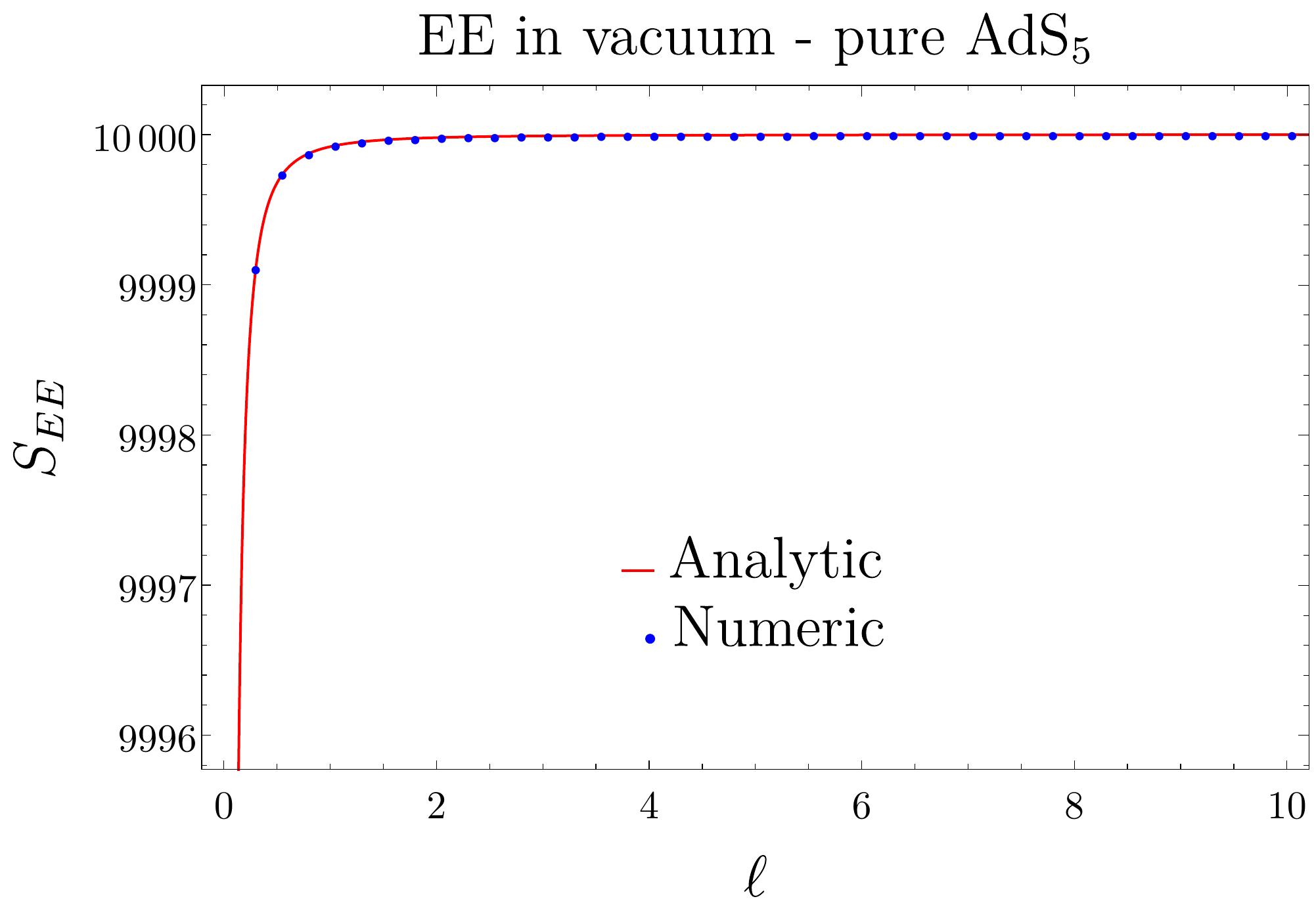}
	\quad
	\includegraphics[height=.22\textheight]{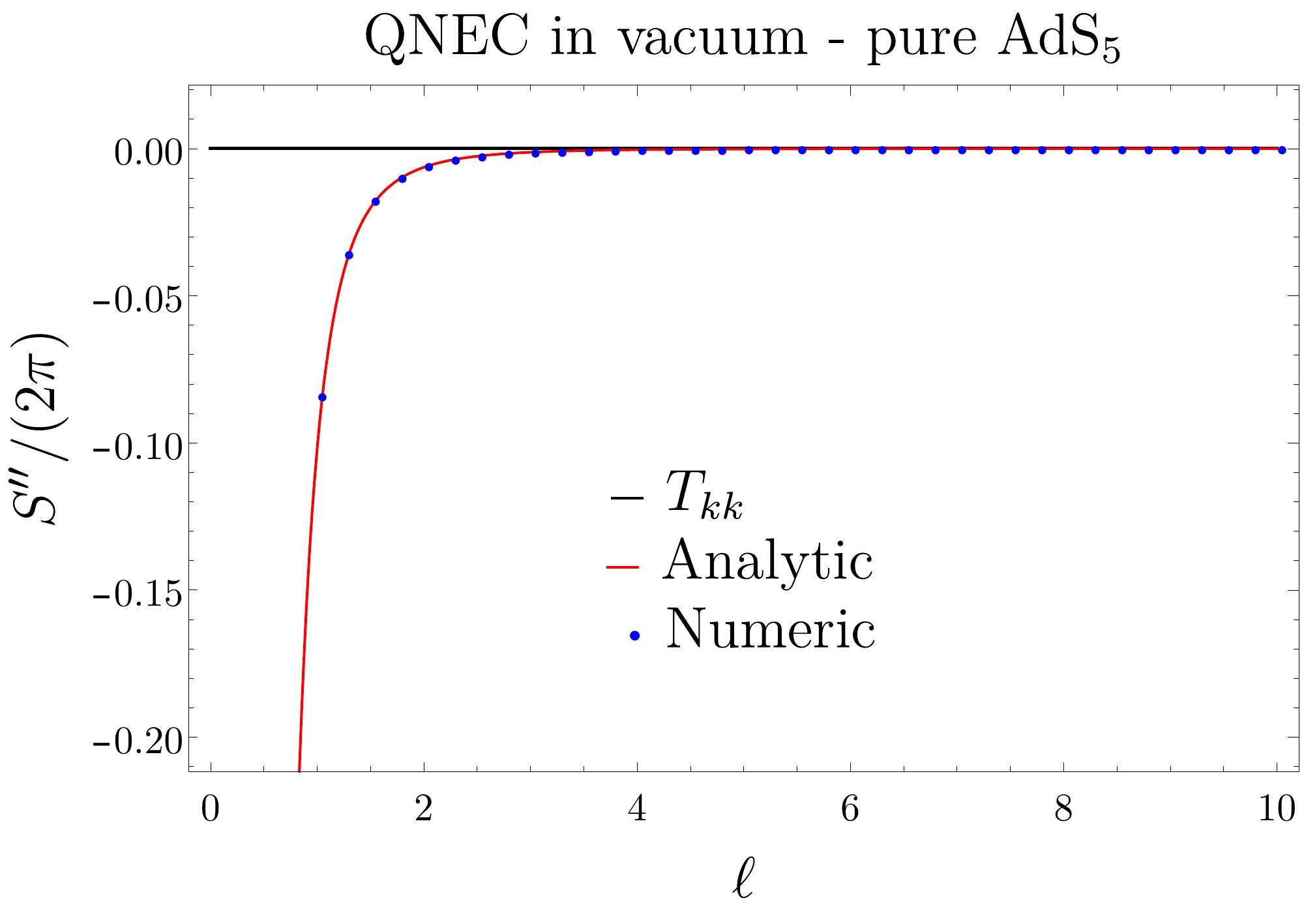}
	\caption{Left panel: Perfect agreement of the numerical computation of EE with the analytic result \eqref{eq:S_vac} (the saturation value of $10000$ is chosen arbitrarily by the cutoff $z_{cut}=5\times10^{-3}$). 
	Right panel: RHS of QNEC ($S''/(2\pi)$) computed with the numeric method explained in section \ref{sec:NumQNEC} in perfect agreement with the analytic result \eqref{eq:QNEC_vac}. As indicated by the black line $T_{kk}=0$, QNEC is of course satisfied and saturates in the limit of large $\ell$.}
	\label{fig:EE_QNEC_vacuum5D}
	\end{center}
\end{figure}

Although this is the most trivial system we can think of [this amounts to the functions in \eqref{eq:genMetric5d} to be $A\!=\!1/z^2,\, B\!=\!0,\, F\!=\!0,\, R\!=\!1/z$], it is a very important one for our work.
Since there exists an exact analytic solution for the EE \cite{Ryu:2006ef}
\begin{equation} \label{eq:S_vac}
    S_{vac} = \frac{1}{4 G_N} \left( \frac{1}{z_\text{cut}^2} - \frac{1}{2c_0^3\ell^2} \right)\,, \qquad c_0=\frac{3\Gamma[1/3]^3}{2^{1/3}(2\pi)^2}\,,
\end{equation}
where $z_{cut}$ is the UV cutoff and $\ell$ is the width of the strip entangling region.
The vacuum state is boost invariant such that the null deformation can be written as $\ell\!=\!\sqrt{(\ell_0\!+\!\lambda)^2\!-\!\lambda^2}$, which leads to QNEC by taking the second derivative with respect to $\lambda$ and subsequently setting $\lambda\!=\!0$
\begin{equation} \label{eq:QNEC_vac}
    \frac{1}{2\pi}\,S''_{vac} = -\frac{1}{2 \pi \, c_0^3 \, \ell_0^4} \approx -\frac{0.102071}{\ell_0^4} \,,
\end{equation}
where Newtons constant $G_N$ is set to unity.
With this result we can verify that the procedure introduced in section \ref{sec:NumQNEC} works to good accuracy.
Obviously the EMT vanishes for the vacuum (and therefore $T_{kk}\!=\!0$), guaranteeing that QNEC is satisfied everywhere.
In the limit of large entangling region ($\ell_0\!\to\!\infty$) it is saturated.
In Figure \ref{fig:EE_QNEC_vacuum5D} we show the numerical results for EE and QNEC as a function of the size of the entangling region $\ell_0$ in the vacuum case together with the known analytic values.
The precise agreement encourages us to continue the path and move to more complicated systems.

\subsection{Thermal State -- Schwarzschild Black Brane}
 \label{sec:AdS5BB}
 
\begin{figure}
	\begin{center}
	\includegraphics[height=.22\textheight]{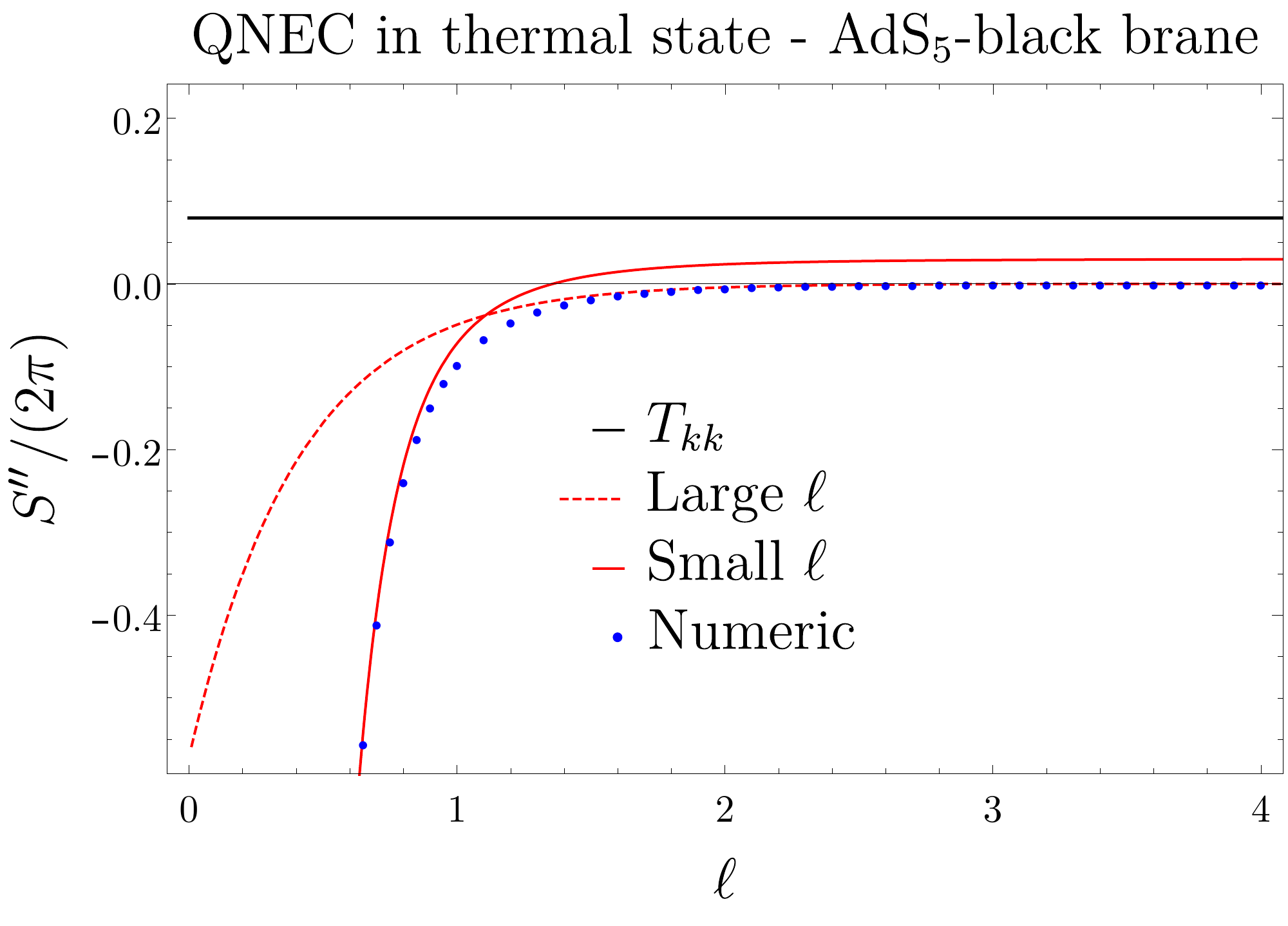}
	\quad
	\includegraphics[height=.22\textheight]{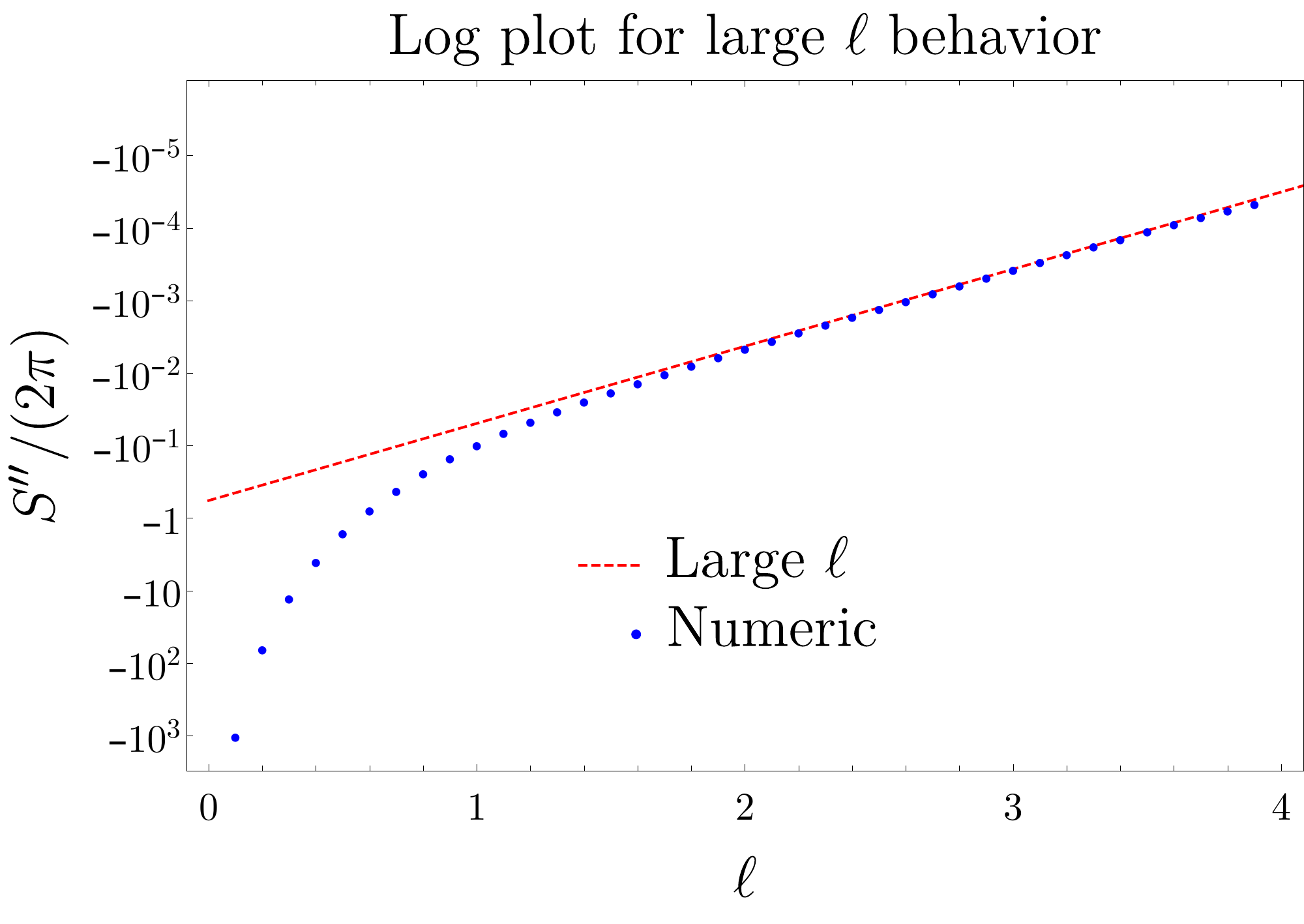}
	\caption{Left panel: One can see how the numerical results (blue dots) deviate from the small $\ell$ result (solid red line) and then approach the large $\ell$ results (dashed red line).
	Right panel: In a logarithmic plot one can see how well the numerics and perturbative result \eqref{eq:AdSBB_largeL} agree in the large $\ell$ regime.}
	\label{fig:QNEC_AdS5BB}
	\end{center}
\end{figure}

The first example where we apply our numerical methods is a thermal state in the CFT dual to a Schwarzschild black brane in AdS.
This system is characterized by a single quantity, the temperature $T$ of the state, related to the mass of the black hole in the dual theory by $M\!=\!(\pi T)^4$.
The metric is determined by setting $A\!=\!1/z^2\!-\!M z^2$, $R\!=\!1/z$ and $B\!=\!F\!=\!0$ in \eqref{eq:genMetric5d}.
In this particularly symmetric geometry, the null projection on both possible directions $k^\mu_\pm\!=\!\{(1,1), (-1,1)\}$ yields the same positive constant for the arbitrary choice $M=1$
\begin{equation}
    T_{\pm\pm} = T_{kk} = \frac{1}{4\pi}\,.
\end{equation}
Later we will encounter other examples where parity symmetry is broken and the direction $k^\mu_\pm$ does make a difference.
Similarly for the variation of EE there are two choices, yielding $S''_\pm$ in general.
The time reversal symmetry of this system leads to the same result for both choices, but this will change as well in more complex systems discussed later on.\\
\\
The exact calculation of QNEC is no longer possible in this system, but using perturbative methods gives us some analytic results.
In the limits of very small and very large entangling regions ($T\ell\!\ll\!1$ or $T\ell\!\gg\!1)$ we found expressions for $S''_\pm\!=\!S''_{therm}$ using series expansions \cite{Ecker:2017jdw}
\begin{align}
    &T\ell \ll 1: \quad \frac{1}{2\pi}\,\mathcal{S}''_{therm} \approx -\frac{0.102071}{\ell^4} + 0.030002 - 0.130203\, \ell^4\,, \label{eq:AdSBB_smallL} \\
    &T\ell \gg 1: \quad \frac{1}{2\pi}\,\mathcal{S}''_{therm} \approx -0.571853\,e^{-\sqrt{6} \ell} \,. \label{eq:AdSBB_largeL}
\end{align}
Details regarding the perturbative calculation can be found in appendix \ref{app:Perturbative}.
In between these limits a numerical calculation is the only way to find the results shown in figure \ref{fig:QNEC_AdS5BB}.
In the left panel one can see the positive null energy $T_{kk}$ together with the numeric results (blue dots) and the perturbative results for small $\ell$ \eqref{eq:AdSBB_smallL} (solid red) and large $\ell$ (dashed red).
The agreement of the numerics with the perturbative results is very good, as can be seen in the right panel in the logarithmic plot of the large $\ell$ behavior.
First of all, we find perfect agreement of the numerical solution with perturbation theory in both limiting cases, which is one more confirmation of the numerical method developed.
In the intermediate regime ($1\!\lesssim\!\ell\!\lesssim\!2$) the numerical solution connects the two asymptotic branches smoothly.
Obviously there is neither saturation nor violation of QNEC in this system.

\subsection{Global Quench -- Vaidya}
 \label{sec:Vaidya5D}
 
\begin{figure}
	\begin{center}
	\includegraphics[height=.22\textheight]{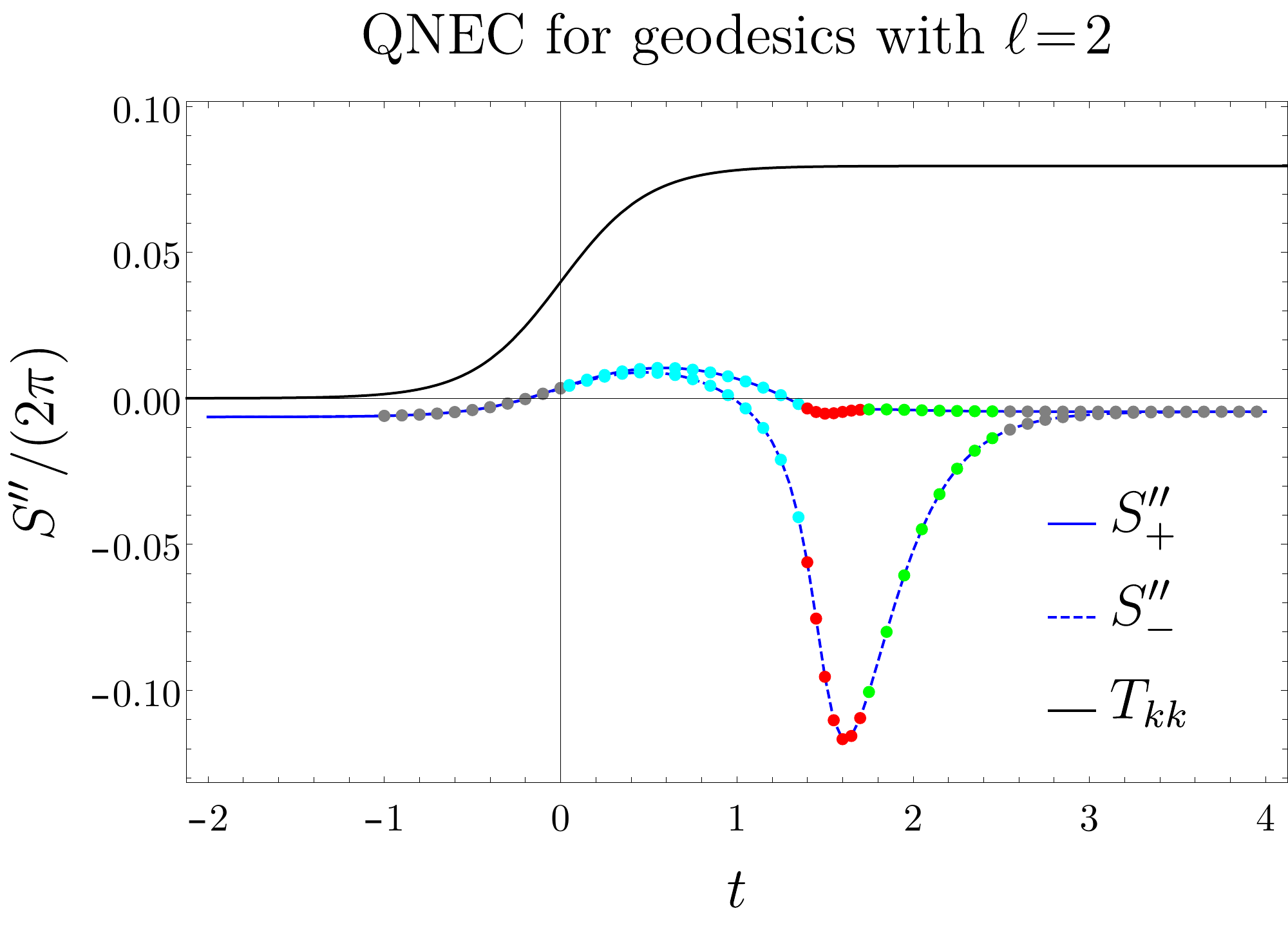}
	\quad
	\includegraphics[height=.22\textheight]{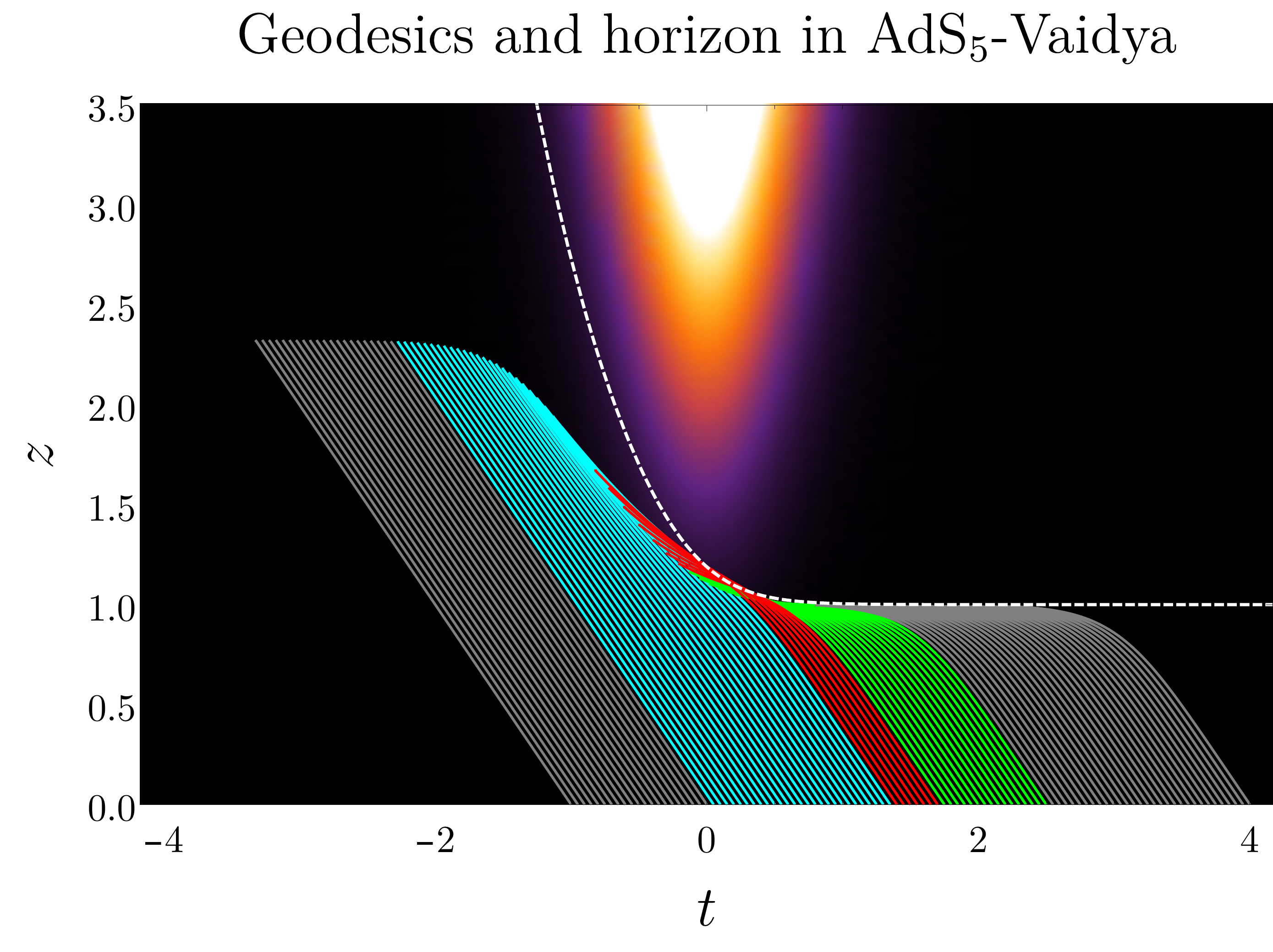}
	\caption{Left panel: Time evolution of QNEC for both null directions. The colored points relate to the right plot.
	Right panel: Central geodesics are influenced by the horizon (white dashed line) and the mass shell (shown as density plot).}
	\label{fig:QNEC_geos_tevo_vaidya5D}
	\end{center}
\end{figure}

Our next step after investigating time independent systems is to consider a global quench in the CFT, which is dual to the so-called AdS-Vaidya geometry, where a homogeneous shell of null dust is injected \cite{AbajoArrastia:2010yt}
\begin{align}
    &A=z^{-2} - M(t) z^2\,,\qquad R=1/z\,,\qquad B=F=0\,, \\
    &M(t)\equiv\tfrac{1}{2}\left(1+\tanh(2\,t)\right)\,.
\label{eq:MVaidya}
\end{align}
In this parity symmetric setup both null projections of the EMT are given by the same expression
\begin{equation}
    T_{kk}=\frac{1}{4\pi}M(t)\,.
\end{equation}
Due to the time dependence of \eqref{eq:MVaidya} we get different results for the variation of EE, depending on the null vector chosen $S''_+$ and $S''_-$.
One deformation is aligned with the null dust, while the other one is perpendicular.\\
\\
The first approach we take to analyse QNEC is looking at the time evolution.
The right panel of figure \ref{fig:QNEC_geos_tevo_vaidya5D} shows the undeformed (central) geodesics with separation $\ell\!=\!2$ and the apparent horizon (white dashed line) in front of a density plot of the mass shell.
The position of the apparent horizon and the bulk EMT are given by
\begin{equation}
    z_h = \frac{1}{M(t)^{1/4}}\,, \qquad\qquad T^{bulk}_{tt} = \frac{3 z^3}{2} M'(t)\,.
\end{equation}
The left plot shows QNEC, computed by deforming these geodesics.
Before the quench at $t\!=\!-\infty$, the system is in its vacuum state where QNEC is known analytically.
At late times $t\!=\!+\infty$, the system is equivalent to the thermal state with $T\!=\!\frac{1}{\pi}$, leading to familiar behavior just discussed in the previous section.
The geodesics in these two regions are colored gray in figure \ref{fig:QNEC_geos_tevo_vaidya5D} and are not influenced by the quench.
In spite of that, the second derivative of their length is slightly different from the expected values at $\pm\infty$.
The interesting part is the behavior of QNEC just after the quench at $t\!=\!0$.
The geodesics marked in light blue and green are pushed away from the horizon, backwards in time and away from the center ($z\!=\!\infty$).
The red curves extend beyond the horizon, but their turning point ($z_{max},T_{min}$) is always outside the horizon.
\begin{figure}
	\begin{center}
	\includegraphics[height=.22\textheight]{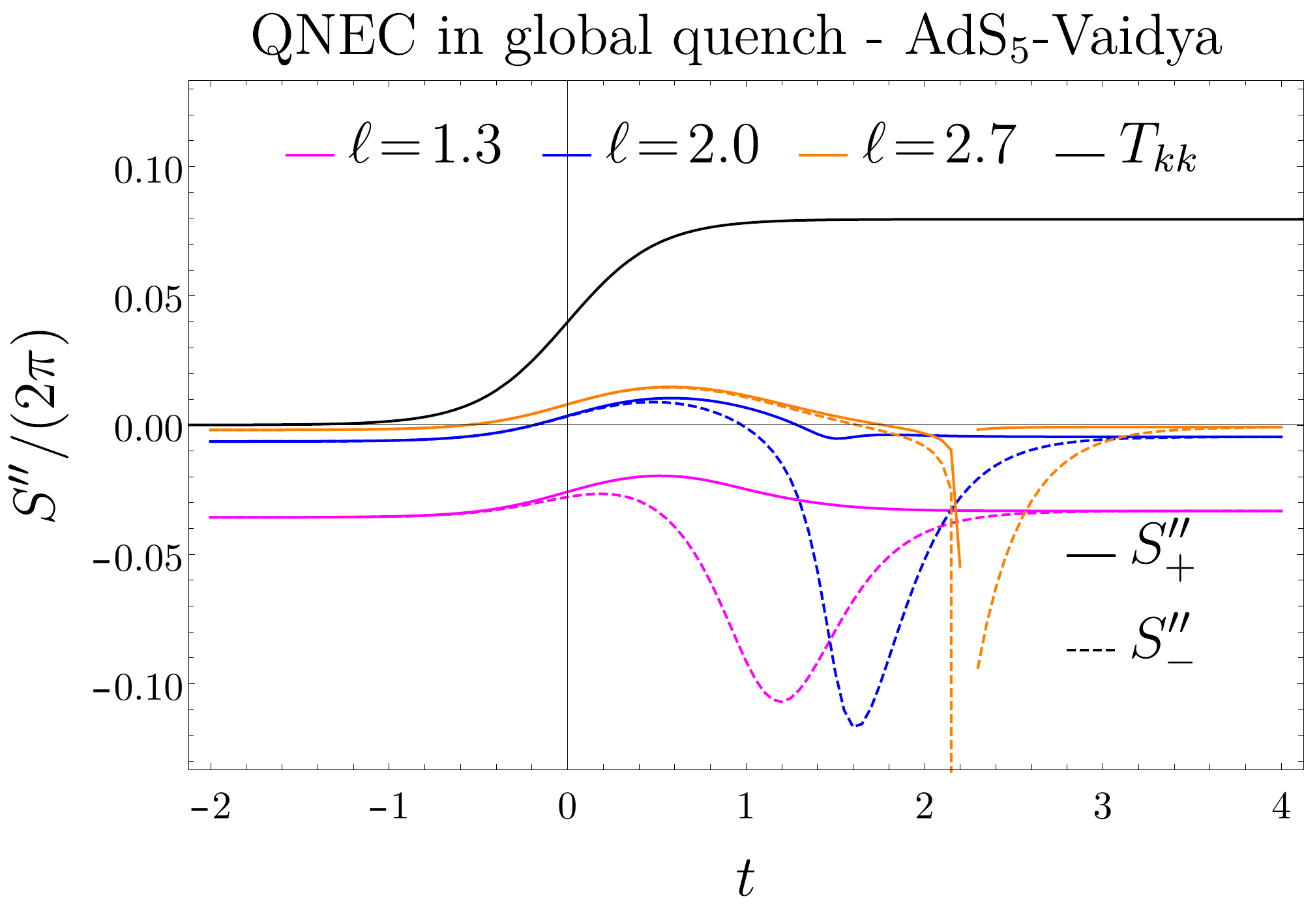}
	\quad
	\includegraphics[height=.22\textheight]{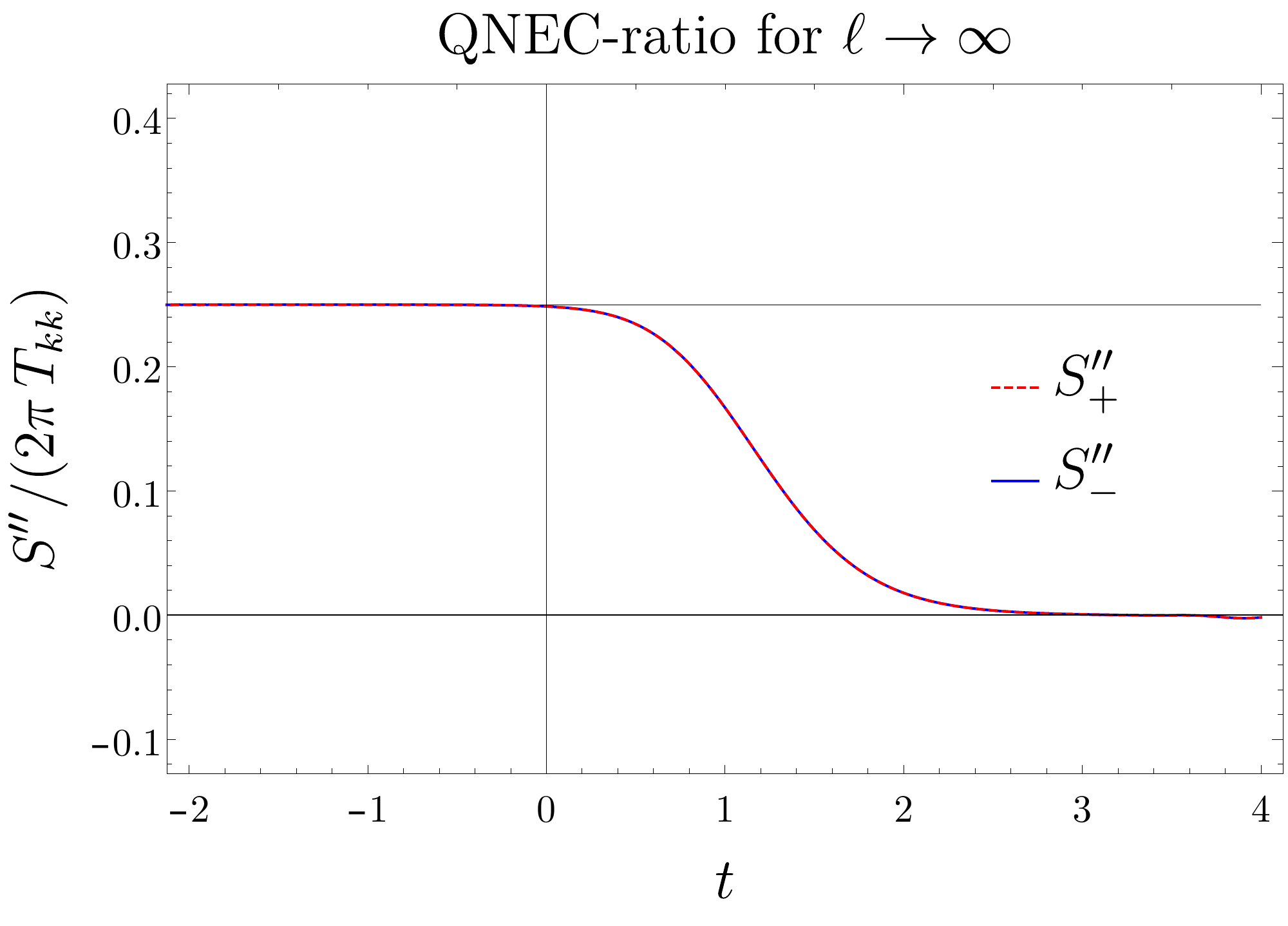}
	\caption{Left panel: Time evolution of QNEC for both null directions and three characteristic separations $\ell$.
	Right panel: The ratio of both sides of QNEC in the limit of large entangling regions.}
	\label{fig:QNEC_tevo_vaidya5D}
	\end{center}
\end{figure} \\
In figure \ref{fig:QNEC_tevo_vaidya5D} we show the time evolution of QNEC for three different separations $\ell$ in the left panel and the $\ell\!\to\!\infty$ limit of the ratio of both sides of QNEC in the right panel.
A common feature of all three cases is that QNEC settles to its late time value considerably later than $T_{kk}$.
Further we observe that for larger separation $S''$ approaches $T_{kk}$ but also the difference between $S''_+$ and $S''_-$ reduces.
While the bump in $S''_+$ stays between $t\!=\!0$ and $t\!=\!1$, the dip in $S''_-$ shifts to later times and becomes almost $\delta$-like.
The small dip in $S''_+$ around $t\!=\!1.5$ in the blue solid line (not visible for $\ell\!=\!1.3$) also develops into a sharp divergence.
Going to larger separation (and to $\ell\!\to\!\infty$ eventually) the only remaining feature is the first bump, such that $S''_+$ and $S''_-$ are identical.
After the quench, when the thermal state is reached, $S''_\pm$ vanishes as expected from the static black brane example.
This behavior is shown in the right panel, where the gray line marks the value $\tfrac{1}{4}$.
This special value is found by taking the ration of both sides of QNEC and is maintained until $S''_\pm$ settles to its thermal value after the quench.
This $\tfrac{1}{4}$-saturation was recently explained as $\tfrac{1}{d}$-saturation (where $d$ is the dimension of the field theory) by Mezei and Virrueta \cite{Mezei:2019sla} who studied quantum quenches and QNEC constraints imposed on them.
\begin{figure}
	\begin{center}
	\includegraphics[height=.21\textheight]{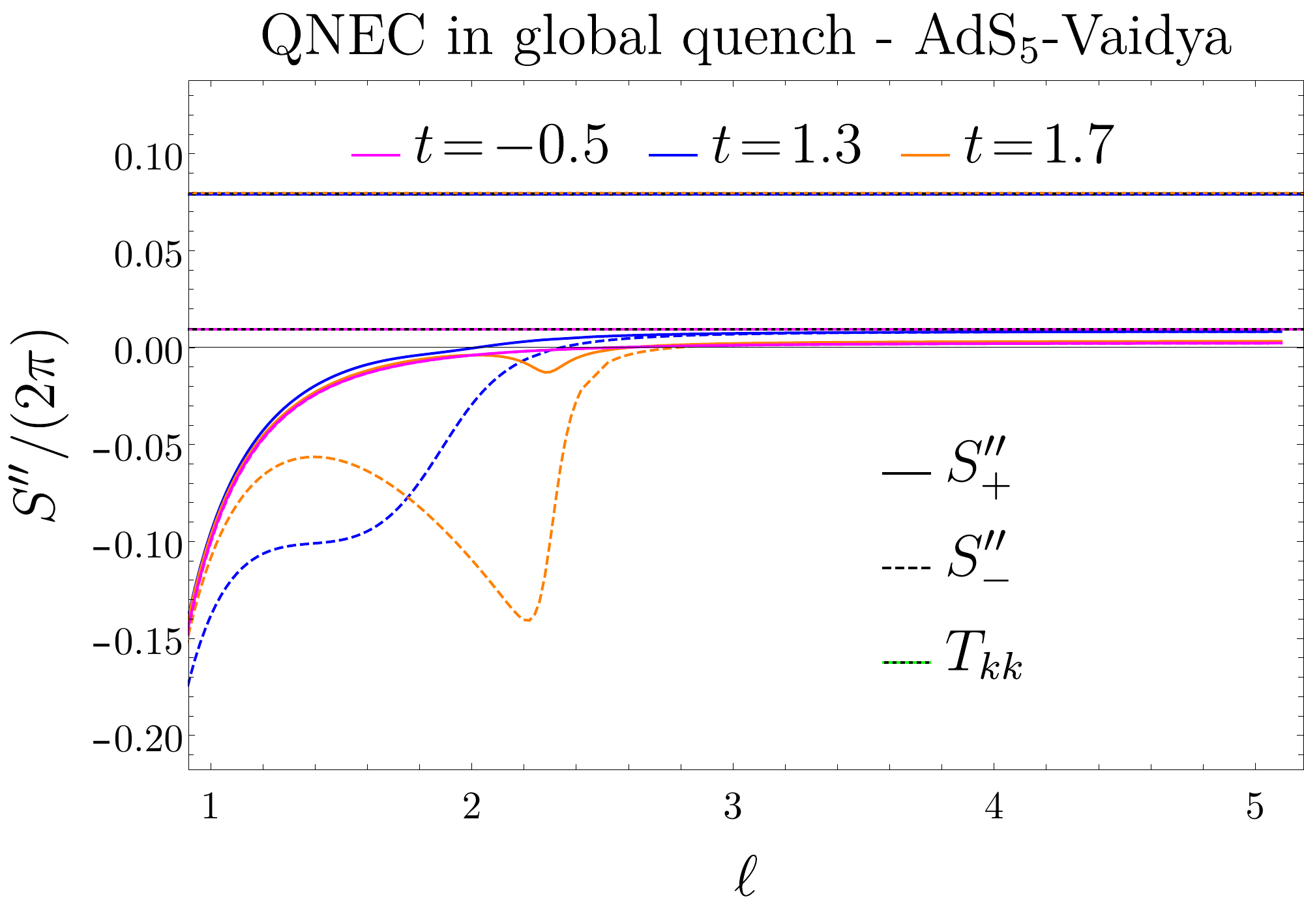}
	\quad
	\includegraphics[height=.21\textheight]{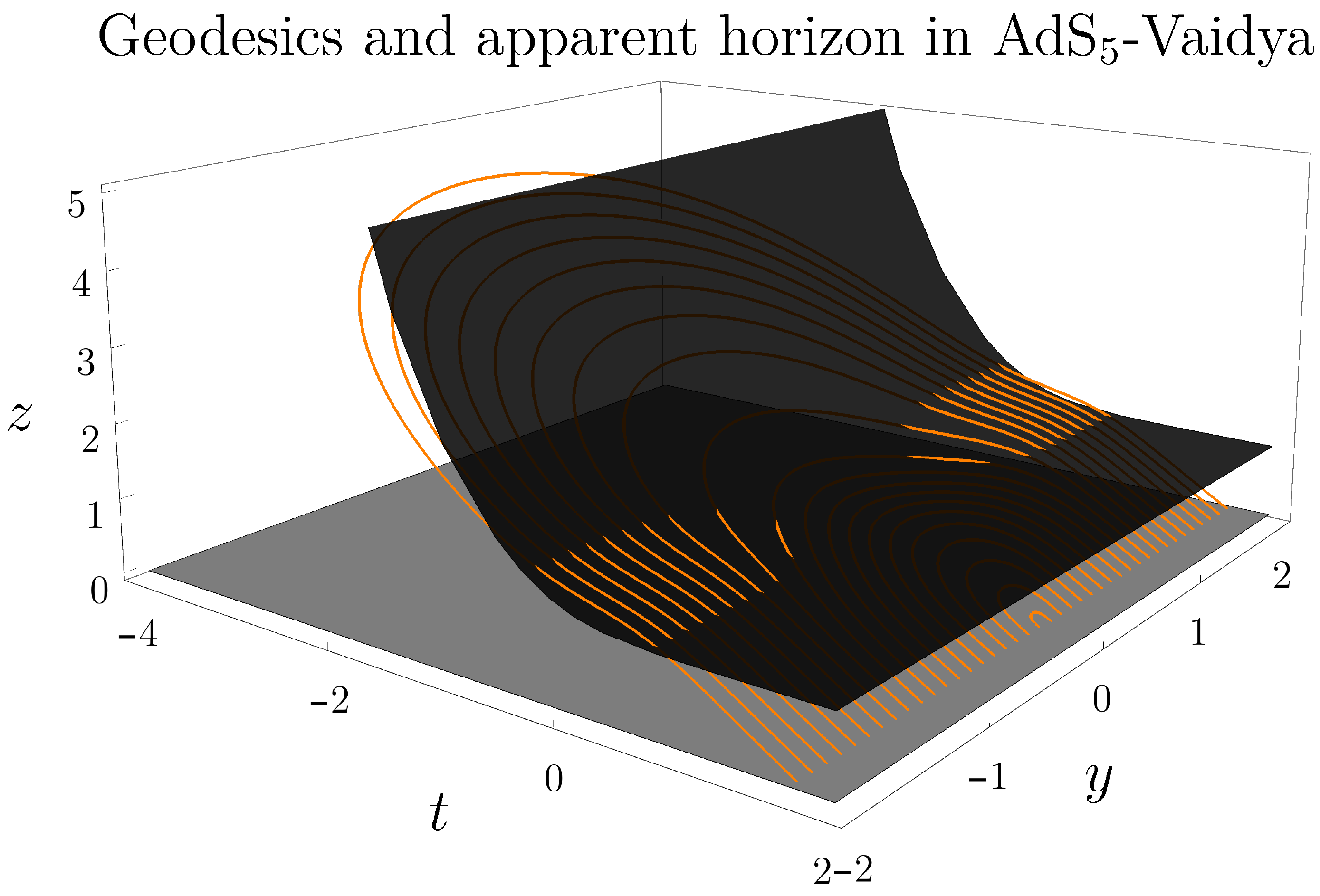}
	\caption{Left panel: Dependence of QNEC on the interval size $\ell$ for both null directions and three characteristic points in time.
	Right panel: Central geodesics for the corresponding orange curve in the left plot intersecting the horizon (black surface).}
	\label{fig:QNEC_levo_vaidya5D}
	\end{center}
\end{figure} \\
The second way we investigate QNEC is its dependence on the size of the entangling region $\ell$.
We do this analysis at three different points in time: before the quench at $t_1\!=\!-0.5$, after the quench at $t_2\!=\!1.3$ and even later at $t_3\!=\!1.7$.
The results are shown in the left panel of figure \ref{fig:QNEC_levo_vaidya5D}, while we show the geodesics (orange curves) corresponding to $t_3$ together with the horizon (black surface) in the right panel.
At early times (pink lines) we cannot distinguish between $S''_\pm$, but we can see a marginally positive value for $S''_\pm$ at larger separations.
This means that even before the quench, QNEC is a stronger condition than the classical NEC.
The blue and orange curves show that after the quench $S''_-$ (dashed lines) is influenced way stronger than $S''_+$ (solid lines).
Similarly to the observation in the time evolution, the dip develops into a $\delta$-like divergence, that is shifted to larger separations for later times.\\
The right panel of figure \ref{fig:QNEC_levo_vaidya5D} shows how the geodesics cross the horizon, keeping the turning point outside.\\
\\
In all cases analysed above, we find that QNEC is satisfied and never violated.
Interestingly, already in this simple setup QNEC poses a stronger restriction on the null energy than the classical NEC.
This is visible especially in the large $\ell$ limit, but can already be seen for fairly small separations and at early times.

\subsection{Heavy Ion Collision -- Colliding Shock Waves}
 \label{sec:HIC}

\begin{figure}
	\begin{center}
	\includegraphics[height=.22\textheight]{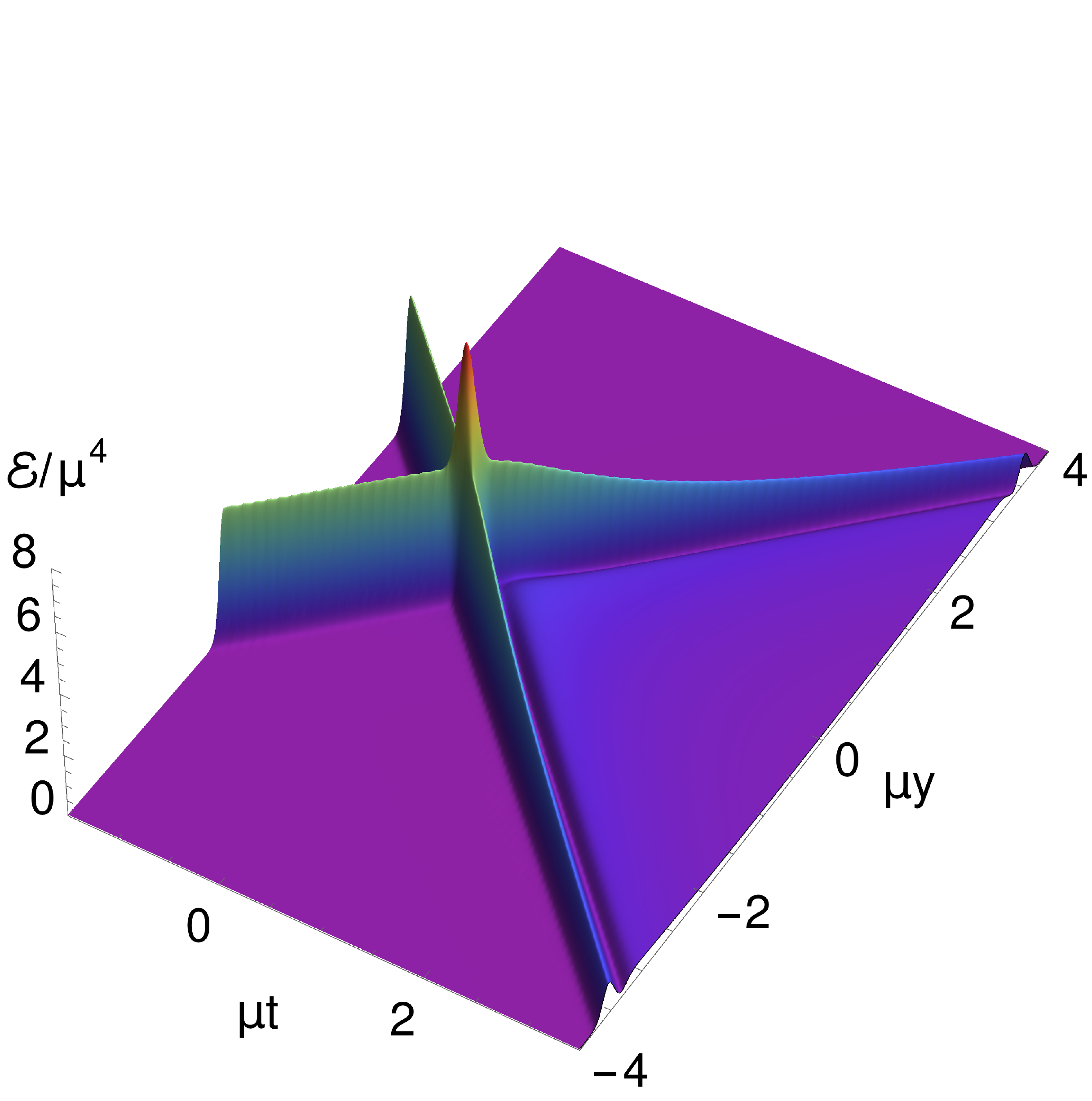}
	\quad
	\includegraphics[height=.22\textheight]{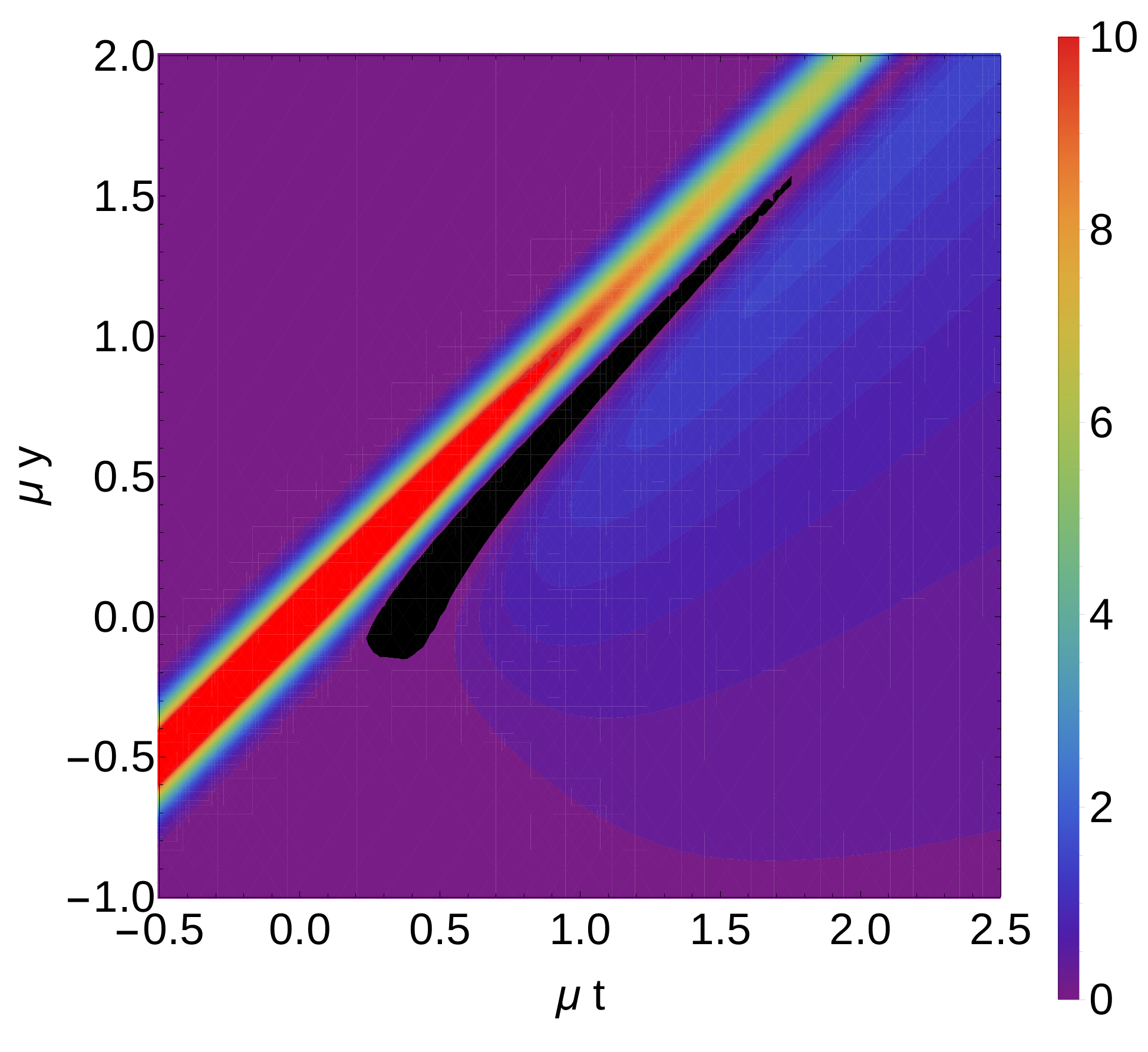}
	\caption{Left panel: Energy density $\mathcal{E}$ in the system of colliding shock waves.
	Right panel: $T_{kk}$ in this system. In the black region the classical NEC is violated.}
	\label{fig:shocks}
	\end{center}
\end{figure}

\begin{figure}
	\begin{center}
	\includegraphics[height=.23\textheight]{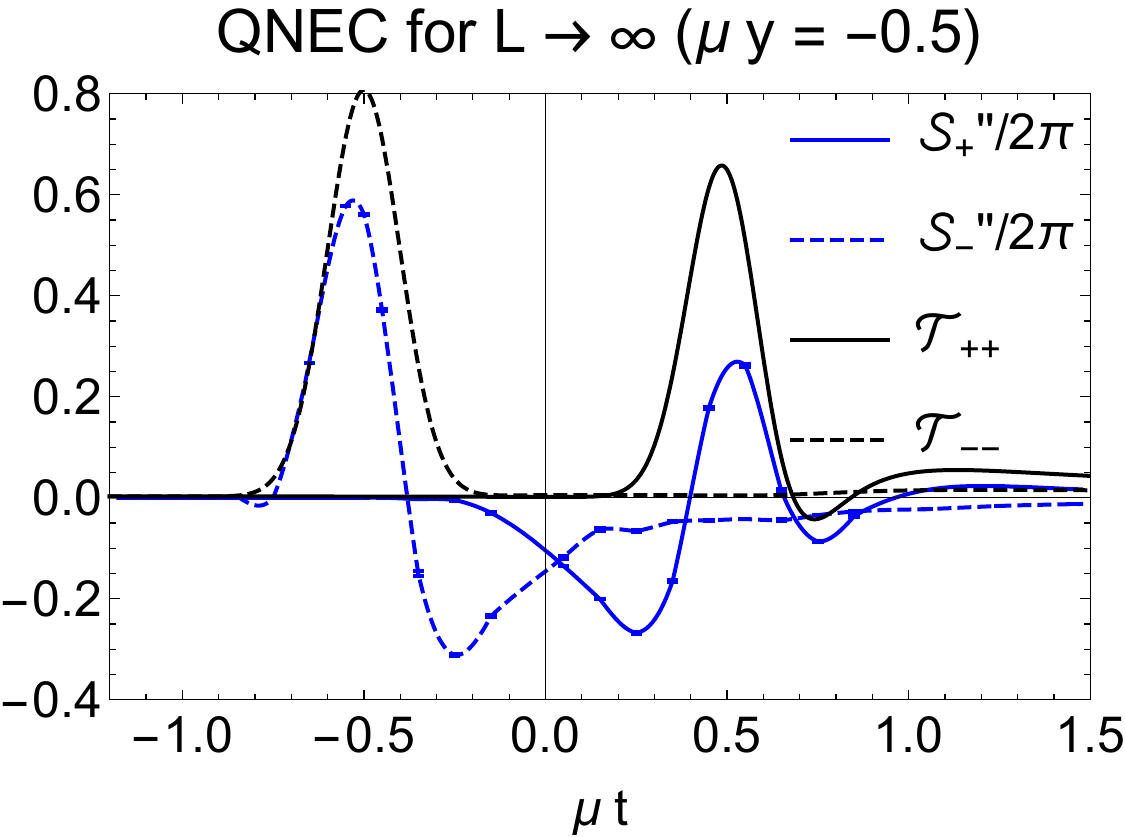}
	\quad
	\includegraphics[height=.23\textheight]{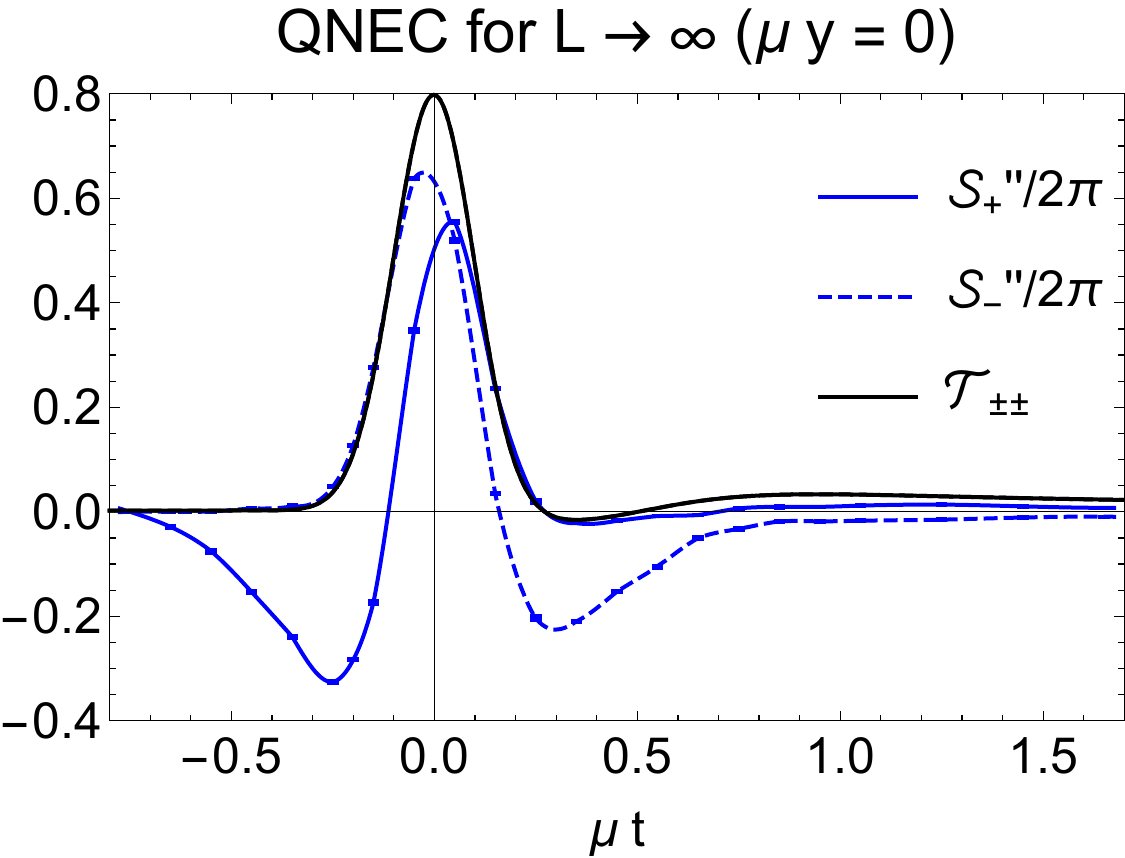}
	\caption{Left Panel: Time evolution of QNEC at $\mu y\!=\!-0.5$ in the limit of large separation.
	Right Panel: Time evolution of QNEC at $\mu y\!=\!0.0$ in the limit of large separation.}
	\label{fig:QNEC_tevo_shocks1}
	\end{center}
\end{figure}

\begin{figure}
	\begin{center}
	\includegraphics[height=.23\textheight]{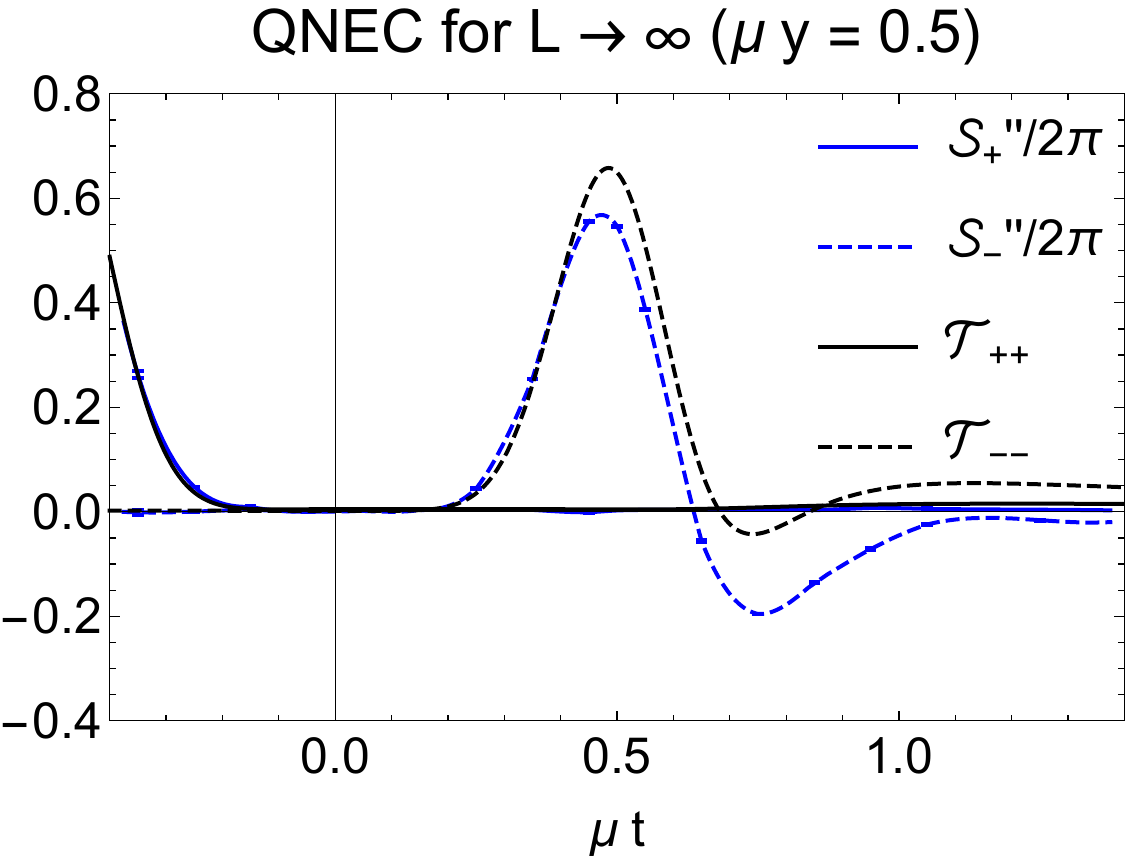}
	\caption{Time evolution of QNEC at $\mu y\!=\!0.5$ in the limit of large separation.}
	\label{fig:QNEC_tevo_shocks2}
	\end{center}
\end{figure}

With all this experience and tools developed in the previous examples, we can now tackle the most complex system with $4$-dimensional boundary theory, colliding lumps of energy in a CFT being a toymodel for HICs.
We studied EE and correlation functions in this system in \cite{Ecker:2016thn} and observed a violation of the classical NEC.
Figure \ref{fig:shocks} shows the energy density (left) and the null projection of the EMT (right).
We observe a region in the forward lightcone of the collision where $T_{kk}$ is negative (black region), which makes it especially interesting to calculate QNEC in that region.
In the left plot one can see a small valley right next to the outgoing shocks.
In this region the energy density is negative, leading in combination with the pressure components (not shown here) to NEC violation.
This happens when the initial conditions are chosen such that the shock waves are very sharp, almost like delta functions \cite{Arnold:2014jva} (the case of delta-like shocks was discussed analytically in \cite{Grumiller:2008va}).
To capture the dynamics of the gravitational shock waves, all functions in \eqref{eq:genMetric5d} are needed and parity as well as time-reversal invariance are broken.
The numerical solution of Einstein's equations to find the metric, determine the boundary EMT as well as the study of holographic EE can be found in our previous work \cite{Ecker:2016thn}.
Using these results allows us to calculate the variations of EE with the relaxation method and the null projections of the EMT evaluate to
\begin{align}
    T_{\pm\pm}&=1/2\pi^2h_\pm(x_\pm)\,, \qquad h_\pm(x_\pm)=\mu^3\exp[-x_\pm^2/2w^2]/\sqrt{2{\pi}w^2}\,, \\
    &x_\pm=t{\pm}y\,, \qquad \qquad \qquad \mu w=0.1\,. 
\end{align}
Since $\mu$ sets the energy scale for the system, we will measure everything in units of $\mu$.
\begin{figure}
	\begin{center}
	\includegraphics[height=.23\textheight]{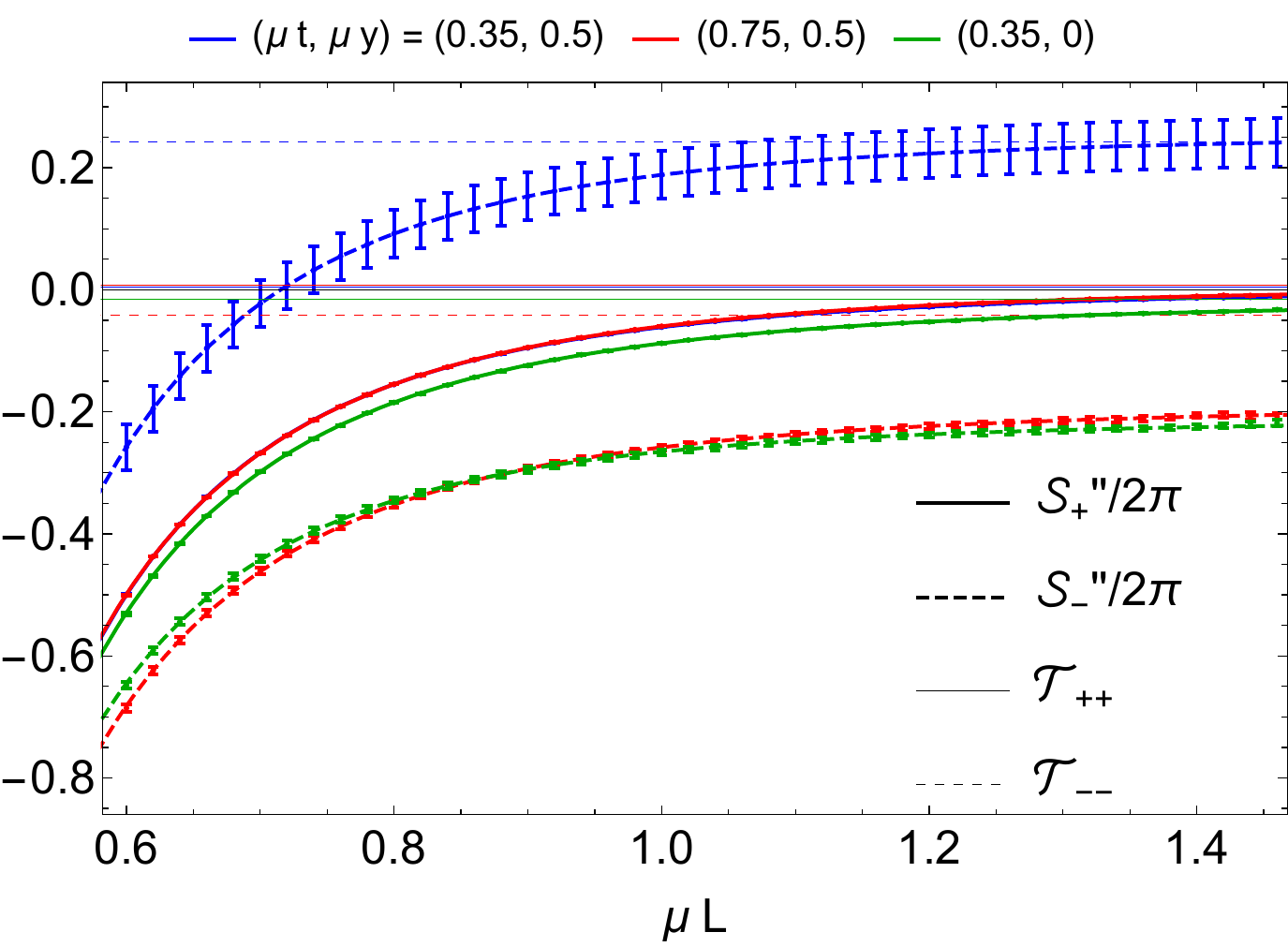}
	\caption{QNEC as function of the separation of the boundary points $\ell\!=\!\mu L$.}
	\label{fig:QNEC_levo_shocks}
	\end{center}
\end{figure}
First we pick three interesting points and evaluate the dependence of QNEC on the boundary separation $\ell\!=\!\mu L$.
In figure \ref{fig:QNEC_levo_shocks} one can see that it depends on the null direction how QNEC behaves.
The blue curves correspond to the first point, chosen to be (almost) on top of the outgoing shock wave in the $k^\mu_-$-direction.
Therefore $T_{--}$ as well as $S_-''$ are positive, while $T_{++}$ is almost zero and $S_+''$ hides behind the solid red curve.
At this point the numerics show QNEC nearly saturated already for rather small intervals, which suggest that it is saturated in the limit of large intervals.
For the next point (red) we picked the position where NEC is violated the most at $T_{--}\!=\!-0.04\mu$.
Here we observe that $S_-''$ is far below that value and saturation is not likely even in the infinity limit of $\ell$.
In the other null direction this point behaves identical to the one shown in blue (solid).
The last point is very interesting as well, since it shows NEC violation in both null directions, lying at the center ($\mu y\!=\!0$) shortly after the collision.
Although $T_{\pm\pm}$ are identical (along the whole $y$-axis, i.e.~the beam axis of the collision), $S_\pm''$ are different.
$S_+''$ is much closer to $T_{\pm\pm}$ than $S_-''$, but saturation at large separations seems unlikely.\\
\\
From these results we can see that QNEC is restricting the null energy stronger, the larger the entangling region gets.
Making use of this, we extract the asymptotic value by fitting the numerical data and extrapolating to $\ell\!\to\!\infty$.
We do this at a big number of points and study the time evolution along three values for $\mu y\!=\!\{-0.5,0.0,0.5\}$, shown in figures \ref{fig:QNEC_tevo_shocks1} and \ref{fig:QNEC_tevo_shocks2}.
We can see that QNEC is satisfied at all points, even if NEC is violated.
The striking result is that QNEC saturates even in highly dynamic out of equilibrium regimes of the system.
In the central plane ($\mu y\!=\!0$) QNEC saturates before (after) the collision for a deformation along $k_-^\mu$ ($k_+^\mu$).
After the collision, a hydrodynamic phase is reached and saturation is not observed anymore.
Away from the central plane we find that QNEC is close to saturation as the outgoing remnants of the collision are passing.
At $\mu y\!=\!0.5$ this is the case for $S_-''$ while $S_+''$ and $T_{++}$ are vanish both.
No saturation is observer at ($\mu y\!=\!-0.5$).
The reason for this difference is that the entangling region always spans towards $+\infty$.
This means in one case it covers all the action between $\pm0.5$, while in the other case it covers only the outside.

\cleardoublepage

\section{QNEC in D=2}
 \label{sec:QNEC_D=2}

Studying QNEC in two dimensions is especially interesting, since the additional term restricts the inequality even more.
Further the tremendous advantage of two dimensions is that the infinite conformal symmetries provide a lot of analytic control.\\
\\
We will proceed as before, verifying the quality of our numerics, using analytic solutions, before applying it to more complex settings.
In addition we present a neat proof for QNEC$_2$ in theories dual to a general class of vacuum solutions to Einstein's equations in three dimensions.
This means that a global quench in $2$ dimensions is a transition from one state where QNEC$_2$ is saturated to another, which we will investigate numerically.
Where possible we augment our numerical results with perturbative calculations in some limiting cases accessible with this methods.
Including matter in the bulk theory leads to excited CFT states, that are introduced via the backreaction of a single scalar particle in global AdS following \cite{Belin:2018juv}.
Finally we consider a self interacting scalar field inducing a phase transition from small to large BHs in the gravitational system similar to \cite{Janik:2015iry}.
This is caused by the choice of potential of a scalar field and can be tuned to first or second order phase transition as well as a crossover.

\subsection{Vacuum and Thermal State -- BTZ Black Hole}
 \label{sec:3DvacBTZ}

Similar to the higher dimensional case, we use the vacuum state to make sure our methods work and give the expected result.
The more interesting case is given by the BTZ black hole in AdS space, which is dual to a thermal CFT$_2$ state.
The special case of a $2$-dimensional CFT allows to calculate EE analytically even for the thermal state and therefore provides an even better example to check our numerics and match them with analytical results.\\
\\
The metric used in the numerical calculation is given by \eqref{eq:genMetric5d}, but without the transversal directions
\begin{equation} \label{eq:genMetric3d}
    \mathrm{d}s^2=-A\mathrm{d}t^2-\frac{2\mathrm{d}t\mathrm{d}z}{z^2}+R^2 e^{-2B}\mathrm{d}y^2\,.
\end{equation}
Further we consider only cases where $R\!=\!\frac{1}{z}$ and $F$ is not required.\\
\\
In this section we use the blackening function $A\!=\!\frac{1}{z^2}(1-M z^2)$, where the vacuum case is covered by $M\!=\!0$ and for the BTZ black hole we chose $M\!=\!1$.
The projection of the boundary EMT is given by
\begin{equation} \label{eq:Tkk_BTZ}
    T_{kk} = \frac{M}{8 \pi \, G_N}\,.
\end{equation}
It is straight forward to apply the relaxation algorithm to these two cases and we find perfect agreement for QNEC$_2$ saturation independent of the size of the entangling region for both cases.

\begin{figure}
	\begin{center}
	\includegraphics[height=.22\textheight]{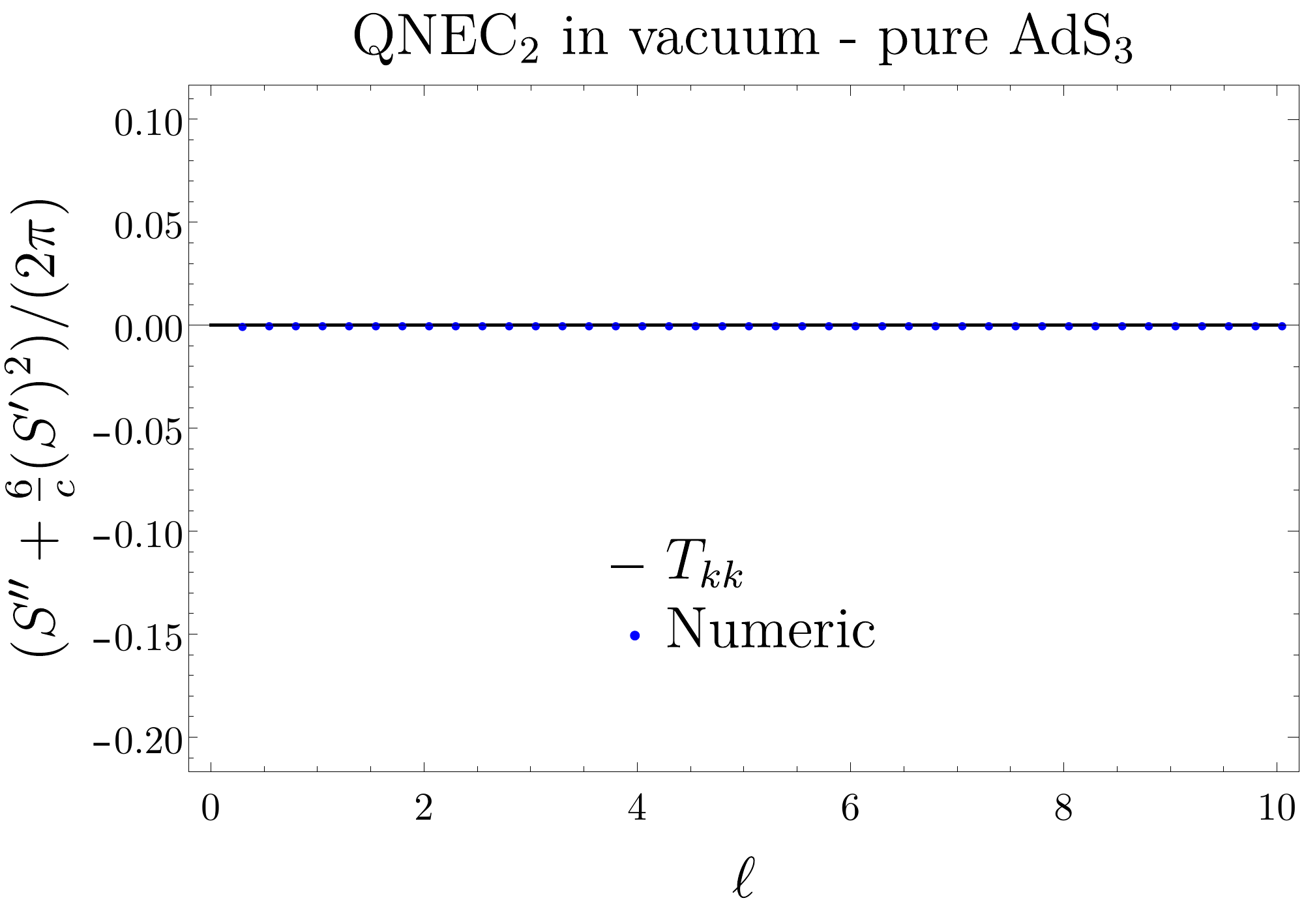}
	\quad
	\includegraphics[height=.22\textheight]{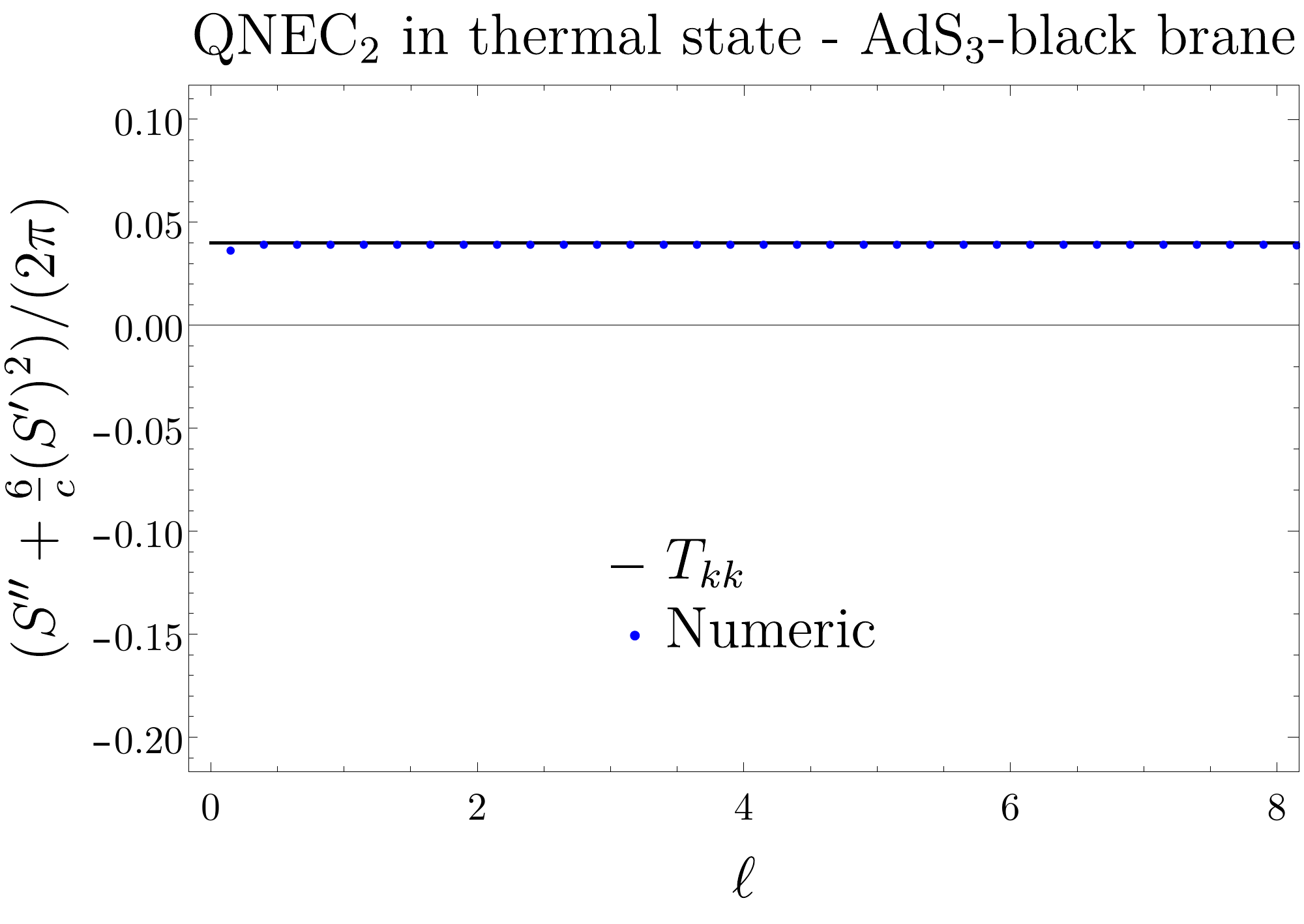}
	\caption{Left panel: Perfect agreement of the numerical computation of QNEC$_2$ with $T_{kk}\!=\!0$. 
	Right panel: Perfect agreement of QNEC$_2$ with \eqref{eq:Tkk_BTZ}.}
	\label{fig:EE_QNEC_vacuum3D}
	\end{center}
\end{figure}

\vspace{25px}
\subsection{Ba\~nados Geometries -- a Proof}
 \label{sec:Banados}

The behavior seen in the previous section can be explained the following way.
The solutions of Einstein gravity with Brown Henneaux boundary conditions in three dimensions are given by the Ba\~nados family of metrics \cite{Banados:1998gg}
\begin{equation} \label{eq:banadosmetric}
\mathrm{d}s^2 = \frac{\mathrm{d}z^2\!-\mathrm{d}x^+\mathrm{d}x^-}{z^2} + \mathcal{L}^+\!(x^+)\left(\mathrm{d}x^+\right)^2\!+ \mathcal{L}^-\!(x^-)\left(\mathrm{d}x^-\right)^2\!- z^2 \mathcal{L}^+\!(x^+)\mathcal{L}^-\!(x^-)\mathrm{d}x^+\mathrm{d}x^-\,.
\end{equation}
If the (anti-) holomorphic functions $\mathcal{L}^\pm$ are positive constants we recover the family of non-extremal BTZ black holes (the extremal limit is obtained if exactly one of these constants vanishes).
Poincar\'e patch AdS$_3$ (we considered as `vacuum' before) corresponds to ${\cal L}^\pm\!=\!0$ and global AdS$_3$ to ${\cal L}^\pm\!=\!-\frac{1}{4}$.
The field theory states dual to the Ba\~nados geometries are given by excited CFT$_2$ states.
Those with non-constant functions $\mathcal{L}^\pm$ are Virasoro descendents of the ones with constant $\mathcal{L}^\pm$.
The null projection of the EMT is fully determined by these functions
\begin{equation} \label{eq:Tkk_banados}
    2\pi\, T_{kk} = 2\pi\,\langle \mathcal{L}^+, \mathcal{L}^- | T_{\pm\pm}(x^\pm) | \mathcal{L}^+, \mathcal{L}^- \rangle = \frac{c}{6}\,\mathcal{L}^\pm(x^\pm)\,,
\end{equation}
where $c$ is the central charge of the CFT.\\
\\
All geometries of the form \eqref{eq:banadosmetric} are locally AdS$_3$ and therefore can be mapped to Poincar\'e patch AdS$_3$.
The suitable coordinate transformation is given by
\begin{equation} \label{eq:banadostrafo}
    x_P^\pm = \int\frac{\mathrm{d}x^\pm}{\psi^{\pm\,2}} - \frac{z^2\psi^{\mp\,\prime}}{\psi^{\pm\,2}\psi^{\mp}(1-z^2/z^2_h)}\,, \qquad\qquad
    z_P = \frac{z}{\psi^+\psi^-(1-z^2/z_h^2)}\,.
\end{equation}
The functions $\psi^\pm$ appearing in the diffeomorphism have the property to solve \textit{Hill's equation}
\begin{equation} \label{eq:hillseq}
    \psi^{\pm\,\prime\prime} - \mathcal{L}^\pm \psi^\pm = 0\,,
\end{equation}
where we dropped the arguments of the functions in favor of more clarity and we normalize the functions to unit Wronskian
\begin{equation}
\psi_1^{\pm}\psi_2^{\pm\,\prime} - \psi_2^{\pm}\psi_1^{\pm\,\prime} = \pm1\,.
\end{equation}
Making use of the coordinate transformation \eqref{eq:banadostrafo}, the HEE for Ba\~nados geometries is given by \cite{Sheikh-Jabbari:2016znt}
\begin{equation}
    S = \frac{c}{6}\,\ln\left( \ell^+(x_1^+,x_2^+) \ell^-(x_1^-,x_2^-)/\epsilon^2 \right)\,,
\end{equation}
with
\begin{equation} \label{eq:ellpm}
    \ell^\pm(x_1^\pm,x_2^\pm) = \psi^\pm_1(x_1^\pm)\psi^\pm_2(x_2^\pm) - \psi^\pm_2(x_1^\pm)\psi^\pm_1(x_2^\pm)\,,
\end{equation}
where $\psi^\pm$ are the appropriate solutions to Hill's equation.
An immediate observation is that HEE separates into holomorphic and anti-holomorphic contributions
\begin{equation}
    S = S^+ +\, S^-\,, \qquad\qquad S^\pm =  \frac{c}{6}\,\ln\left(\ell^\pm(x_1^\pm,x_2^\pm)/\epsilon\right)\,,
\end{equation}
as well as the symmetry of exchanging $x_1$ and $x_2$.
Further this result is universal and recovers all known special cases.\\
\\
Inspired by the transformation behavior of EE \eqref{eq:EEtrafo} and the analogy to vertex operators we define
\begin{equation} \label{eq:defineverexoperator}
    V := e^{\frac{6}{c} S} = \frac{\ell^+(x_1^+,x_2^+) \ell^-(x_1^-,x_2^-)}{\epsilon^2} = V^+ V^-\,, \qquad
    V^\pm := \frac{\ell^\pm(x_1^\pm,x_2^\pm)}{\epsilon}\,.
\end{equation}
Now we consider its second derivative with respect to $x^+$, using \eqref{eq:ellpm}
\begin{align} \label{eq:Vpp}
    V'' &= V^{+\,\prime\prime} V^-\nonumber \\
    &= \frac{\left(\psi_1^{+\,\prime\prime}(x_1^+)\psi_2^+(x_2^+) - \psi_2^{+\,\prime\prime}(x_1^+)\psi_1^+(x_2^+)\right)\,\left(\psi_1^-(x_1^-)\psi_2^-(x_2^-) - \psi_2^-(x_1^-)\psi_1^-(x_2^-)\right)}{\epsilon^2} \nonumber \\
    &= \mathcal{L}^+\,V\,,
\end{align}
where we made use of \eqref{eq:hillseq} in the last step.
On the other hand, keeping $S$ in the equations we find
\begin{equation}
    \frac{V''}{V} = \frac{6}{c} \,\left(S''+\frac{6}{c}\,\left(S^\prime\right)^2\right)\,,
\end{equation}
which combines with \eqref{eq:Vpp} and the result for $T_{kk}$ \eqref{eq:Tkk_banados} to
\begin{equation}
    S'' + \frac{6}{c}\,\left(S^\prime\right)^2 = \frac{c}{6}\,\mathcal{L}^+ = 2\pi \,\langle T_{++} \rangle\,,
\end{equation}
which is nothing else than saturation of QNEC$_2$.
Therefore QNEC$_2$ saturates for all states dual to Ba\~nados geometries.

\subsection{Global Quench -- Vaidya}

\begin{figure}
	\begin{center}
	\includegraphics[height=.22\textheight]{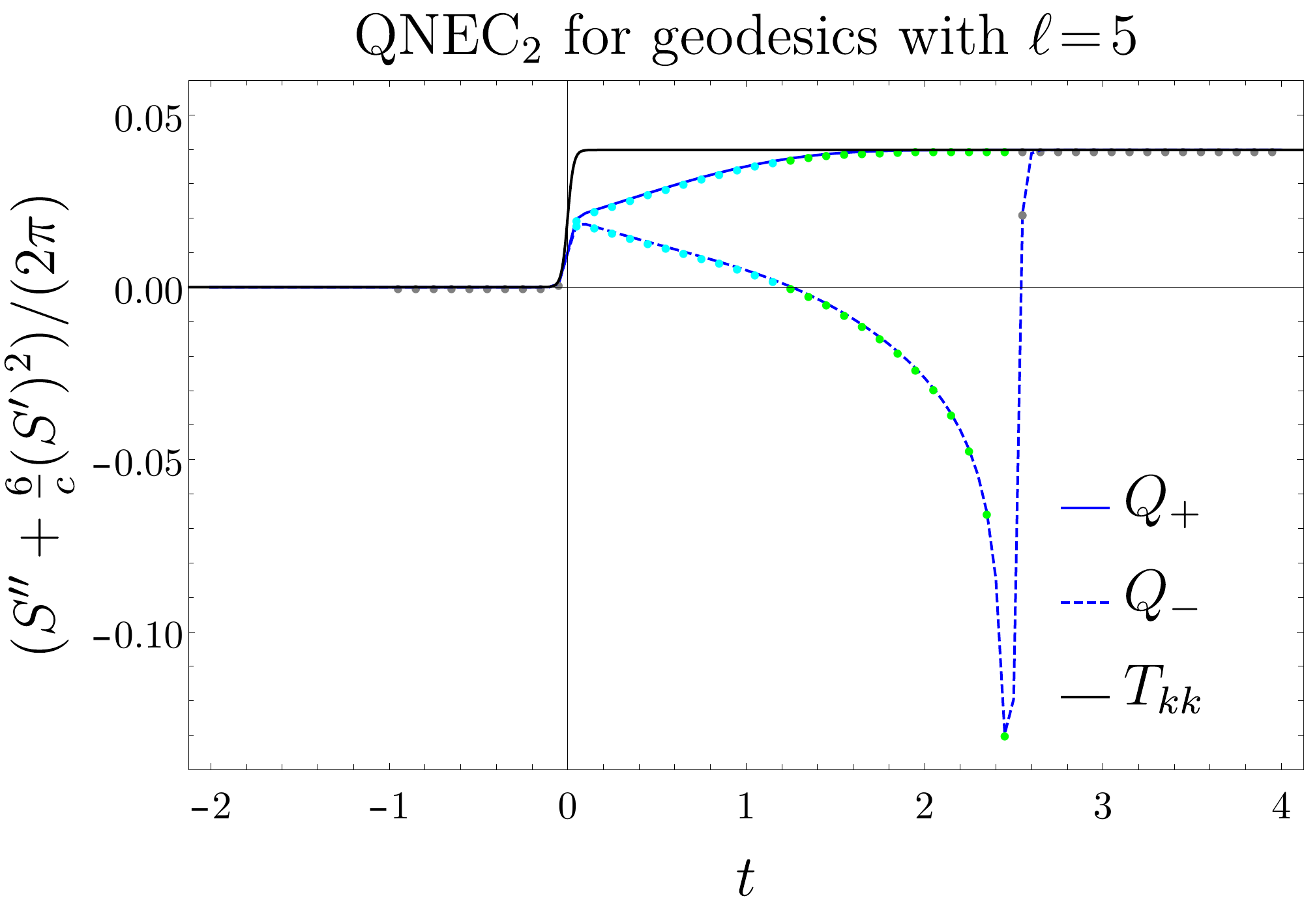}
	\quad
	\includegraphics[height=.22\textheight]{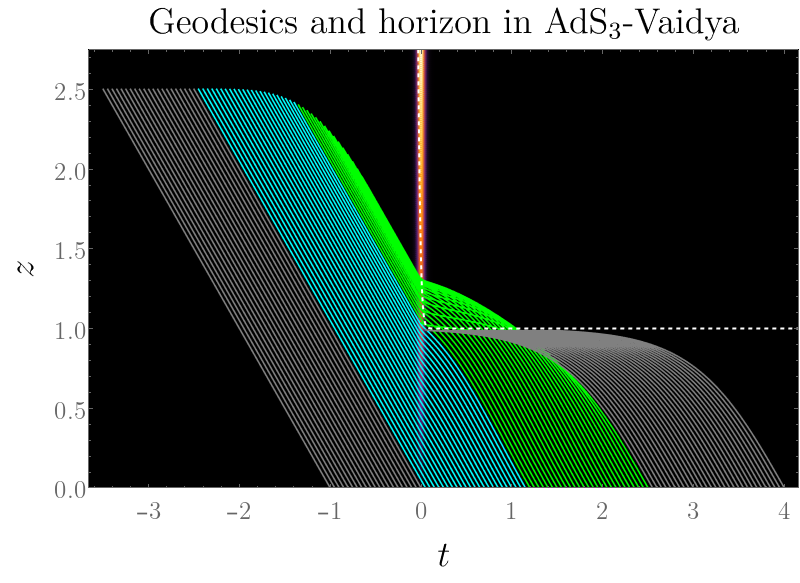}
	\caption{Left panel:Time evolution of QNEC$_2$ for both null directions. The colored points relate the value of $Q_\pm$ to the form of the geodesics in the right plot.
	Right panel: Central geodesics are influenced by the horizon (white dashed line) and the mass shell (shown as density plot).}
	\label{fig:geos_horizon_AdS3Vaidya}
	\end{center}
\end{figure}

\begin{figure}
	\begin{center}
	\includegraphics[height=.23\textheight]{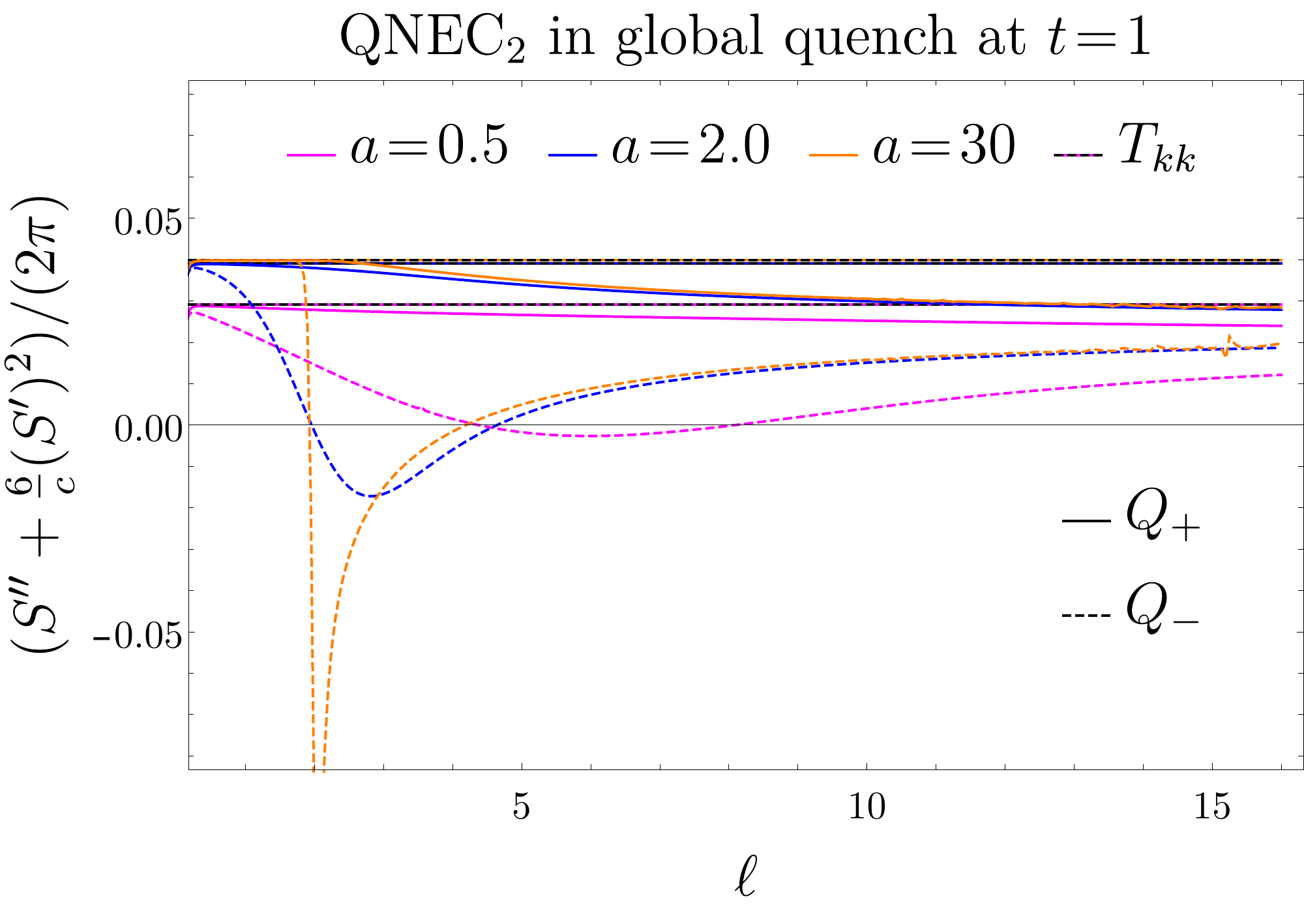}
	\caption{Dependence of QNEC$_2$ on the quench parameter $a$ in addition to the size of the entangling region $\ell$.}
	\label{fig:QNEC_adep_levo_AdS3Vaidya}
	\end{center}
\end{figure}

After obtaining QNEC$_2$ saturation for both, the vacuum and thermal state, we turn to a globally quenched system once more.
Similar to section \ref{sec:Vaidya5D} we can observe the change in QNEC$_2$ while the state is changed from vacuum to a thermal one, both belonging to the Ba\~nados family.
To implement this, we use the time dependent blackening function in \eqref{eq:genMetric3d}
\begin{equation} \label{eq:vaidya3D}
    A(t) = \frac{1}{z^2}(1-M(t)z^2)\,, \qquad\qquad M(t) = \frac{1}{2}\left(1+\tanh(a\,t)\right)\,,
\end{equation}
where $M(t)$ is normalized, such that it approaches the case discussed in section \ref{sec:3DvacBTZ} at late times with $M\!=\!1$ and the quench happens at $t\!=\!0$.
This time we control the sharpness of the quench with the parameter $a$ and investigate its influence on QNEC$_2$ as well.
Again, both null projections of the EMT are given by
\begin{equation}
    T_{kk}=\frac{M(t)}{8 \pi \, G_N}\,,
\end{equation}
while the time dependence leads to different variations of EE for the two null directions $k^\mu_\pm$.
We will denote the QNEC$_2$ combination with $Q_\pm$ to indicate the null vector used.
First we choose a very sharp quench ($a\!=\!30$) and take a closer look at the shape and the behavior of the geodesics in the right figure \ref{fig:geos_horizon_AdS3Vaidya}.
The position of the apparent horizon (white dashed curve) and the bulk EMT (density plot in the background) are given by
\begin{equation}
    z_h = \frac{1}{\sqrt{M(t)}}\,, \qquad\qquad T^{bulk}_{tt} = \frac{z}{2} M'(t)\,.
\end{equation}
The change in shape of a geodesic with fixed boundary separation $\ell$ over time allows us to determine three different regions.
At early times $t<0$ we find the vacuum solutions as expected, while at late times $t\!>\!\ell/2$ we get the same curves as for the BTZ black hole, both shown in gray.
The interesting region is the third one.
Here the geodesics come in contact with bulk matter (light blue), which allows them to cross the apparent horizon (green).
This behavior of the geodesics can be related to the time evolution of QNEC$_2$ in the left panel of figure \ref{fig:geos_horizon_AdS3Vaidya} via the colored dots.
At early and late times (gray), QNEC$_2$ is saturated as expected for vacuum and static BTZ BHs.
The transition from one state to the other, where QNEC$_2$ is not saturated, happens in between, where the geodesics are marked light blue and green.
The non-saturation of QNEC$_2$ is caused by the geodesics (or RT surfaces in general) crossing the bulk region where matter is present.
\begin{figure}
	\begin{center}
	\includegraphics[height=.22\textheight]{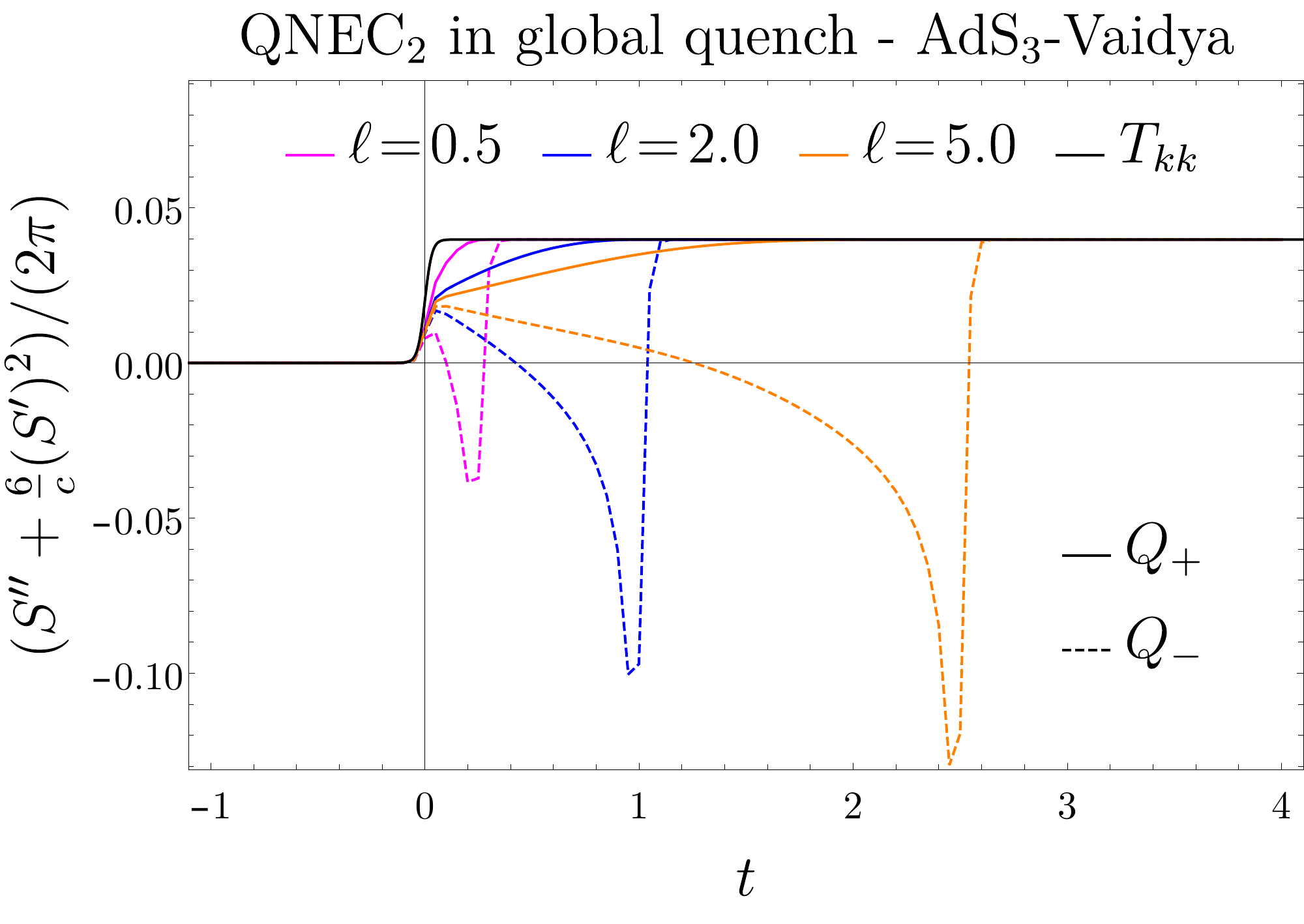}
	\quad
	\includegraphics[height=.22\textheight]{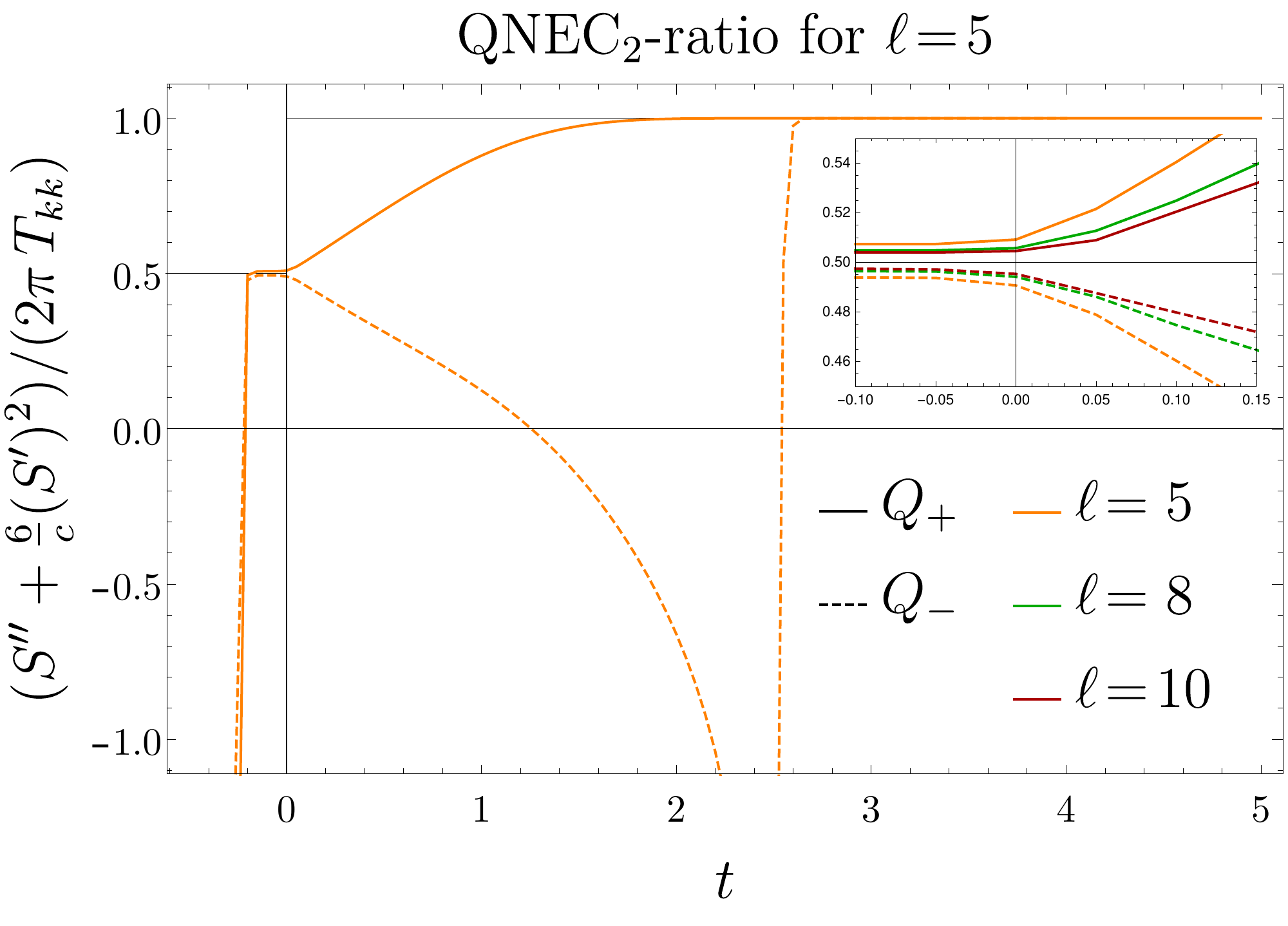}
	\caption{Left panel: Time evolution of QNEC$_2$ for both null directions and three characteristic separation $\ell$.
	Right panel: The ratio of both sides of QNEC$_2$ for $\ell\!=\!5$. The inset shows that half saturation is approached as $\ell$ grows.}
	\label{fig:QNEC_tevo_vaidya3D}
	\end{center}
\end{figure} \\
The difference between $Q_+$ and $Q_-$ can also be seen in figure \ref{fig:QNEC_tevo_vaidya3D}, where we show its time evolution for three different separations in the left panel.
We immediately observe that $Q_-$ develops a dip, that moves to later times for larger separations.
This has the simple reason, that geodesics with larger separation reach deeper into the bulk and further backwards in time.\enlargethispage{\baselineskip}
Further we see that $Q_+$ has a very smooth and monotonic behavior in the quench region.
This can be attributed to the deformation vector $k^\mu_+$ being aligned with the shell of null dust in the bulk, while for $Q_-$ the deformation is perpendicular.
The right panel of figure \ref{fig:QNEC_tevo_vaidya3D} shows the ratio of the QNEC$_2$ combination and the null-projection of the EMT for the $\ell\!=\!5$ case from the left plot.
Around $t\!=\!0$ the ratio for $Q_\pm$ is very close to $\frac{1}{2}$ before the branches split up, mirroring the behavior of the left plot, ending up saturating QNEC$_2$.
The value $\frac{1}{2}$ is maintained for some time $t\!<\!0$, before it seems to diverge.
This is just an artefact of evaluating $\frac{0_-}{0_+}$ numerically.
The inset of this plot shows, that for growing $\ell$, the half saturation is approached.
This perfectly fits the $\frac{1}{d}$ saturation, mentioned in section \ref{sec:Vaidya5D}, found in \cite{Mezei:2019sla}.
\begin{figure}
	\begin{center}
	\includegraphics[height=.22\textheight]{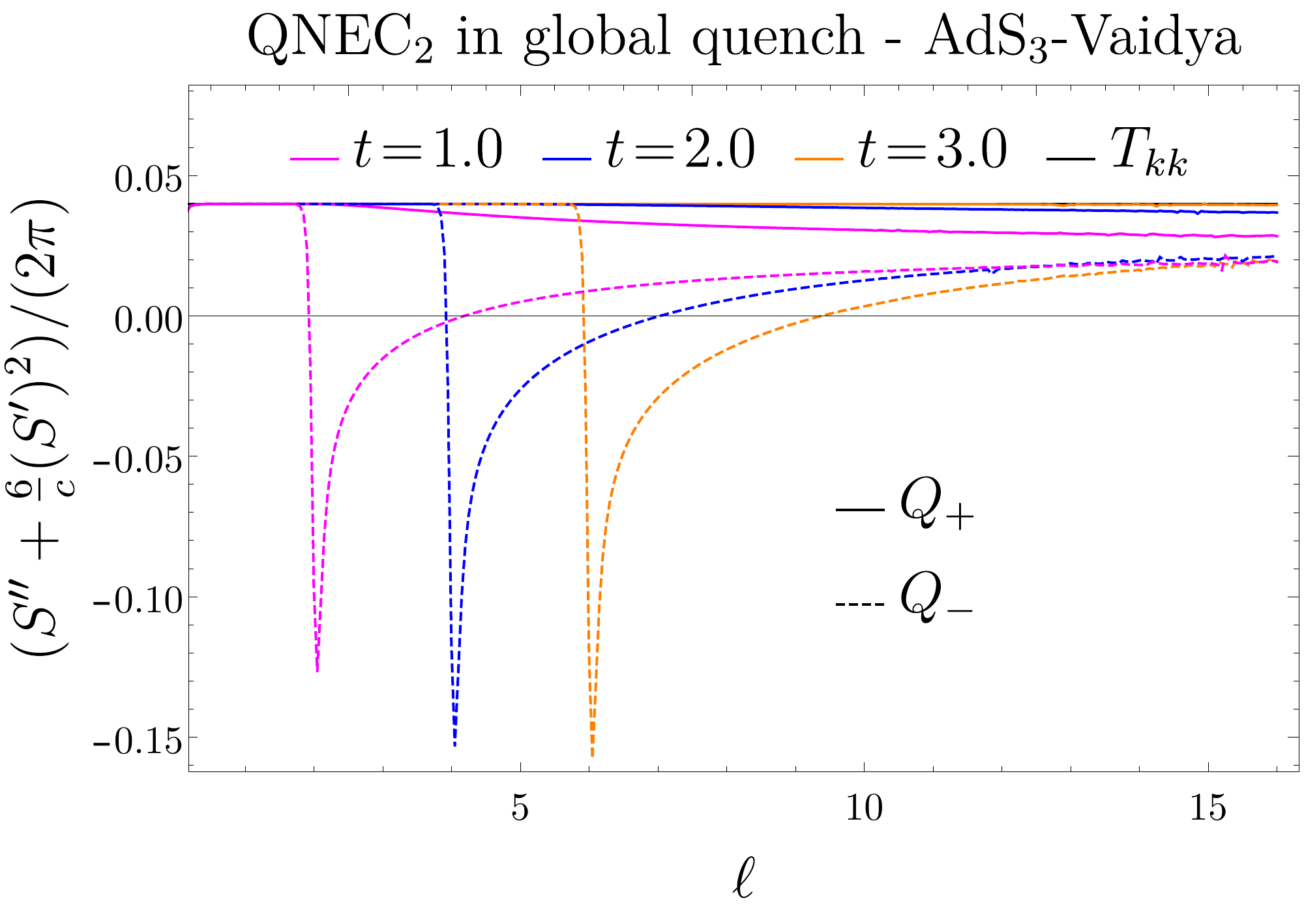}
	\quad
	\includegraphics[height=.22\textheight]{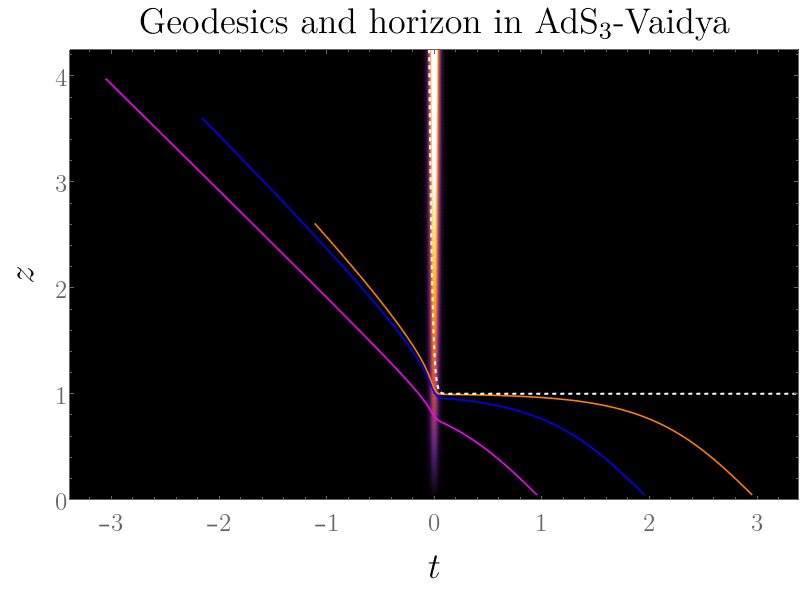}
	\caption{Left panel: Dependence of QNEC on the interval size $\ell$ for both null directions and three characteristic points in time.
	Right panel: The tips (central turning points) of the geodesics used in the left plot (color coded) always stay outside the horizon.}
	\label{fig:QNEC_levo_Vaidya3D}
	\end{center}
\end{figure} \\
Following the same pattern as in the higher dimensional case, we study the dependence of QNEC$_2$ on the size of the entangling region next.
This time we restrict ourselves to boundary times $t\!={1,2,3}$ after the quench, since the saturation has been proven for the vacuum before it.
Figure \ref{fig:QNEC_levo_Vaidya3D} shows the expected behavior of the dip in $Q_-$, that is shifted to larger values of $\ell$ for later times.
$Q_+$ starts to deviate from $T_{kk}$ at the same time as $Q_-$, but only slowly decreases.
The numerical results suggest that in the limit $\ell\!\to\!\infty$, both $Q_\pm$ reach the same value, similar to what was observed in section \ref{sec:Vaidya5D}.
The right plot of figure \ref{fig:QNEC_levo_Vaidya3D} shows the tip (or turning point) of the geodesics calculated for the left plot.
Choosing the pink curve as example, one can nicely see that QNEC$_2$ is saturated for $\ell\!<\!2t\!<\!2$.
The geodesic with $\ell\!=\!2$ is the first one to come into contact with the matter shell.
For larger separations the geodesics always cross the matter shell and QNEC$_2$ can never be saturated again.
The same is true for the blue and orange curves.
As we have seen in figures \ref{fig:QNEC_geos_tevo_vaidya5D}, \ref{fig:QNEC_levo_vaidya5D} and \ref{fig:geos_horizon_AdS3Vaidya}, the RT-surfaces can reach behind the apparent horizon, but their tip will always remain outside, as visualized in figure \ref{fig:geos_horizon_AdS3Vaidya}.
\begin{figure}
	\begin{center}
	\includegraphics[height=.22\textheight]{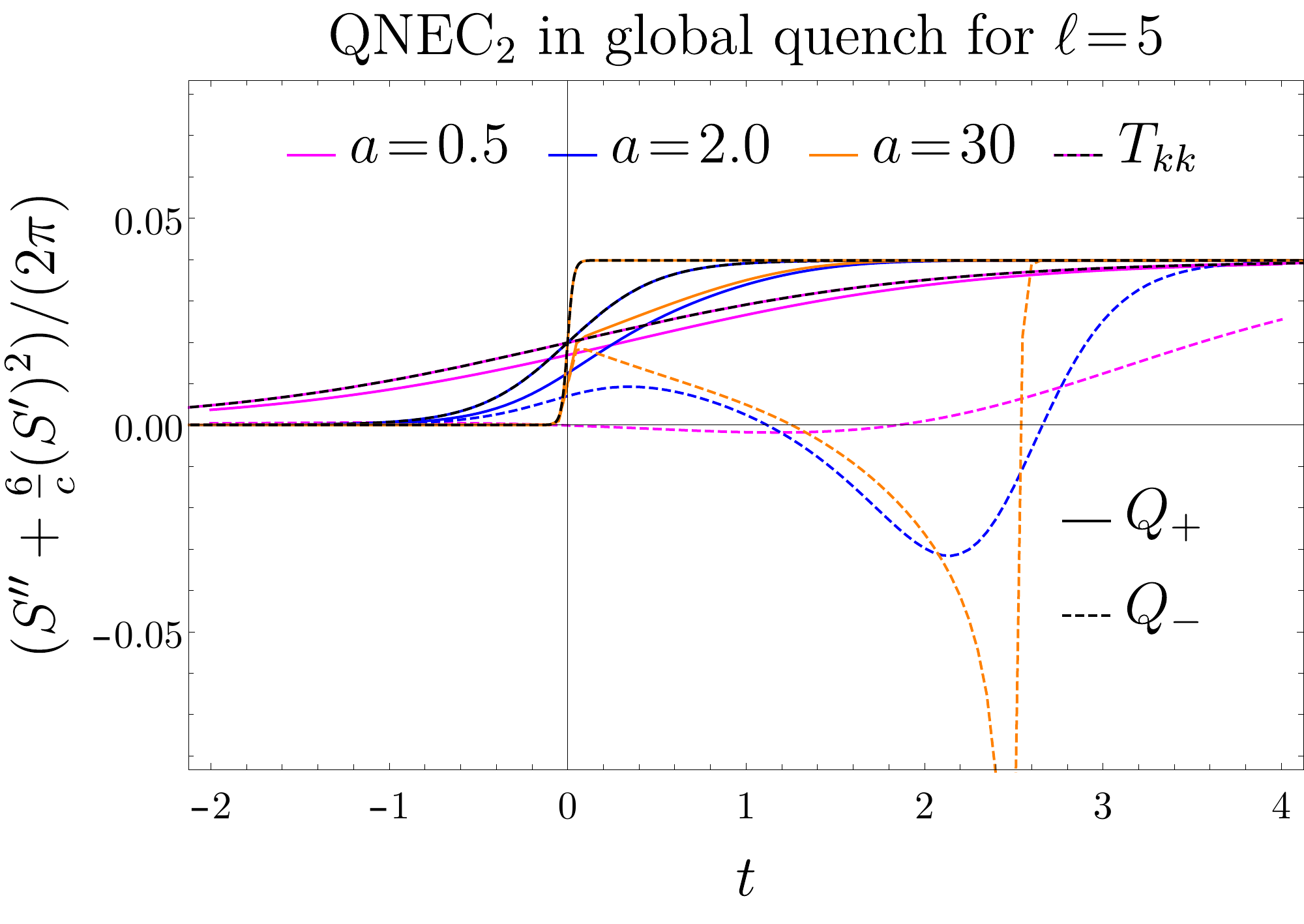}
	\quad
	\includegraphics[height=.22\textheight]{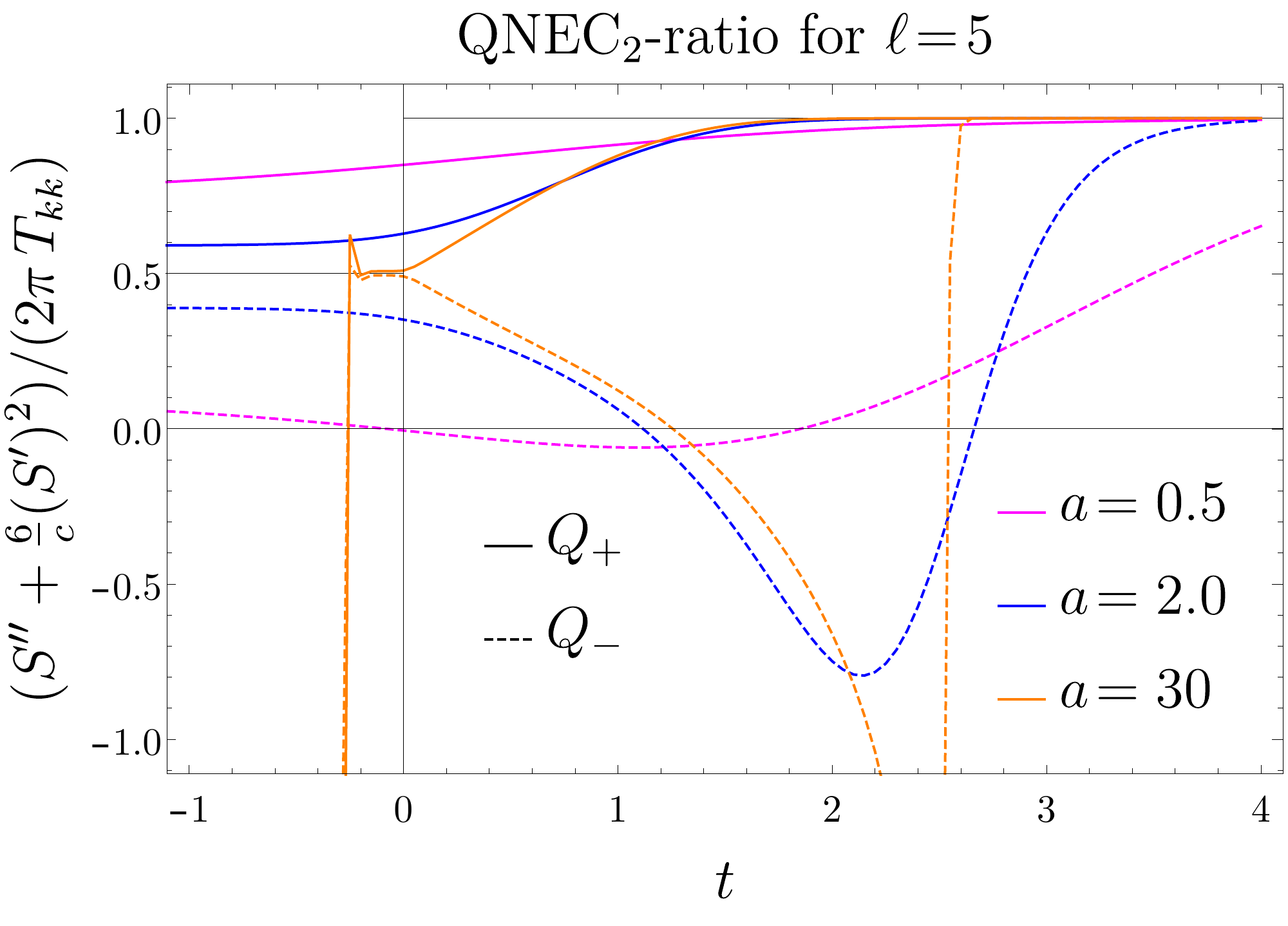}
	\caption{Dependence of QNEC$_2$ on the quench parameter $a$. Left panel: Time evolution of QNEC$_2$ for both null directions.
	Right panel: Ratio of both sides of QNEC$_2$ for both null directions.}
	\label{fig:QNEC_adep_tevo_AdS3Vaidya}
	\end{center}
\end{figure}
Compared to the AdS$_5$-Vaidya case with $a\!=\!2$, the difference between $Q_+$ and $Q_-$ is much more pronounced in the case discussed above.
Therefore we study how QNEC$_2$ is influenced by the quench parameter $a$ of equation \eqref{eq:vaidya3D}, determining the sharpness of the transition from vacuum to BTZ BH.\\
\\
First we investigate the time evolution once more in figure \ref{fig:QNEC_adep_tevo_AdS3Vaidya}.
The left plot shows QNEC$_2$ for $\ell\!=\!5$ for three different quench parameters.
$a\!=\!0.5$ is a very soft (or slow) transition, $a\!=\!2$ was used for the quench in CFT$_4$ and despite being further away from the last value $a\!=\!30$, its effect is much closer to the sharp quench than the slow transition.
The according projection of the EMTs are marked by dots on the black curves corresponding to the color of the QNEC$_2$ curves.
The effect on $Q_+$ is not spectacular, only stretching the time interval of the transition from $0$ to $T_{kk}$.
$Q_-$ on the other hand develops an increasingly pronounced minimum, that turns into a delta-like divergence for $a\!\to\!\infty$.
Similarly in the right panel of figure \ref{fig:QNEC_adep_tevo_AdS3Vaidya}, the ratio of both sides of QNEC$_2$ reflects this feature.
The aforementioned half saturation can be observed only for the sharp quench.
The reason for this is that $\ell\!=\!5$ is not large enough (see inset of right figure \ref{fig:QNEC_tevo_vaidya3D}) and the smeared out quench reinforces this effect.
This is obvious from figure \ref{fig:QNEC_adep_levo_AdS3Vaidya}, where we can see that for smaller $a$, $Q_\pm$ approaches the asymptotic value much slower.
A perturbative study of the two limits mentioned \cite{Ecker:2019ocp} confirms the numerical results.

\subsection{Backreaction of Bulk Matter in Global AdS}
 \label{sec:backreaction}
 
In this section we take a step aside and consider AdS$_3$ in global coordinates \eqref{eq:globalAdS} with a minimally coupled scalar field of mass $m^2\!=\!4h(h\!-\!1)$.
We are interested in excited CFT$_2$ states that are obtained by the action of conformal primaries with weight $h$ on the vacuum.
These states are described by the backreaction of a single scalar particle in the center of AdS$_3$ on the geometry
\begin{equation} \label{eq:backreactionmetric}
    \mathrm{d}s^2 = -(r^2+G_1(r)^2)\, \mathrm{d} t^2 + \frac{\mathrm{d} r^2}{r^2+G_2(r)^2} + r^2\, \mathrm{d}\varphi^2\,, \qquad \qquad \varphi\sim\varphi+2\pi\,,
\end{equation}
where the functions $G_1(r)$ and $G_2(r)$ are determined by solving Einstein's equations \eqref{eq:Einsteinequations}.
The EMT and the scalar field are given by
\begin{align}
    T_{\mu\nu}= \partial_\mu \phi \partial_\nu \phi -\frac{1}{2}g_{\mu\nu}\left((\nabla\phi)^2+m^2\phi^2\right)\,, \\
    \phi = \frac{a}{\sqrt{2\pi}}\,\frac{e^{-2iht}}{\big(1+r^2\big)^h} +  \frac{a^\dagger}{\sqrt{2\pi}}\,\frac{e^{2iht}}{\big(1+r^2\big)^h}\,,
\end{align}
where $a$ and $a^\dagger$ are the usual annihilation and creation operators.
Solving Einstein's equations \eqref{eq:Einsteinequations} with the EMT for a single particle state $|\psi\rangle\!=\!a^\dagger|0\rangle$ leads to the result
\begin{align}
    G_1(r) &= 1 - 8 G_N h + \mathcal{O}(G_N^2)\,, \\
    G_2(r) &= 1 - 8 G_N h\, \bigg(1- \frac{1}{(r^2+1)^{2h-1}}\bigg) + \mathcal{O}\big(G_N^2\big)\,.
\end{align}
Since the lower bound for the conformal weight is $h\!\geq\!\frac{1}{2}$, we observe that the second term in the parenthesis is subleading in a Fefferman-Graham expansion and leads to the asymptotic form of the metric
\begin{equation}
    \mathrm{d} s^2 = -(r^2+G^2)\, \mathrm{d} t^2 + \frac{\mathrm{d} r^2}{r^2+G^2} + r^2\, \mathrm{d}\varphi^2 + \dots\,, \qquad\qquad G=1-8 G_N h\,.
\end{equation}
This metric describes a geometry with conical deficit of $16\pi G_N h$, which is (contrary to the general form \eqref{eq:backreactionmetric}) part of the Ba\~nados family, characterized by $\mathcal{L}_\pm\!=-\frac{1}{4}G^2$.
Using equation \eqref{eq:Tkk_banados} we know the null projection of the EMT (i.e.~the left hand side of QNEC$_2$) in the boundary theory to leading and subleading order in $\frac{1}{c}$
\begin{equation} \label{eq:Tkk_backreaction}
    2\pi\langle T_{\pm\pm}\rangle = -\frac{c}{24}\, G^2= -\frac{c}{24} + h+ {\cal O}(h^2/c)\,.
\end{equation}
To incorporate finite $c$ corrections in the evaluation of the right hand side of QNEC$_2$ as well, we need the according corrections to holographic EE proposed in \cite{Faulkner:2013ana,Barrella:2013wja}
\begin{equation} \label{eq:EEcorrections}
    S = \frac{\mathcal{A}}{4G_N} + \frac{\delta\mathcal{A}}{4G_N} + S_{\textrm{\tiny bulk}}\,.
\end{equation}
The first term is the well known large $c$ result, the second term is the change in area of the RT surface due to the backreaction on the geometry and $S_{\textrm{\tiny bulk}}$ is the entanglement in the bulk across the extremal surface.
While there exists an exact expression for the second term, the bulk term can only be evaluated perturbatively in certain limits (or numerically).

\subsubsection{RT Contribution to QNEC\texorpdfstring{$_{\boldsymbol{2}}$}{2}}

Applying the procedure to calculate QNEC$_2$ with paper and pencil, explained in appendix \ref{app:Integrals}, to the geometry \eqref{eq:backreactionmetric}, we are able to find exact expressions for $\mathcal{A}$ and $\delta\mathcal{A}$.
The area of the extremal surface in the quantum corrected geometry is given by the integral
\begin{equation}
    \mathcal{A}+\delta\mathcal{A} = \int\limits_{0}^{\Delta\varphi+\lambda} \mathrm{d} \varphi\, \mathcal{L}(z,\,\dot z,\,\dot t) =\int\limits_{0}^{\Delta\varphi+\lambda} \mathrm{d} \varphi\, \frac{1}{z}\,\sqrt{1+\frac{\dot z^2}{f_2(z)}-\dot t^2f_1(z)}\,,
\end{equation}
where $\lambda$ parametrizes the deformation of the entangling region, we employed the coordinate transformation $r\!\to\!\frac{1}{z}$ and the functions $f_1$ and $f_2$ are given by
\begin{align}
    f_1(z)& = 1 + z^2 \,\big(1 - 8 G_N h\big)^2\,, \\
    f_2(z)& = 1 + z^2 \,\Big[ 1 - 8 G_N h \Big(1- \left(1+z^{-2}\right)^{1-2h}\Big) \Big]^2\,.
\end{align}
The integral can be organized in powers of $\lambda$, where we only keep the contributions up to including order $\lambda^2$
\begin{equation}
    \mathcal{A}+\delta\mathcal{A} = 2\ln{\frac{z_\ast}{z_{\textrm{\tiny cut}}}}+\int\limits_0^1 \mathrm{d} x \,I_{\mathcal{A}}^{(0)} +\int\limits_0^1 \mathrm{d} x \,I_{\mathcal{A}}^{(1)} + {\cal O}(\lambda^3)\,,
\end{equation}
with the integrand
\begin{equation}
    I_{\mathcal{A}}^{(0)} = \frac{2(S(x)\!-\!1)}{x} - \frac{2\lambda S(x) x\tan\!\frac{\Delta\varphi}{2}}{1\!+\!x^2\!+\!(1\!-\!x^2)\cos\!\Delta\varphi} + \frac{\lambda^2 S(x) x \big((1\!-\!x^2)\cos\!\Delta\varphi\!-\!2\!+\!x^2\big)}{\big(1\!+\!x^2\!+\!(1\!-\!x^2)\cos\!\Delta\varphi\big)^2}
\end{equation}
where
\begin{equation}
    S(x):=\big(1-x^2\big)^{-1/2}\Big(1+x^2\tan^2\frac{\Delta\varphi}{2}\Big)^{-1/2}\,.
\end{equation}
The order $G_N$ integrand $I_{\mathcal{A}}^{(1)}$ is of similar structure but too long to be displayed here.
The integrals can be solved in closed form and recover the result for EE \cite{Belin:2018juv} in the limit $\lambda\!\to\!0$.
For the QNEC$_2$ combination we find (for the RT contribution $S_{RT} = \frac{\mathcal{A}+\delta\mathcal{A}}{4G_N}$)
\begin{equation} \label{eq:RTcontributionQNEC}
    S_{RT}'' + \frac{6}{c}\,\big(S_{RT}'\big)^2 = -\frac{c}{24} + h - \frac{h\sqrt{\pi}\,\Gamma[2h+2]}{4\Gamma[2h+\tfrac32]}\,\sin^{4h-2}\frac{\Delta\varphi}{2}\, +\mathcal{O}(1/c)\,,
\end{equation}
valid for positive (half-)integer weights $h$.

\subsubsection{QNEC\texorpdfstring{$_{\boldsymbol{2}}$}{2} for Small Intervals}

\begin{figure}
	\begin{center}
	\includegraphics[height=.22\textheight]{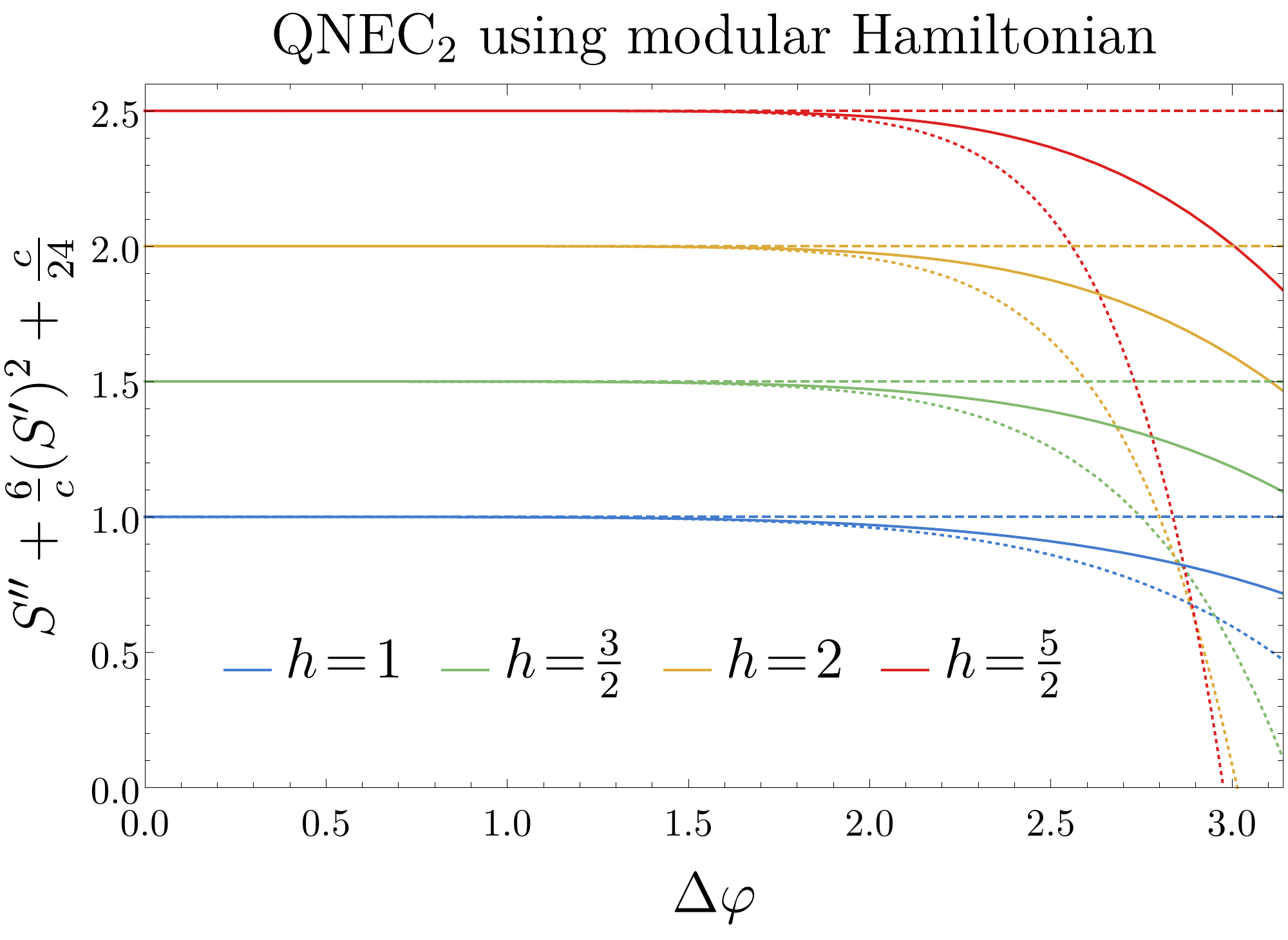}
	\quad
	\includegraphics[height=.22\textheight]{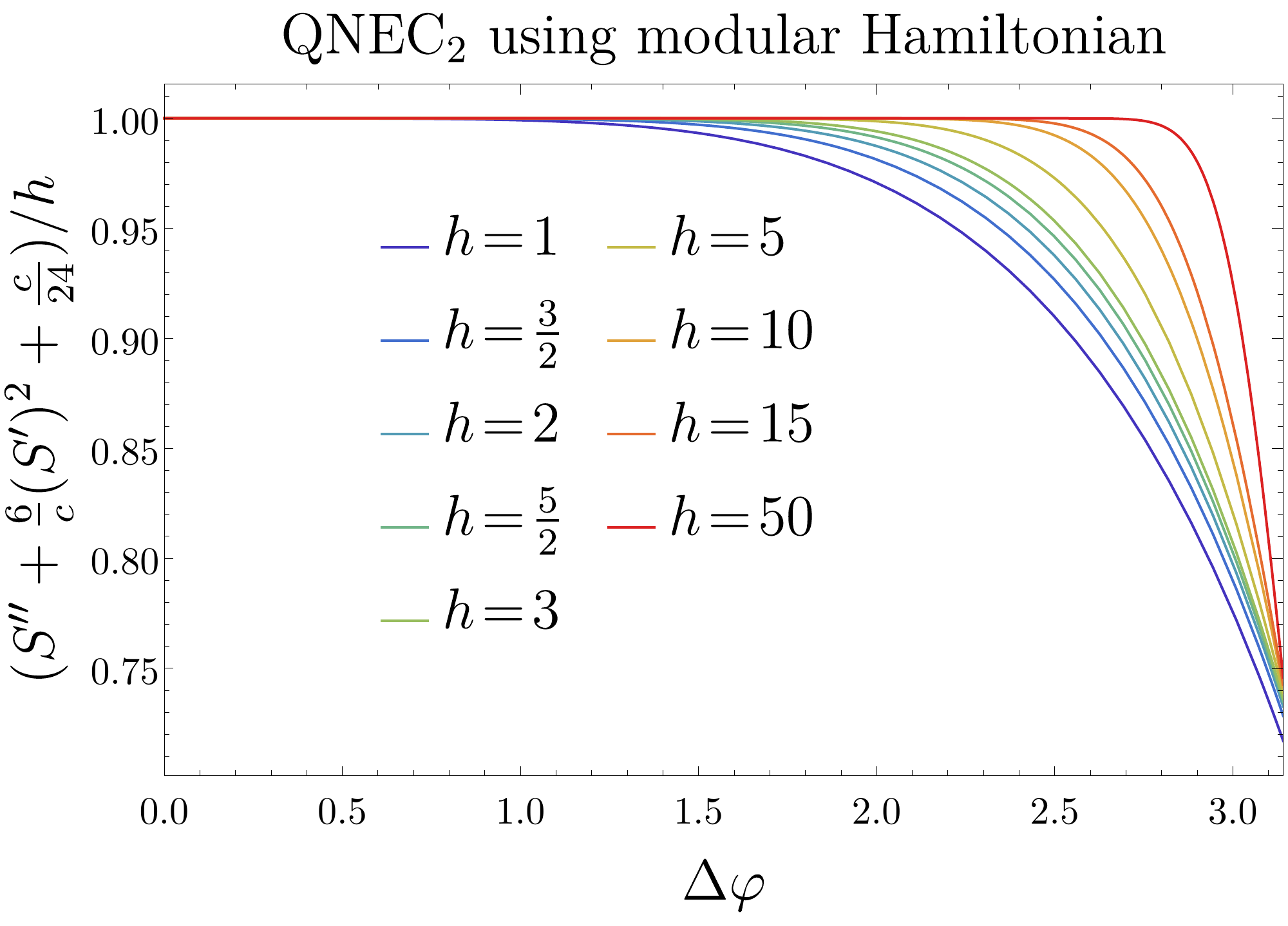}
	\caption{Left Panel: The solid lines are the numerical results (including the RT contribution) for the finite $c$ corrections to QNEC$_2$, the dotted lines are the small interval $\Delta\varphi$ results and the dashed lines are $T_{kk}$ from equation \eqref{eq:Tkk_backreaction}.
	Right Panel: Rescaling the correction by $h$ shows that only close to $\Delta\varphi\!=\!\pi$ QNEC$_2$ is not saturated. The larger the weight $h$, the smaller this region gets, confirming the perturbative results of section \ref{sec:backreactionlargeh}.}
	\label{fig:numericbackreaction}
	\end{center}
\end{figure}

To find the RT corrections to QNEC$_2$ in the limit of small intervals $\Delta\phi$, we need the expansion of \eqref{eq:RTcontributionQNEC}
\begin{align}
    S_{RT}'' &+ \frac{6}{c}\,\big(S_{RT}'\big)^2 = -\frac{c}{24} + h - \frac{h\sqrt{\pi}\,\Gamma[2h+2]}{4\Gamma[2h+\tfrac32]}\,(\Delta\varphi)^{4h} \times \nonumber \\
    &\Big( \frac{2^{2-4h}}{\Delta\varphi^2} - \frac{2^{-4h}}{3}(-1+2h) + \frac{2^{2-4h}}{45} (3-11h+10h^2)\Delta\varphi^2 +\mathcal{O}(\Delta\varphi^4) \Big) \nonumber \\
    &+\mathcal{O}(1/c)\,, \label{eq:RTcontributionQNEC_small}
\end{align}
that has to be evaluated for every $h$ separately.\\
\\
The contribution of $S_{bulk}$ can be computed, using the expectation value of the modular Hamiltonian
\begin{equation}
    \Delta S = S_{bulk} - S_0 = 2\pi\,\Delta \langle H_0 \rangle\,,
\end{equation}
where $S_0$ is the EE of the vacuum and the vacuum modular Hamiltonian is defined via the density matrix $\rho_{vac}\!=\!e^{-2\pi H_0}$.
Since we are interested in the null deformation, similar to section \ref{sec:QNEC_D=4} we calculate the modular Hamiltonian in a boosted frame.
This is given by the integral of the expectation value of the bulk EMT over the entanglement wedge $\Sigma_A$ \cite{Belin:2018juv,Casini:2011kv}
\begin{equation} \label{eq:modHamiltonianIntegral}
     \Delta\langle H_0\rangle = \int_{\Sigma_A} \mathrm{d}\Sigma_{A} \sqrt{|g_{\Sigma_A}|}\, \xi^\nu n^\mu\langle\psi | T_{\mu \nu}|\psi\rangle\,,
\end{equation}
where $g_{\Sigma_A}$ is the induced metric, $\xi^{\nu}\!=\!(1,0,0)$ is the Killing vector generating Rindler-time translations and $n^{\mu}\!=\!((\rho^2\!-\!1)^{-1/2},0,0)$ is the normal vector to $\Sigma_{A}$ (all in Rindler coordinates, see \cite{Belin:2018juv} for the explicit transformation to apply to the boosted $T_{\mu \nu}$).
We solve the integral numerically for arbitrary $\Delta\varphi$ and several values of $h$ before turning to the small interval limit.
In figure \ref{fig:numericbackreaction} we show only the finite $c$ corrections to the leading term.
The left panel shows the agreement of the full numeric result (including the RT contribution) with the small interval expansion calculated below.
The right panel shows the same term rescaled by $h$ such that we can explore the large $h$ limit discussed in the following section \ref{sec:backreactionlargeh}.\\
\\
Employing the expansion in $\Delta\varphi$ once again and choosing the weight $h$, the integral can be solved, yielding
\begin{align}
    &h=\tfrac{1}{2}: \qquad S_{bulk}'' + \frac{6}{c}\,\big(S_{bulk}'\big)^2 = \frac{1}{3}+\mathcal{O}\left(\Delta \varphi ^2\right)\,, \\
    &h=1: \qquad S_{bulk}'' + \frac{6}{c}\,\big(S_{bulk}'\big)^2 = \frac{\Delta \varphi ^2}{5}-\frac{43 \Delta \varphi ^4}{2520}+\frac{\Delta \varphi ^6}{21600}+\mathcal{O}\left(\Delta \varphi ^8\right)\,,  \nonumber \\
    &h=\tfrac{3}{2}: \qquad S_{bulk}'' + \frac{6}{c}\,\big(S_{bulk}'\big)^2 = \frac{3 \Delta \varphi ^4}{35}-\frac{73 \Delta \varphi ^6}{5040}+\frac{1571 \Delta \varphi ^8}{1663200}+\mathcal{O}\left(\Delta \varphi ^{10}\right)\,,  \nonumber \\
    &h=2: \qquad S_{bulk}'' + \frac{6}{c}\,\big(S_{bulk}'\big)^2 = \frac{2 \Delta \varphi ^6}{63}-\frac{667\Delta \varphi ^8}{83160}+\frac{2789\Delta \varphi ^{10}}{3088800}+\mathcal{O}\left(\Delta \varphi ^{12}\right)\,. \nonumber
\end{align}
We display the first four values for $h$, which is sufficient to find the following pattern.
Inserting the same values for $h$ into \eqref{eq:RTcontributionQNEC_small}, it turns out that the leading order terms $\Delta\varphi^{4h-2}$ exactly cancel and the right hand side of QNEC$_2$ is given by
\begin{equation}
    S'' + \frac{6}{c}\,\big(S'\big)^2 = -\frac{c}{24} + h + \mathcal{O}(\Delta\varphi^4h)\,.
\end{equation}
Here $S\!=\!S_{RT}\!+\!S_{bulk}$ contains all corrections according to \eqref{eq:EEcorrections} and can be compared with the null energy \eqref{eq:Tkk_backreaction}.
We see that in the limit of small intervals QNEC$_2$ is saturated up to order $\Delta\varphi^{4h}$ and the plus sign indicates that it is not violated
\begin{equation}
    2\pi\,\langle T_{\pm\pm}\rangle - S'' - \frac{6}{c}\,\big(S'\big)^2 = +\mathcal{O}(\Delta\varphi^{4h})\,.
\end{equation}

\subsubsection{QNEC\texorpdfstring{$_{\boldsymbol{2}}$}{2} for Large Weight and Half-Interval}
 \label{sec:backreactionlargeh}
 
Since we are interested not only in small intervals (where the RT surfaces are close to the boundary and not really affected by the bulk particle at the center), we consider $\frac{1}{h}$ as another small parameter for perturbation theory.
In order to do this and keep the quantum correction small at the same time, we consider the limit $c\!\gg\!h\!\gg\!1$.
As we have seen from the numerical solution in figure \ref{fig:numericbackreaction}, at any interval $\Delta\varphi\!<\!\pi$, QNEC$_2$ is saturated up to tiny correction in the large $h$ limit.
This changes once we consider the case $\varphi\!=\!\pi$, which is the maximum interval possible (since $\varphi$ is $2\pi$ periodic).
First we perform the expansion of \eqref{eq:RTcontributionQNEC} to find the RT contribution in our new limit
\begin{equation} \label{eq:RTcontributionLarge}
    S_{RT}''+\frac{6}{c}\,\big(S_{RT}'\big)^2 = -\frac{c}{24} + h - \frac{h}{4}\,\sqrt{2\pi\,h} + \dots\,,
\end{equation}
where the ellipsis represent terms that grow slower than linear in $h$ or vanish.
In the large $h$ limit, the last term dominates the expression.\\
\\
We calculate the bulk contribution once more using the integral for the modular Hamiltonian \eqref{eq:modHamiltonianIntegral}, which simplifies surprisingly in our limit.
To find its contribution to QNEC$_2$ we need to boost the system once more to introduce the parameter $\lambda$ for the null deformation.
For the bulk contribution we find
\begin{align}
    \Delta\langle H_0\rangle &= \frac{\sqrt{\pi } \Gamma (2 h+1)}{\Gamma \left(2 h+\frac{1}{2}\right)}-\pi  h |\lambda| + \label{eq:bulkcorrectionslarge}\\
    &\Big[\pi^{-3/2}\,h\, \Gamma (2 h) \Big(\frac{\pi ^2 h}{\Gamma \left(2 h+\frac{1}{2}\right)}+\frac{8 h (h+1)-3}{\Gamma \left(2 h+\frac{5}{2}\right)}\Big)-\frac{h}{2}\Big]\, \lambda^2+O\left(\lambda^3\right)\,. \nonumber
\end{align}
Extracting $S_{bulk}$ and its derivatives from the equation above together with the RT contribution \eqref{eq:RTcontributionLarge} results in the quantum corrected r.h.s.~of QNEC$_2$
\begin{equation}
    S'' + \frac6c \big(S^\prime\big)^2 \Big|_{\Delta\varphi=\pi} = -\frac{c}{24} + \frac{3h}{4} + \mathcal{O}(\sqrt{h}) + \mathcal{O}(1/c)\,,
\end{equation}
where $S$ includes all corrections according to \eqref{eq:EEcorrections}.
In \cite{Ecker:2019ocp} we present another way to calculate QNEC$_2$ in this limit (as well as the small interval limit), which gives the same leading order corrections different subleading $\mathcal{O}(\sqrt{h})$-terms.
Nonetheless, the subleading terms in \eqref{eq:bulkcorrectionslarge} agree perfectly with the numeric solution of the integral shown in figure \ref{fig:numericbackreaction}.\\
\\
In the large $h$-limit at $\Delta\varphi\!=\!\pi$ we find that QNEC$_2$ is not saturated, but gapped by $\frac{h}{4}$
\begin{equation}
    2\pi\,\langle T_{\pm\pm}\rangle - S'' - \frac{6}{c}\,\big(S'\big)^2 = +\frac{h}{4} + \dots\,,
\end{equation}
where the plus sign guarantees that QNEC$_2$ is not violated.

\subsection{Black Hole Phase Transition in AdS with Bulk Matter}
 \label{sec:phasetransition}
 
\begin{figure}
	\begin{center}
	\includegraphics[height=.3\textheight]{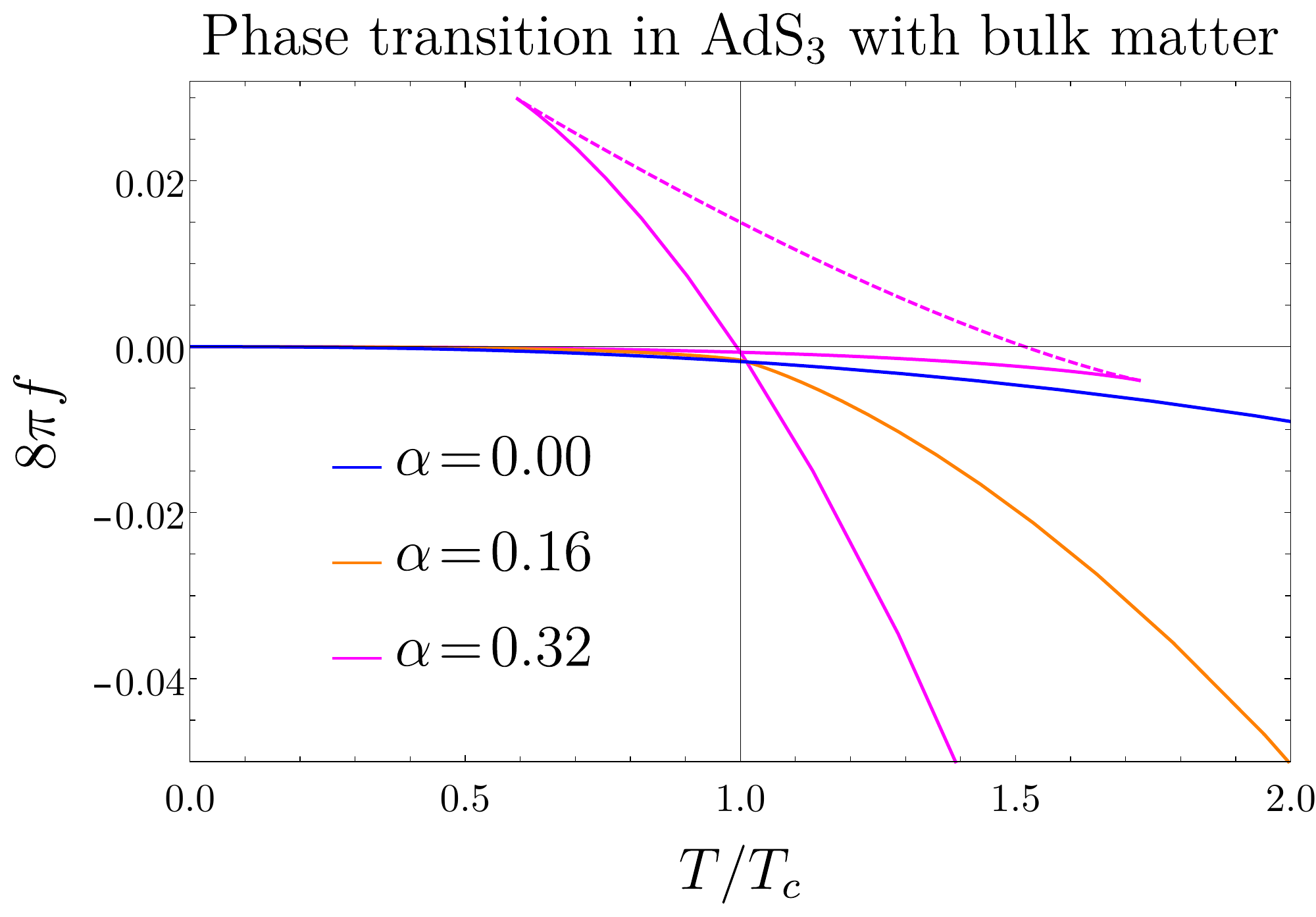}
	\caption{The free energy density $f$ for three choices of $\alpha$. The type of the transition from the left to the right branch depends on this parameter, crossover (blue), second order phase transition (orange) and first order phase transition (purple). The gray line marks the critical temperature where the transition happens and the thermodynamically stable state is a mixed state.}
	\label{fig:freeEDphasetransition}
	\end{center}
\end{figure}

The last system we study is a deformed holographic CFTs, dual to Einstein gravity plus a massive scalar field in AdS$_3$ \cite{Ecker2020}.
The free parameter in the self interaction potential leads to a non-trivial phase structure.
After discussing the model and its thermodynamics we will investigate QNEC$_2$ in ground states (i.e.~domain wall solutions) and thermal states (i.e.~black hole solutions).\\
\\
With the approach described in section \ref{sec:HamiltonianApproach} and the potential $V(\phi)$
\begin{equation}
    V(\phi)=-\frac{1}{2}W(\phi)^2+\frac{1}{2}W'(\phi)^2\,, \qquad W(\phi)=-2-\frac{1}{4}\,\phi^2-\frac{\alpha}{8}\,\phi^4\,, \label{eq:potential}
\end{equation}
our model is fully defined (where the cosmological constant is set to unity and absorbed into the potential).
We restrict to potentials that can be written in terms of the superpotential $W(\phi)$, characterized by a single parameter $\alpha$.
Our choice for the superpotential reveals a rich phase structure and restricts the conformal weight of the dual operator via its relation to the mass of the scalar field $\Delta(\Delta\!-\!2)\!=\!m^2\!=\!-\frac{3}{4}$.
We consider only $\Delta\!=\!\frac{3}{2}$.
The use of the superpotential formalism has the technical advantage of simplifying the equations for finding the domain wall solutions and guarantees the absence of logarithmic terms in the near boundary solution for the thermal states.\\
\\
Using the domain wall parametrization of the metric
\begin{equation}
    \mathrm{d} s^2=\mathrm{d}\rho^2+e^{2A(\rho)}\left(-\mathrm{d} t^2+ \mathrm{d} x^2\right)\,,
\end{equation}
the equations of motion simplify to first order equations
\begin{equation}
    \frac{\mathrm{d} A(\rho)}{\mathrm{d}\rho}=-\frac{1}{2}W(\phi(\rho))\,, \qquad \qquad \frac{\mathrm{d}\phi(\rho)}{\mathrm{d}\rho}=\frac{\mathrm{d} W(\phi(\rho))}{\mathrm{d}\phi(\rho)}\,, \label{1stO-eqs}
\end{equation}
which are solved by
\begin{align}
    \phi(\rho)&=\frac{j e^{-\rho/2}}{\sqrt{1-\alpha j^2 e^{-\rho}}}\,, \label{eq:domainWallSolution1}\\
    A(\rho)&=\Big(1-\frac{1}{16 \alpha}\Big) \rho-\frac{j^2}{16 \left(e^\rho-\alpha j^2\right)}+\frac{\log\!\left(e^\rho-\alpha j^2\right)}{16 \alpha}\,, \label{eq:domainWallSolution2}
\end{align}
where $j$ is the source of the scalar field.
From the near boundary expansion we find that the EMT and the expectation value for the dual operator vanish
\begin{equation} \label{eq:domainwallEMT}
    \langle T_{ij} \rangle=0=\langle {O_{\phi}} \rangle\,.
\end{equation}
Since the free energy is given by the EMT $F\!=\!-\langle T_{xx}\rangle$, the solution \eqref{eq:domainWallSolution1} and \eqref{eq:domainWallSolution2} is dual to the ground state of the field theory.\\
\\
To find solutions corresponding to thermal states, we make a slightly different ansatz than \eqref{eq:genMetric3d}, namely
\begin{equation}
    \mathrm{d} s^2=e^{2A}\left(-H \mathrm{d} t^2+\mathrm{d} x^2\right)+e^{2B}\,\frac{\mathrm{d} r^2}{H}\,,
\end{equation}
where all functions depend on $r$ only.
The residual gauge freedom is fixed by using Gubser gauge \cite{Gubser:2008ny}, where the radial coordinate is identified with the value of the scalar field $r=\phi(r)$.
In this gauge the equations of motion can be cast in the form of a single master equation.
The metric functions can be expressed as integrals of the solution of the master equation.
For special choices of the scalar field potential they can be solved in closed form \cite{Gubser:2008ny}, but for our potential \eqref{eq:potential} we need to solve it numerically \cite{Ecker2020}.
The resulting phase structure depending on the free parameter of the superpotential $\alpha$ can be seen via the free energy density $f$ in figure \ref{fig:freeEDphasetransition}.
While more details of the thermodynamics, EE and $c$-functions are discussed in \cite{Ecker2020}, we will focus only on QNEC$_2$ in these systems in the following two sections.

\subsubsection{QNEC\texorpdfstring{$_{\boldsymbol{2}}$}{2} in Ground States}

\begin{figure}
	\begin{center}
	\includegraphics[height=0.21\textheight]{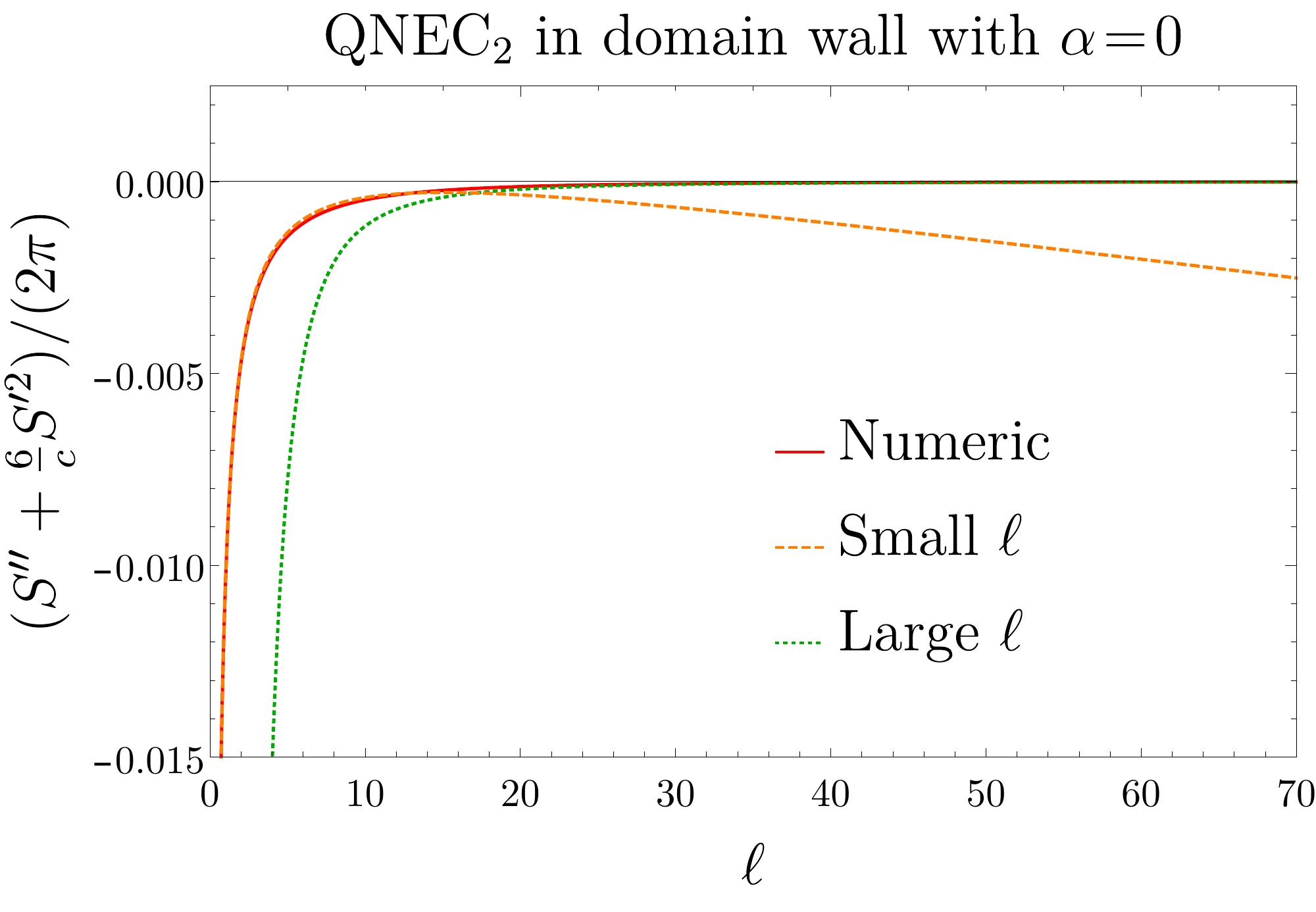}
	\quad
	\includegraphics[height=0.21\textheight]{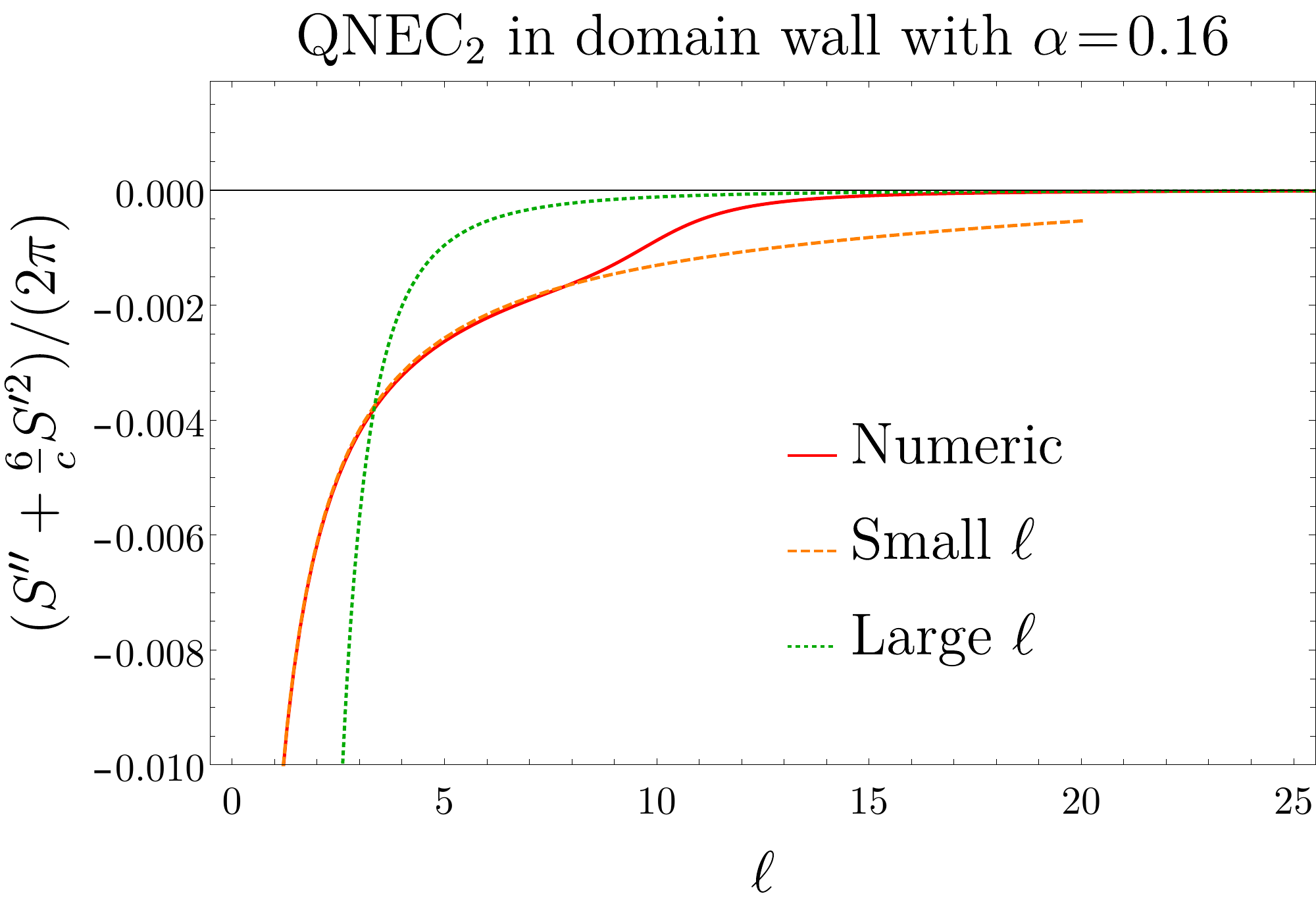}
		\caption{Numerical (red) and perturbative small (orange) and large (green) $\ell$ results for QNEC$_2$ as function of interval size $\ell$. $T_{kk}\!=\!0$ is indicated by the black line.
		}
	\label{fig:QNEC_domainwall1}
	\end{center}
\end{figure}

There are three qualitatively different possibilities for the parameter $\alpha$ in the potential \eqref{eq:potential}: Case 0, where $\alpha\!=\!0$, Case I, where $0\!<\!\alpha\!<\!1$ and Case II, where $\alpha\!<\!0$.
While the geometry for Case 0 and Case I develops a curvature singularity at the values of the radial coordinate $\rho\!\to\-\infty$ and $\rho\!=\!\ln\alpha$ respectively, Case II has a second asymptotic AdS region.
This is interesting for the discussion of $c$-functions in \cite{Ecker2020}, but for (EE and) QNEC negative $\alpha$ shows the same behavior as Case 0.
Therefore we cover Case 0 and Case I by choosing the three values $\alpha\!=\!\{0,0.16,0.32\}$, representing the crossover as well as first and second order phase transitions, shown in figure \ref{fig:freeEDphasetransition}.\\
\\
Employing the methods developed in appendix \ref{app:PnP}, we find perturbative expressions in the limit of small and large entangling regions.
In the small $\ell$ regime we find the universal result, valid for all three Cases 0, I and II
\begin{align}
    \lim_{\ell\ll 1}\Big(S^{\prime\prime} + \frac{6}{c}\,\big(S^\prime\big)^2\Big) = -\frac{c\pi}{64\ell} + \frac{c(128-3\pi^2)}{18432} - \frac{53\pi+9\pi^3}{1179648}\,c\,\ell \nonumber \\
- \frac{c\alpha}{24} + \frac{229\pi}{98304}\,c\alpha\,\ell  - \frac{5\pi}{512}\,c\alpha^2\ell+ \mathcal{O}(\ell^2)\,,
\label{eq:QNECsmall}
\end{align}
where the second line vanishes for Case 0.
To obtain a large $\ell$ results we need to discriminate between the different cases.
For $\alpha\!=\!0$ we find
\begin{equation}
    \lim_{\ell\gg 1}\Big(S^{\prime\prime} + \frac{6}{c}\,\big(S^\prime\big)^2\Big) = -\frac{2c}{3\ell^2\,\ln\ell} - \frac{c(6\ln2-1)}{3\ell^2\,\ln^2\ell} + {\cal O}(1/(\ell^2\,\ln^3\ell))\,,
\end{equation}
for $0\!<\!\alpha\!<\!1$ is given by
\begin{equation}
    \lim_{\ell\gg 1}\Big(S^{\prime\prime} + \frac{6}{c}\,\big(S^\prime\big)^2\Big) = -\frac{c}{24\alpha\,\ell^2\,\ln^2\ell} + {\cal O}(1/(\ell^2\,\ln^3\ell))\,,
\end{equation}
and $\alpha\!<\!0$ can be found in \cite{Ecker2020}, since it is not relevant for our purpose.\\
\\
For a general solution for arbitrary interval we need to employ the numerical methods discussed in sections \ref{sec:GeoEQ} and \ref{sec:NumQNEC} again.
Since we work with the ground state of the CFT where the EMT vanishes \eqref{eq:domainwallEMT}, the null projection is zero as well and QNEC$_2$ is always satisfied and saturated in the large $\ell$ limit.
In figure \ref{fig:QNEC_domainwall1} we show in the left panel the numerical together with the perturbative results for QNEC$_2$ in the cross over ($\alpha\!=\!0$).
The right panel of the same figure shows the results for the second order phase transition ($\alpha\!=\!0.16$).
In contrast to the crossover, the transition is clearly located around $\ell\!=\!10$.
\begin{figure}
	\begin{center}
	\includegraphics[height=0.33\textheight]{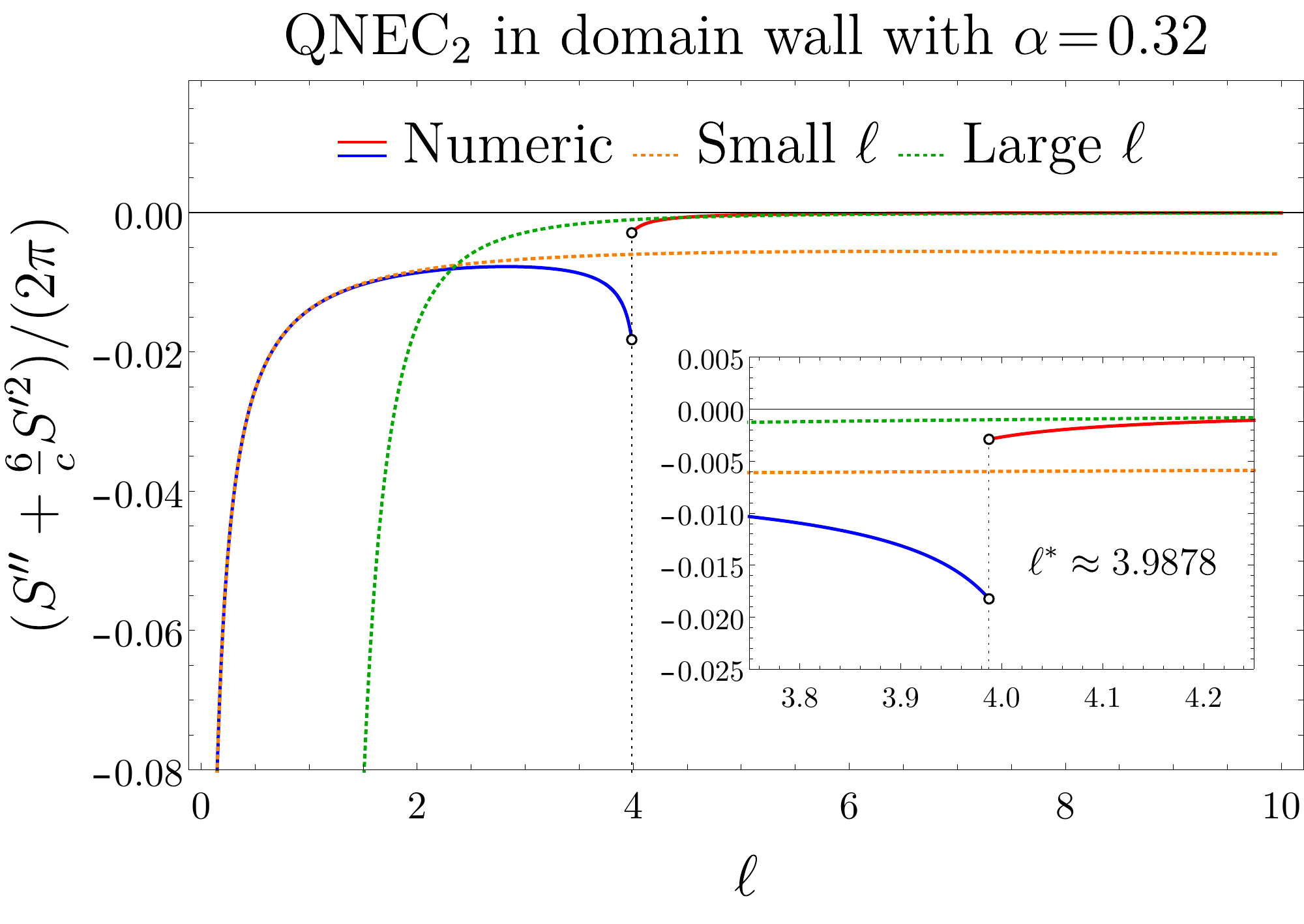}
	\caption{Numerical (blue and red) and perturbative small (orange) and large (green) $\ell$ results for QNEC$_2$ as function of interval size $\ell$. $T_{kk}\!=\!0$ is indicated by the black line. The vertical dashed line marks $\ell^*$ where the multiple extremal surfaces exchange dominance and cause the jump.
	}
	\label{fig:QNEC_domainwall2}
	\end{center}
\end{figure} \\
The most interesting case is the first order phase transition ($\alpha\!=\!0.32$) where QNEC$_2$ has a jump at $\ell^*\!\approx\!3.9878$, shown in figure \ref{fig:QNEC_domainwall2}.
The reason for this jump is a feature of the geometry that is already visible in EE \cite{Ecker2020}, namely the existence of two different branches of extremal surfaces.
In a region around $\ell^*$ there exist two solutions to the geodesic equation with the same boundary condition, but different length.
Only the one that minimizes the length can be interpreted as RT surface and therefore EE.
The exchange of dominance at $\ell^*$ leads to a kink in EE and therefore the jump in its derivatives.

\subsubsection{QNEC\texorpdfstring{$_{\boldsymbol{2}}$}{2} in Thermal States}

\begin{figure}
	\begin{center}
     \includegraphics[height=0.22\textheight]{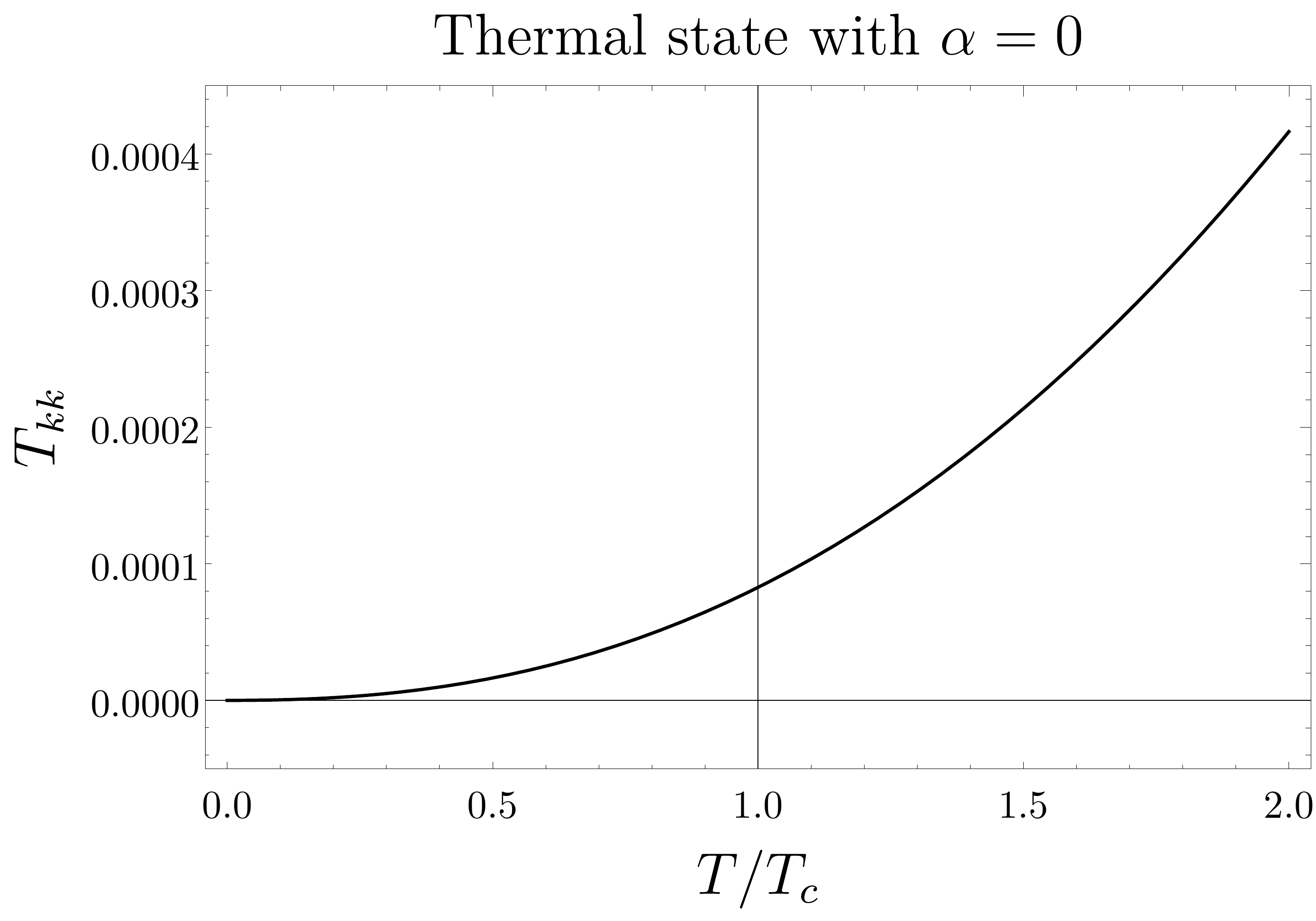}	\includegraphics[height=0.22\textheight]{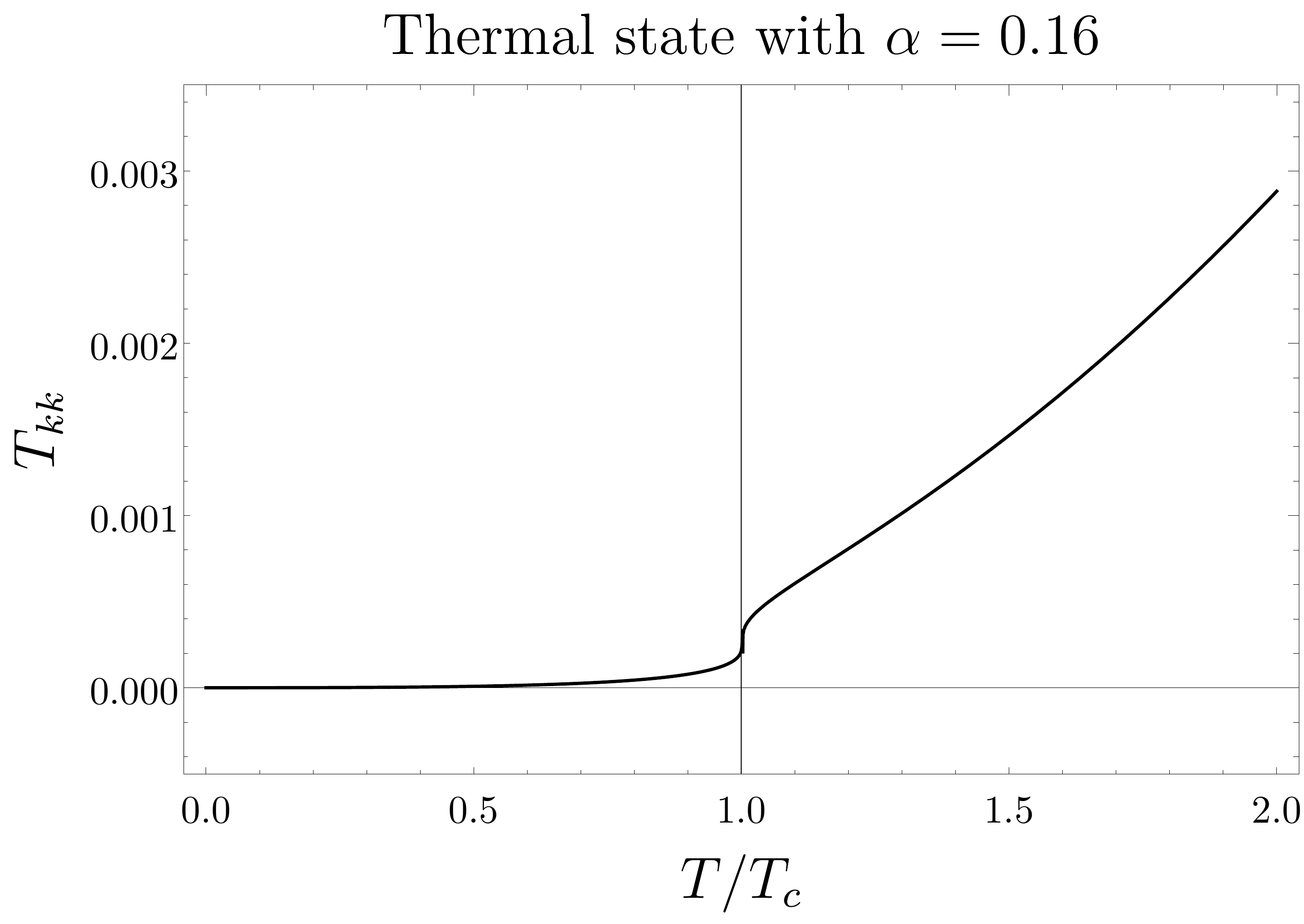} \includegraphics[height=0.22\textheight]{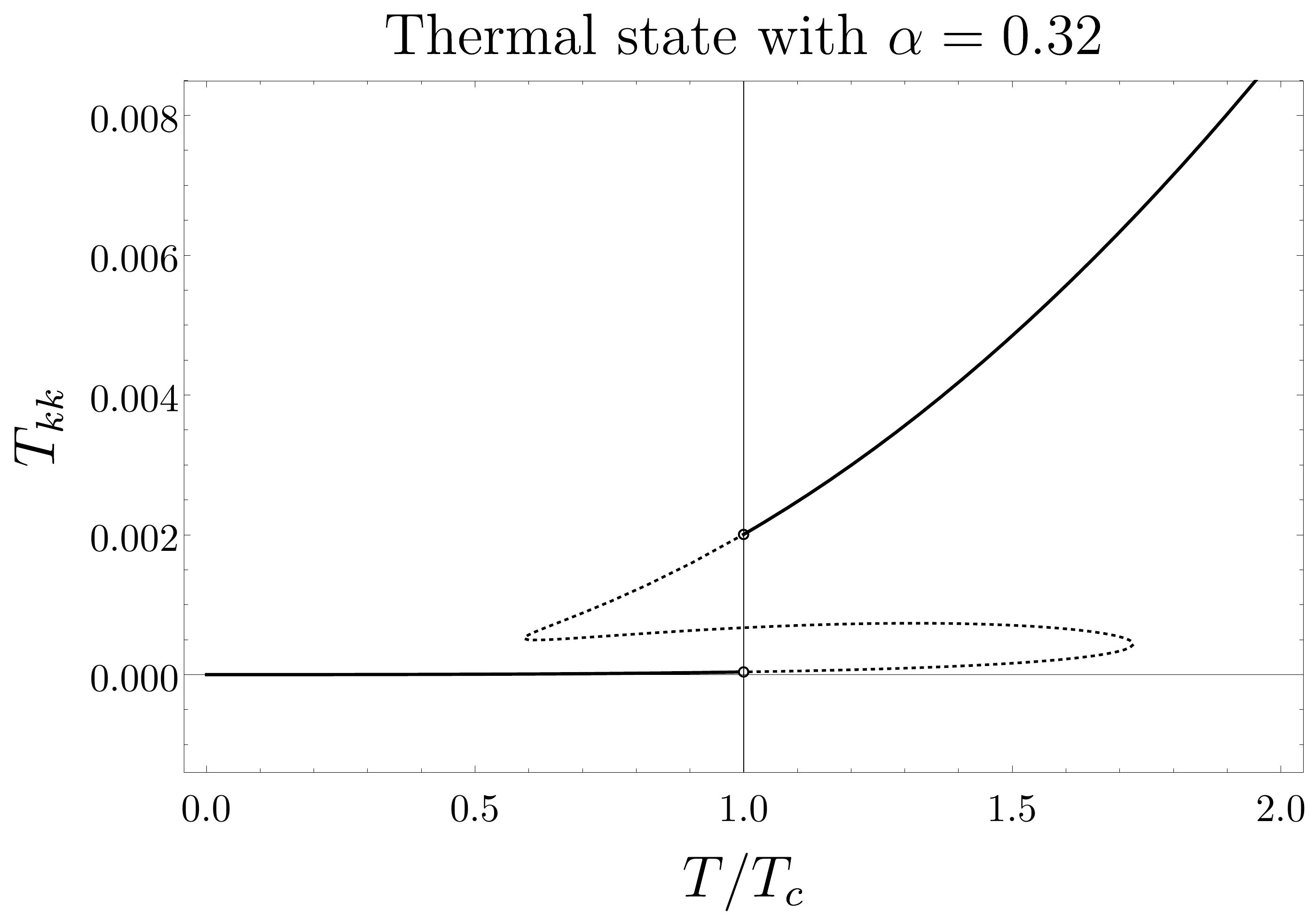}		
	 \caption{$T_{kk}$ as function of temperature for different values of $\alpha$.}
	\label{fig:Tkk_phasetransition}
	\end{center}
\end{figure}

\begin{figure}
	\begin{center}
	\includegraphics[height=0.2\textheight]{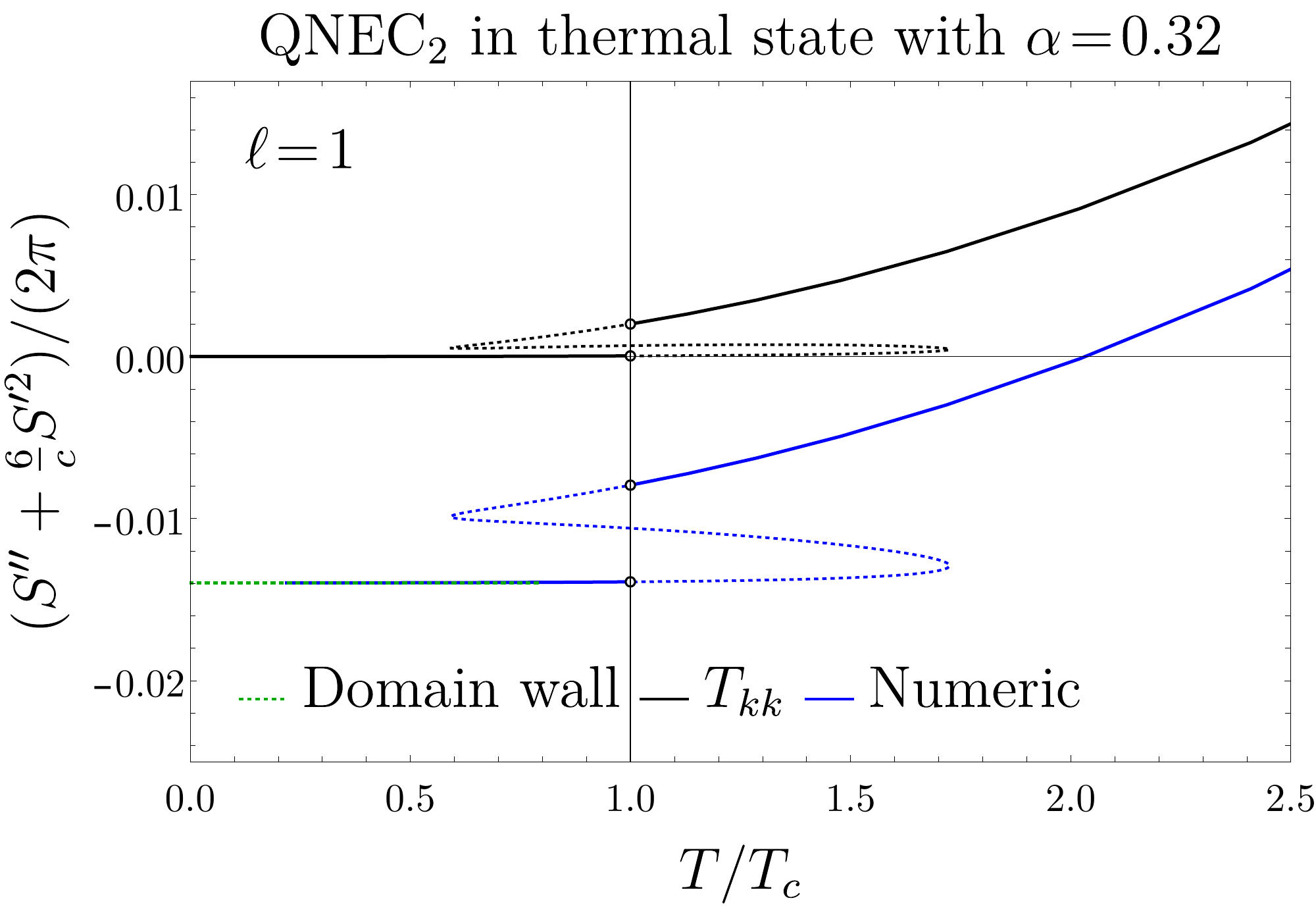}
	\quad
	\includegraphics[height=0.2\textheight]{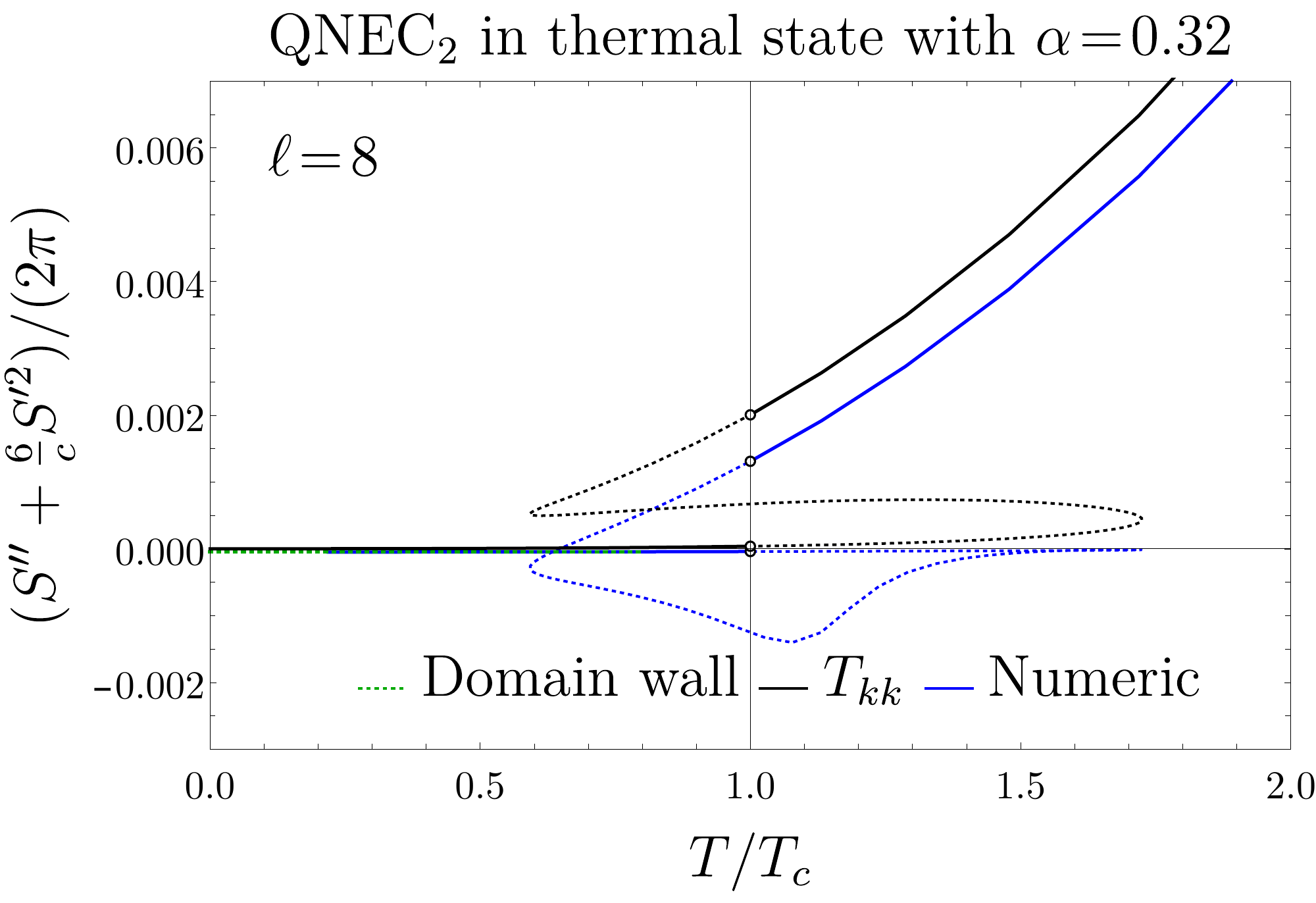}
		\caption{QNEC$_2$ for the black brane solution with $\alpha=0.32$. For separations $\ell\!=\!1$ (left) and $\ell\!=\!8$, well below and above the region with multiple RT-surfaces, QNEC$_2$ shows only the discontinuity at the phase transition at $T_c$. For $\ell\!=\!8$ QNEC$_2$ of the disfavoured state has a surprising shape.}
	\label{fig:QNEC_thermalState3}
	\end{center}
\end{figure}

When investigating QNEC$_2$ for the thermal states we first evaluate $T_{kk}$ for the same three values $\alpha\!=\!\{0,0.16,0.32\}$ as before.
Figure \ref{fig:Tkk_phasetransition} shows the different behavior of $T_{kk}$ as a positive function of temperature in units of the critical temperature for the respective model.
For the crossover (top left, $\alpha\!=\!0$) we see the expected featureless smooth monotonic curve, while for the second order phase transition (top right, $\alpha\!=\!0.16$) $T_{kk}$ increases suddenly at $T\!=\!Tc$.
The most interesting behavior shows the first order phase transition in the plot at the bottom with $\alpha\!=\!0.32$.
In this case $T_{kk}$ is no longer a single valued function of $T$.
From the free energy density (shown in figure \ref{fig:freeEDphasetransition}) we know the thermodynamically preferred phases (solid lines).
In this case we label the preferred phase with $T\!<\!T_c$ \textit{small BH} and the one with $T\!>\!T_c$ \textit{large BH}, where the small/large is to be understood in relation to the AdS radius.
Right at $T_c$ the stable state is actually an inhomogeneous mix of both phases, whose investigation is left for future work.
\begin{figure}
	\begin{center}
	\includegraphics[height=0.22\textheight]{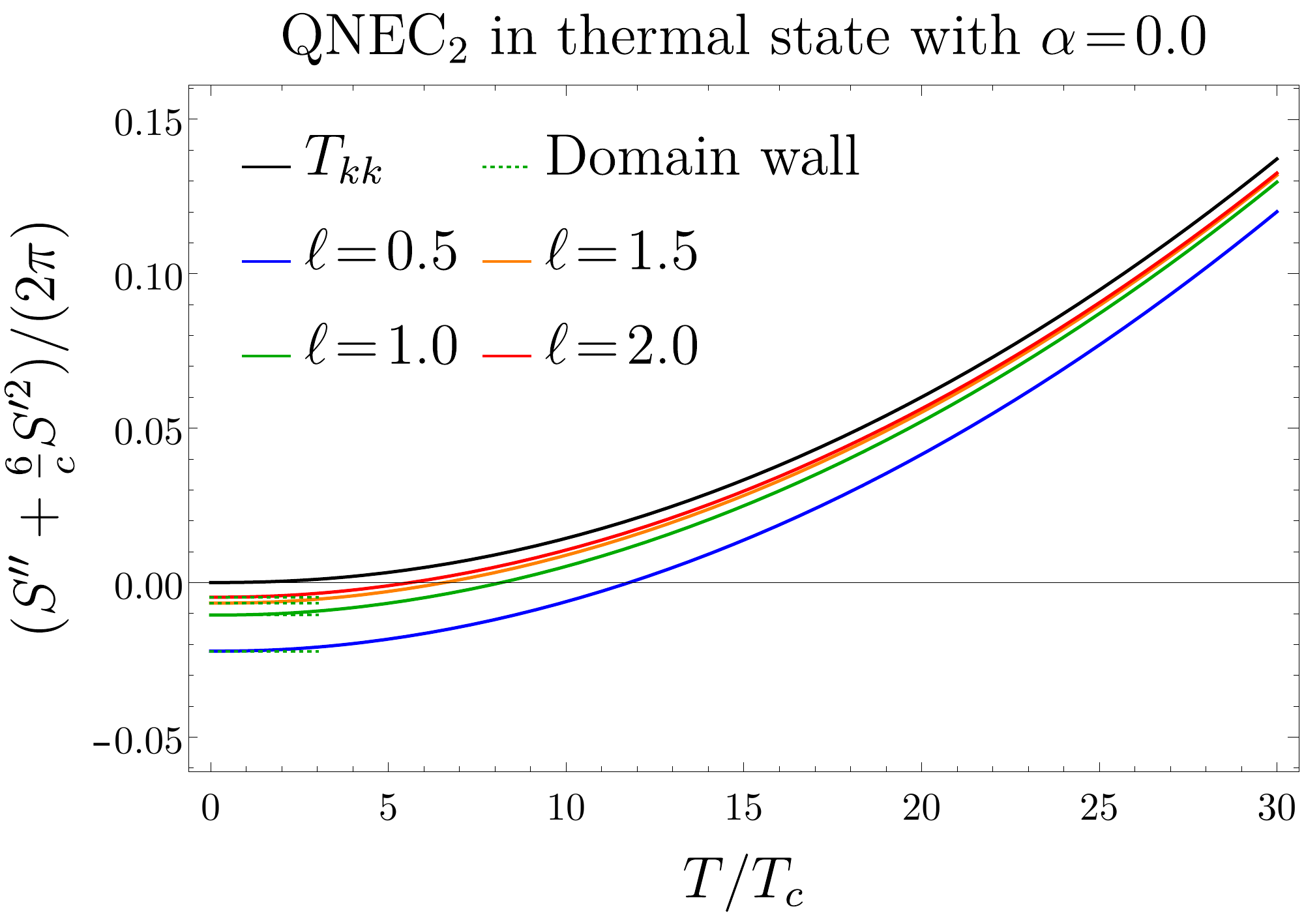}
	\quad
	\includegraphics[height=0.22\textheight]{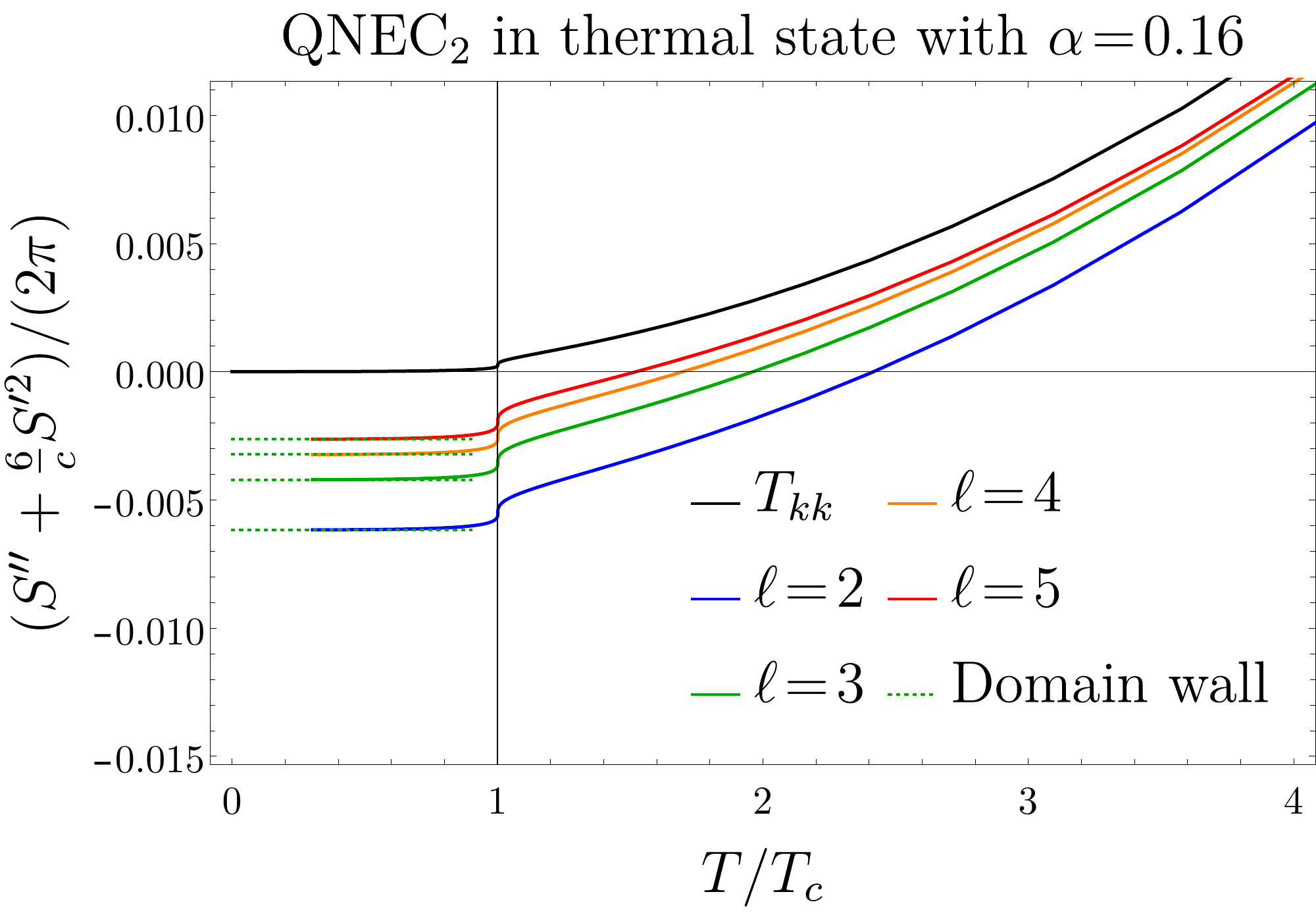}
		\caption{QNEC$_2$ for the black brane solution with $\alpha=0$ (left) and for $\alpha=0.16$ (right) for different values of $\ell$.}
	\label{fig:QNEC_thermalState1}
	\end{center}
\end{figure} \\
Once more we employ our numerical machinery to calculate QNEC$_2$, first for the crossover and the second order phase transition.
The results are shown in figure \ref{fig:QNEC_thermalState1} for several values of $\ell$ together with $T_{kk}$.
The dotted green lines indicate the domain wall solution with corresponding $\alpha$, shown in the previous section, which is the $T\!\to\!0$ limit of the thermal state.
\begin{figure}
	\begin{center}
	\includegraphics[height=0.33\textheight]{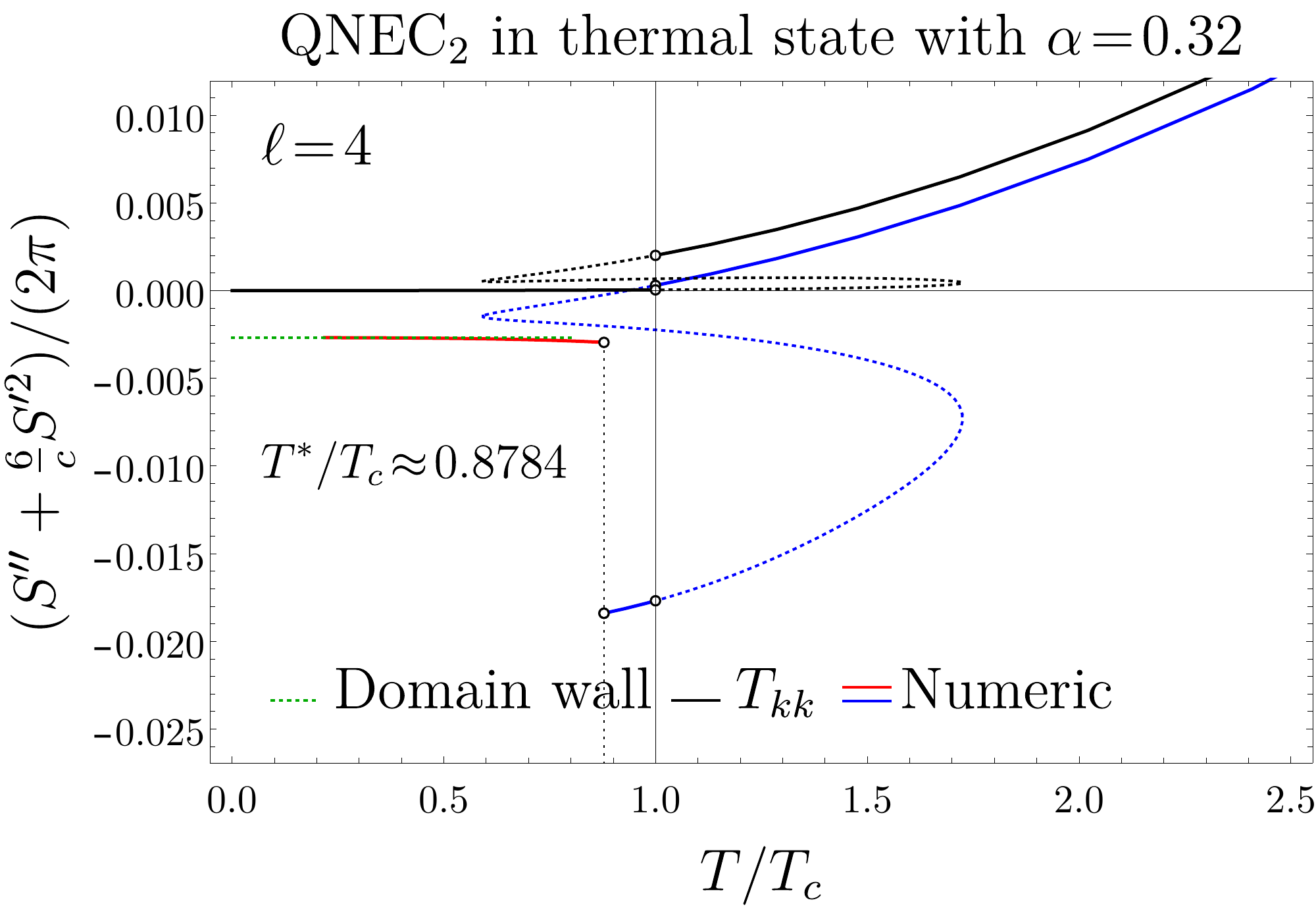}
		\caption{QNEC$_2$ for the black brane solution with $\alpha=0.32$. For a separation around $\ell\!=\!4$ an additional jump in the small BH region from one stable phase to another (red and blue solid lines) happens. The dashed blue (and black) line indicates the disfavored state of QNEC$_2$ ($T_{kk}$). The large BH branch shows no remarkable features.}
	\label{fig:QNEC_thermalState2}
	\end{center}
\end{figure} \\
Similar to the ground state, also the thermal state with $\alpha\!=\!0.32$ provides the richest example.
Choosing the interval $\ell\!=\!4$ very close to the critical value observed in the domain wall solution, we find a jump in the small BH branch in addition to the expected one from the phase transition.
Figure \ref{fig:QNEC_thermalState2} shows our numerical result for that case in red and blue.
The dashed green line indicates the corresponding domain wall solution and $T_{kk}$ is shown in black.
The dashed blue and black sections belong to the thermodynamically disfavoured states.
The discontinuity at $T^*/T_c\!\approx\!0.8784$ is caused by the presence of multiple extremal surfaces as explained above.
The second jump at the critical temperature is the actual phase transition between small and large BHs.
When we study the same system with different interval size, we find only one discontinuity at $T_c$.
The results for $\ell\!=\!1$ and $\ell\!=\!8$ are shown in figure \ref{fig:QNEC_thermalState3}, covering intervals smaller and larger than $\ell^*$.
In the left plot we see the picture we naively expected in the first place, where the QNEC$_2$ combination shows the same behavior as $T_{kk}$ and approaches the ground state in the limit $T\!\to\!0$.
For large separations the stable branches show the expected behavior and QNEC$_2$ is almost saturated for small BHs.
The surprising part is the shape of the dashed line which has a barely visible loop at the right turning point.

\cleardoublepage

\section{Closing remarks}
 \label{sec:End}

\subsection{Summary}

The connection between gravity and gauge theories known as holography, provides a powerful tool in the search for a consistent theory of quantum gravity.
It allows us to study strongly coupled field theories as well as properties of spacetime in regions of strong gravity.
Further, this duality provides a tool to compute entanglement entropy (EE) easily.
One of many applications of EE is the newly developed quantum null energy condition (QNEC), which provides a decisive improvement of the classical energy conditions.
As elaborated on in section \ref{sec:Intro}, the central objective of the thesis was to investigate QNEC, its properties and applications.\\
\\
After reviewing all the necessary theoretical and numerical ingredients in sections \ref{sec:TheoBG} and \ref{sec:NumBG}, QNEC was introduced in section \ref{sec:QNEC}.
Its precise definition and origin in the quantum focussing conjecture are reviewed as well as the holographic proof, followed by the general proof in $d\!\geq\!3$ dimensions.
In two dimensional CFTs QNEC takes a stronger form involving an additional term that ensures the same transformation behavior as the energy momentum tensor.
The final part of this section was dedicated to the numerical method we developed for calculating QNEC.
Using the holograhpic description of EE, we moved one boundary point of the extremal surfaces (i.e.~geodesics) along a null direction in order to achieve the required deformation of the entangling region.
This way we could obtain EE as a function of the shift along the null direction and calculate its derivatives.
A similar approach to calculate QNEC with paper and pencil, using a variable boundary for the area integral, is shown in the appendix \ref{app:PnP}.\\
\\
In section \ref{sec:QNEC_D=4} the focus was on $4$-dimensional QFTs, where we confirmed the validity of our numerical tools with the analytic results of the vacuum state.
For the thermal states we found that QNEC is satisfied but not saturated.
The perturbative results obtained for small and large entangling regions agreed perfectly with the full numerical solution.
The behavior of QNEC in a global quench from vacuum to a thermal state already showed many interesting features.
While long before and after the quench, QNEC approached the behavior of vacuum or thermal states respectively, in the intermediate regime QNEC developed a dip and diverged for a certain parameter range.
The dependence on the null direction was explained by the alignment of the deformation with (or perpendicular to) the mass shell in the Vaidya geometry.
Further we found striking numerical evidence for $\tfrac{1}{4}$-saturation before and around the quench for large entangling regions.
The toy model for heavy ion collisions where NEC is violated, provided a very interesting system for QNEC studies.
We found regions where QNEC is stronger than NEC, QNEC is weaker than NEC and where QNEC is saturated in spite of the far from thermal equilibrium character.
This was the only system in four dimensions where we found saturation.\\
\\
Section \ref{sec:QNEC_D=2} was dedicated to the stronger inequality QNEC$_2$.
Starting along the same lines as the $4$-dimensional case, in vacuum and thermal states, QNEC$_2$ saturated.
This was expected since they are dual to Ba\~nados geometries and we found a proof for QNEC$_2$ saturation in these states.
We showed striking numerical evidence for $\tfrac{1}{2}$-saturation before and around the quench for large entangling regions.
In general the quench taught us that the presence of bulk matter leads to non-saturation of QNEC$_2$.
This happened only if the extremal surfaces pass through regions where the matter is present.\\
The next step when investigating the influence of bulk matter was adding a massive scalar field to the gravitational theory.
First we studied the backreaction of a single particle at the center of global AdS, which required quantum corrections to EE to be taken into account for calculating QNEC$_2$.
For the maximal entangling region (the semi-circle) we considered the limit of large conformal weight of the scalar field and found that QNEC$_2$ is gapped by $\tfrac{h}{4}$, depending on the weight $h$ of the scalar field.\\
Finally we considered a massive self-interacting scalar field in the bulk.
Already the ground states showed interesting features in QNEC$_2$, depending on the shape of the potential.
For a specific parameter range we found multiple extremal surfaces for the same boundary region, leading to a kink in EE and a jump in QNEC$_2$.
Thermal states showed a crossover, first order phase transition or second order phase transition from small to large black holes, depending on the choice of potential.
Again we found the jump in QNEC$_2$ caused by multiple branches of extremal surfaces in addition to the jump caused by the second order phase transition.

\cleardoublepage

\subsection{Conclusion}

First of all we can identify three cases: \textit{satisfaction, saturation} and \textit{violation} of QNEC.
In view of all the proofs listed in section \ref{sec:QNECproofs} the last scenario has to come with something considered unphysical like non-unitary field theories.
Nonetheless it is an interesting result, when we change our point of view.\vspace{-5px}
\paragraph{Satisfaction/Violation.}
QNEC (and probably also QNEC$_2$) are unavoidable convexity conditions in unitary relativistic QFTs and can serve as validators of semi-classical toy models.
The model for heavy ion collisions considered in section \ref{sec:HIC} passes this test.
A different example that should be treated with caution is a holographic model of Hawking evaporation of a black hole, where QNEC is violated in the dual field theory.\\
\\
In physical situations satisfaction of QNEC alone does not provide a lot of information, but its (non-)saturation behavior is determined by bulk matter.\vspace{-5px}
\paragraph{Saturation.}
In the absence of bulk matter QNEC$_2$ always saturates.
This is shown by the proof in section \ref{sec:Banados} using the properties of Ba\~nados geometries.\vspace{-5px}
\paragraph{Non-saturation I.}
In the quenched systems we find $\tfrac{1}{2}$- and $\tfrac{1}{4}$-saturation numerically for the $2$- and $4$-dimensional systems respectively.
This occurs in the region before and around the quench, where bulk matter has the most influence.
This behavior was later confirmed by a holographic calculation as the factor \textit{one over boundary spacetime dimension} by \cite{Mezei:2019sla}.\vspace{-5px}
\paragraph{Non-saturation II.}
When taking into account quantum corrections from a massive scalar particle, we find that QNEC$_2$ is gapped by $\tfrac{h}{4}$, depending on the weight of the scalar field $h$.
This happens only in the case of the maximal entangling region where the geometric probes are close to the particle at the center of spacetime.\\
\\
Finally when investigating QNEC$_2$ in the system with phase transitions, we see that QNEC$_2$ can never be saturated due to the presence of the matter field in the whole spacetime.
Nonetheless we find characteristic features that can be seen in EE and QNEC$_2$, but not in thermodynamic quantities of the system.\vspace{-5px}
\paragraph{Prediction.}
For some scalar field potentials, the features of QNEC$_2$ in the ground state provide enough information to predict the thermal phase structure of the system, in particular the existence of first order phase transitions.

\cleardoublepage

\subsection{Future Directions}

While some questions have been answered and a suitable method to calculate QNEC numerically was found, this only opens the door for a number of new questions and directions.\\
\\
A general extension of this work would be the inclusion of quantum corrections beyond the supergravity approximation and including $\tfrac{1}{N_c}$ corrections to the CFT.
Another project for the future is to include a gauge field on the gravity side and continue the investigation of QNEC and its saturation behavior in the dual theories.
Besides improving the model toward more complexity, finding more systems and states to study QNEC is equally important.
Such states may include violations of the classical energy conditions, dynamical matter fields in the bulk or even curved boundary theories.
A different direction in the search for new models would be an extension towards holographic Bondi-Metzner-Sachs field theories or warped CFTs along the lines of \cite{Grumiller:2019xna,Detournay:2020vrd}.\\
\\
After these general remarks, a more specific goal is improving QNEC itself.
In spite of being the only known consistent quantum EC, we have seen evidence that QNEC cannot always be saturated.
Besides the precise gaps of $\tfrac{1}{d}$ for the quenched states or $\tfrac{h}{4}$ for the backreacted system, in our latest work \cite{Ecker2020} we found an even stricter inequality than QNEC$_2$ involving the Casini-Huerta $c$-function.
All this can be seen as hints towards more general quantum energy conditions, that reduce to QNEC whenever it is saturated.
The search for these unknown quantum energy conditions is certainly an exciting project for the future.\\
\\
In the very end I want to mention another project, that combines my personal interest in the numerical aspect with the generalization of all the previous work.
Extending the numerical algorithm from computing geodesics to higher dimensional extremal surfaces is a challenging task.
If implemented, such a tool would allow us to remove the restriction to strip-entangling regions and perform the null-deformation locally even in higher dimensional systems and for arbitrary entangling regions.

\cleardoublepage

\begin{appendix}
 \section{Application of characteristic formulation}
 \label{app:Application}

In this appendix I show how the characteristic formulation, explained in section \ref{sec:NumRelApproaches} is used to solve Einstein's equations for the system of colliding shock waves, discussed in section \ref{sec:HIC}.
I will explain the steps necessary to calculate the numerical solution to Einstein's equations for this particular example, making a general ansatz for the metric in Eddington-Finkelstein gauge \eqref{eq:genMetric5d}
\begin{equation} \label{eq:metric_z}
     \mathrm{d}s^2 = - A \, \mathrm{d}v^2 + 2 \, \mathrm{d}r \mathrm{d}v + 2 \, F \, \mathrm{d}v \mathrm{d}y + R^2 \, e^{-2B} \mathrm{d}y^2 + R^2 \, e^B \left( \mathrm{d}x_1^2 + \mathrm{d}x_2^2 \right),
\end{equation}
where $A$, $B$, $F$ and $R$ are functions of the coordinates $(r, v, y)$, but not $x_i$.
The gravitational shocks propagate in the $y$-direction while spacetime is homogeneous and isotropic in the $x_i$-directions.
Often it is more convenient to work with an inverted radial coordinate $z\!=\!1/r$ since the boundary is then located at $z\!=\!0$ instead of $r\!=\!\infty$.
For more clarity of the equations I will stick with $r$ for now and perform the coordinate transformation later on.
Inserting the above metric into Einstein's equations for vacuum \eqref{eq:EEQ_vac} leads to the following set of equations of motion
\begin{align} \label{eq:EEQ_Shocks}
 R'' &= -\frac{1}{2} \, R (B')^2, \\
 R^2 \, F'' &= R \, (6 \tilde{R} \, B' + 4 \, \tilde{R}' + 3 \, F' \, R') + R^2 \, (3 \, \tilde{B} \, B' + 2 \, \tilde{B}') - 4 \, \tilde{R} \, R', \\
 12 \, R^3 \, \dot{R}' &= e^{2 B} \, (R^2 \, (4 \, \tilde{B} \, F' - 7 \, \tilde{B}^2 - 4 \, \tilde{\tilde{B}} + 2 \, \tilde{F}' + (F')^2) \nonumber \\
    &\quad + 2 \, R \, (\tilde{R} \, (F' - 8 \, \tilde{B}) - 4 \, \tilde{\tilde{R}}) + 4 \, \tilde{R}^2) + 24 \, R^2 \, (R^2 - \dot{R} \, R'), \\
 6 \, R^4 \, \dot{B}' &= e^{2 B} \, (R^2 \, (-\tilde{B} \, F' + \tilde{B}^2 + \tilde{\tilde{B}} - 2 \, \tilde{F}' - (F')^2) \nonumber \\
    &\quad + R \, (\tilde{R} \, (\tilde{B} + 4 \, F') + 2 \, \tilde{\tilde{R}}) - 4 \, \tilde{R}^2) - 9 \, R^3 \, (\dot{R} \, B' + \dot{B} \, R'), \\
 2 \, R^4 \, A'' &= e^{2 B} \, (R^2 \, (7 \, \tilde{B}^2 + 4 \, \tilde{\tilde{B}} - (F')^2) + 8 \, R \, (2 \, \tilde{B} \, \tilde{R} + \tilde{\tilde{R}}) - 4 \, \tilde{R}^2) \nonumber \\
    &\quad -2 \, R^4 \, (3 \, \dot{B} \, B' + 4) + 24 \, \dot{R} \, R^2 \, R', \\
 6 \, R^2 \, \dot{F}' &= 3 \, (R^2 \, (-(2 \, B' \, (\tilde{A} + 2 \, \dot{F}) + 2 \, \tilde{A}' + 6 \, \dot{B} \, \tilde{B} + 4 \, \tilde{\dot{B}} + A' \, F')) \nonumber \\
    &\quad +2 \, R(R' \, (\tilde{A} + 2 \, \dot{F}) - 6 \, \dot{B} \, \tilde{R} - 4 \, \tilde{\dot{R}} - 3 \, \dot{R} \, F') + 8 \, \dot{R} \, \tilde{R}), \\
 6 \, R^2 \, \ddot{R} &= e^{2 B} \, (R \, (2 \, \tilde{B} \, (\tilde{A} + 2 \, \dot{F}) + \tilde{\tilde{A}} + 2 \, \tilde{\dot{F}}) + \tilde{R} \, (\tilde{A} + 2 \, \dot{F})) \nonumber \\
    &\quad +3 \, R^2 \, (\dot{R} \, A'-\dot{B}^2 \, R).
\end{align}
This seems a bit confusing, but defining derivatives along in- and outgoing radial null geodesics as well as in the longitudinal direction $y$, explains the symbols used
\begin{align}
    h'= \partial_r h, \qquad \dot{h} = \partial_v h - \frac{1}{2} \, A \, \partial_r h, \qquad \tilde{h} = \partial_y h + F \, \partial_r h.
\end{align}
Despite the complexity of the problem, the characteristic formulation produced this set of equations which is quite simple and can be solved with an methodical algorithm.
Given some initial condition for $B$ \cite{Chesler:2008hg,vanderSchee:2014qwa} the equations can be solved for the other functions on one characteristic.
This can be done using spectral methods, see section \ref{sec:NumRelMethods}.
Once all functions are known on the characteristic, it is possible to evolve $B$ to the next slice, using a Runge-Kutta algorithm (see section \ref{sec:NumRelMethods}).
With $B$ on the next slice, the procedure starts all over again until the metric is known in a sufficiently large domain for the problem at hand.\\
\\
In order to close the loop to section \ref{sec:HoloRen}, as a side remark I want to make the relation to section \ref{sec:HoloRen}.
The numerical solutions can be used to find the non-renormalizable coefficients of the near boundary FG expansion \eqref{eq:FGexpansion}.
With this information it is possible to construct the EMT and find the results for energy density and the NEC \eqref{eq:NEC} in the dual field theory shown in figure \ref{fig:shocks}.

\paragraph{Numerical simplifications}
can be achieved by the coordinate transformation \mbox{$r\!\to\!z\!=\!\frac{1}{r}$}, despite making equations slightly more difficult.
It allows to replace the (otherwise diverging) functions with their finite parts, securing precise numerics up to the AdS boundary
\begin{align}
 &A = \frac{1}{z^2} + \frac{2 \, \xi}{z} + \xi^2 - 2 \, \partial_v \xi + z^2 \, A_{reg}, \label{Areg_SW} \\
 &B = z^4 \, B_{reg}, \label{Breg_SW} \\
 &F = \partial_y \xi + z^2 \, F_{reg}, \label{Freg_SW} \\
 &R = \frac{1}{z} + \xi + z^4 \, R_{reg}, \label{SIGMAreg_SW} \\
 &\dot B = z^3 \, \dot B_{reg}, \label{BDOTreg_SW} \\
 &\dot R = \frac{1}{2 \, z^2} + \frac{\xi}{z} + \frac{\xi^2}{2} + z^2 \, \dot R_{reg}, \label{SIGMADOTreg_SW}
\end{align}
where the divergent terms are known from the near boundary expansion and solving Einstein's equations order by order.
Replacing the functions in \eqref{eq:EEQ_Shocks} with these ones, one can follow the same procedure of solving the system on one characteristic and then evolve to the next one for the regular functions.
After finishing the computation, the replacement can be reversed.

\cleardoublepage

\section{QNEC with Paper and Pencil}
 \label{app:PnP}
  
In this appendix I want to show how EE and QNEC can be tackled by paper and pencil (or rather \textit{Mathematica}).
Although the main focus of this thesis lies on numerical computation, it is often useful to have these tools available.
First I will establish the general procedure for calculating extremal surfaces and their deformations, finding a set of three integrals.
Since in most cases these integrals cannot be solved in closed form, perturbative approaches are necessary and I will give one example in the second part.
Although I will stick to AdS$_3$ here, the methods remain the same for higher dimensions, if we consider the simplification to strip regions in section \ref{sec:GeoEQ}.
One or another variant of the methods shown below are applied in sections \ref{sec:AdS5BB}, \ref{sec:backreaction} and \ref{sec:phasetransition}.
  
\subsection{Integrals}
 \label{app:Integrals}

The main ingredient of QNEC is entanglement entropy, which is related to some extremal surface via holography (see sections \ref{sec:EE} and \ref{sec:HEE}).
In the following we will restrict ourselves to systems relevant in this thesis, that are $3$-dimensional bulk theories (or systems that can be reduced to such) where the metric depends only on the radial (\textit{holographic}) coordinate.
The area (which we use synonymous for the length of a geodesic) is then given by the integral of the geodesic Lagrangian
\begin{equation} \label{eq:areaintegral1}
    \mathcal{A}(\lambda,\,\ell,\,z_{cut}) = 2 \int\limits_0^{(\ell+\lambda)/2-\omega}\mathrm{d} x\,\mathcal{L}(\dot t,\,\dot z,\,z)\,,
\end{equation}
and depends on the shift in the null direction $\lambda$, the size of the entangling interval $\ell$ and a cutoff in the radial direction $z_{cut}$ ($\omega$ is a specific function of this cutoff).
Making use of the symmetry we compensate integrating from the central (turning) point of the geodesic to the boundary by a factor of $2$.
To express the corresponding Lagrangian in terms of the metric components, we use the spatial coordinate $x$ as affine parameter (and denote derivatives w.r.t.~it with a dot)
\begin{equation}
    \mathcal{L}(\dot t,\,\dot z,\,z) = \sqrt{g_{tt}(z)\dot t^2+g_{zz}(z)\dot z^2+g_{xx}(z)+g_{tz}(z)\dot t\dot z+g_{tx}(z)\dot t+g_{zx}(z)\dot z}\,.
\end{equation}
In order to simplify this integral we make use of the symmetries of the geometry and calculate the Noether charges
\begin{equation}
Q_1 = \dot z\,\frac{\partial\mathcal{L}}{\partial\dot z} +\dot t\,\frac{\partial\mathcal{L}}{\partial\dot t} - \mathcal{L}\,, \qquad Q_2 = \frac{\partial\mathcal{L}}{\partial\dot t}\,.
\end{equation}
Evaluating $Q_1$ at the turning point of the geodesic $z\!=\!z_*$ and defining $\Lambda$ as a specific combination of $Q_1$ and $Q_2$, allows to express $\dot t$ and $\dot z$ as
\begin{equation} \label{eq:dots}
    \dot t = \Lambda\,h(z,\,z_*,\,\Lambda) \qquad\qquad \dot z=f(z,\,z_*,\,\Lambda)\,,
\end{equation}
where $f$ and $h$ contain components of the metric.\\
\\
We can now write down the integral over the time direction along the geodesic
\begin{equation} \label{eq:tempint}
    \frac{\lambda}{2} = \int\limits_0^{\lambda/2}\mathrm{d} t = \int\limits_{z_*}^0\mathrm{d} z\,\frac{\dot t}{\dot z} = \Lambda\,\int\limits_{z_*}^0\mathrm{d} z\,\frac{h(z,\,z_*,\,\Lambda)}{f(z,\,z_*,\,\Lambda)}\,.
\end{equation}
Similarly we can rewrite the spatial part as
\begin{equation} \label{eq:regionint}
    \frac{\ell+\lambda}{2} = \int\limits_0^{(L+\lambda)/2}\mathrm{d} x = \int\limits_{z_*}^0\frac{\mathrm{d} z}{\dot z} = \int\limits_{z_*}^0\frac{\mathrm{d} z}{f(z,\,z_*,\,\Lambda)}\,.
\end{equation}
Following the same pattern, \eqref{eq:areaintegral1} can be rewritten as well
\begin{equation} \label{eq:areaintegral2}
    \mathcal{A} = 2 \int_{z_*}^{z_{cut}}\mathrm{d} z\,\frac{\mathcal{L}(\Lambda\,h(z,\,z_*,\,\Lambda),\,f(z,\,z_*,\,\Lambda),\,z)}{f(z,\,z_*,\,\Lambda)}\,.
\end{equation}
The last integral diverges when the cutoff is removed, which makes it convenient to renormalize it with an $\lambda$ and $\ell$ independent counter-term.
In order to find EE and QNEC, the first integral \eqref{eq:tempint} is solved and provides an expression for $\Lambda$ in terms of $\lambda$ and $z_*$.
This expression can then be used to solve the second integral \eqref{eq:regionint} to provide $z_*$ as function of $\lambda$ and $\ell$.
Since we are interested in the second derivatives w.r.t.~$\lambda$, we can expand the integrands accordingly, which simplifies the calculation a lot. 
These two results allow to express the area integral \eqref{eq:areaintegral2} in terms of $\lambda$ and $\ell$ only.
Performing the integration of \eqref{eq:areaintegral2} then yields EE and consequently QNEC.

\subsection{Perturbative Solutions}
 \label{app:Perturbative}

Although the procedure to find EE and QNEC just presented above seems pretty straight forward, usually the integrals cannot be solved in closed form.
The reason for this is the complicated form of the functions $f$ and $h$, even if the metric is known analytically.
Often the only way to get analytic results is solve the integrals in some limit perturbatively.
In this thesis mainly small or large $\ell$ expansions were considered.
The first application of this procedure is the small and large $\ell$ result for QNEC \eqref{eq:AdSBB_smallL} and \eqref{eq:AdSBB_largeL} in section \ref{sec:AdS5BB}.
Here I will show the calculation in the small interval limit for the black brane in AdS$_5$, as example for all other applications of this perturbative method.
For a consistent treatment we need the chain of inequalities to be $0\!<\tfrac{\lambda}{\ell}\!\ll\!T\ell\!\ll\!1$.\\
\\
This example was reduced from finding higher dimensional minimal surfaces to solving the geodesic equation in an effectively $3$-dimensional auxiliary spacetime (see section \ref{sec:GeoEQ}) by choosing a strip as entangling region.
For the area functional \eqref{eq:area_functional_5D} for the metric \eqref{eq:genMetric5d} with choice $A\!=\!1/z^2\!-\!(\pi T)^4 z^2$, $R\!=\!1/z$ and $B\!=\!F\!=\!0$ reads
\begin{equation} \label{eq:areaexample}
    \mathcal{A}(\lambda,\,\ell,\,z_{cut}) = 2 \int_0^{(\ell+\lambda)/2-\omega}\mathrm{d} y \frac{1}{z^3} \sqrt{1+\frac{\dot z^2}{f(z)}-\dot t^2 f(z)}\,,
\end{equation}
where $f(z)\!=\!1-\!M z^4$ and the only change compared to \eqref{eq:areaintegral1} is the additional conformal factor $\tfrac{1}{z^3}$ and the change of parameter from $x$ to $y$.
The next step is to find the Noether charges
\begin{equation}
    Q_1 = \frac{1}{z^3\sqrt{1+\dot z^2/f(z)-\dot t^2 f(z)}}\,, \qquad Q_2 = -\frac{f(z)}{z^3\sqrt{1+\dot z^2/f(z)-\dot t^2 f(z)}}\,.
\end{equation}
Evaluating $Q_1$ at the turning point $z_*$ and defining the constants $\Lambda$ and $N_*$ yields
\begin{equation}
    \Lambda := \frac{-Q_2}{Q_1} = \dot t f(z)\,, \qquad Q_1^* = \frac{1}{z_*^3 N_*}\,, \qquad N_* := \sqrt{1-\frac{\Lambda^2}{/f(z_*)}}\,.
\end{equation}
As indicated in \eqref{eq:dots} can write $\dot t$ and $\dot z$ as
\begin{equation} \label{eq:dotsexample}
    \dot t = \frac{\Lambda}{f(z)}\,, \qquad \qquad \dot z = \sqrt{\Lambda^2-f(z)+\frac{N_*^2z_*^6f(z)}{z^6}}
\end{equation}
Instead of determining $\Lambda$ directly from \eqref{eq:tempint}, it turns out that for this system it is easier to combing it with \eqref{eq:regionint} to
\begin{equation}
    \frac{\lambda}{2}-\frac{\ell+\lambda}{2} = \int_{z_*}^0\mathrm{d} z\,\left(\frac{\dot t}{\dot z} - \frac{1}{\dot z}\right)\,.
\end{equation}
Inserting \eqref{eq:dotsexample} and rescaling the integration variable to $x\!=\!\tfrac{z}{z_*}$ allows us to express $\lambda$ via the following integral
\begin{equation}
    \Lambda = \frac{\lambda}{\ell+\lambda+2z_*I_\Delta}\,, \quad I_\Delta = \int^1_0 \mathrm{d}x \frac{1}{\sqrt{\Lambda^2-f(z_* x)+\tfrac{N_*^2 f(z_*x)}{x^6}}} \left( \frac{1}{f(z_*x)} -1 
\right)\,.
\end{equation}
Evaluating this integral perturbatively yields
\begin{align}
    \Lambda = &\frac{\lambda}{\ell+\lambda}  - (\pi T z_*)^4\,\frac{4\pi c_0\lambda z_*}{15 \sqrt{3}(\ell+\lambda)^2} \\
    &+ (\pi T z_*)^8\, \left(\frac{16\pi^2 c_0^2\lambda z_*^2}{675(\ell+\lambda)^3} -\frac{2\lambda z_*}{3(\ell+\lambda)^2}\right) + \mathcal{O}((T z_*)^{12}) + \mathcal{O}(\lambda^3/\ell^3)\,, \nonumber
\end{align}
where $c_0\!=\!\tfrac{3\Gamma[1/3]^3}{2^{1/3}(2\pi)^2}$.
With this result we can continue the procedure and evaluate \eqref{eq:regionint} to obtain a series expansion for $z_*$
\begin{align}
    \frac{z_*}{c_0\ell} =& 1 + (\pi T\ell)^4\,\frac{2\pi c_0^6}{15\sqrt{3}} + (\pi T\ell)^8\, \left(\frac{4\pi^2 c_0^{12}}{135}-\frac{c_0^9}{6}\right)\\
    &+\frac{\lambda}{\ell}\, \left(1 - (\pi T\ell)^4\,\frac{2\pi c_0^6}{3\sqrt{3}} + (\pi T\ell)^8\, \left(\frac{4\pi^2c_0^{12}}{15}-\frac{3c_0^9}{2}\right)\right) \nonumber \\
    &+\frac{\lambda^2}{\ell^2}\, \left(-\frac12 + (\pi T\ell)^4\,\left(\frac{c_0^4}{6} - \frac{49\pi c_0^6}{45\sqrt{3}}\right) \right. \nonumber \\
    &+ \left. (\pi T\ell)^8\, \left(\frac{c_0^8}{6} - \frac{71 c_0^9}{12} - \frac{c_0^{10} \pi}{5 \sqrt{3}} + \frac{2074 c_0^{12}\pi^2}{2025} \right) \right) + \mathcal{O}((T\ell)^{12})+\mathcal{O}(\lambda^3/\ell^3)\,. \nonumber
\end{align}
The area is then given by integral \eqref{eq:areaexample}, inserting $f(z)$, \eqref{eq:dotsexample} and changing the integration parameter from $y$ to $z$.
Instead of subtracting counter-terms we decide to expand in powers of the cutoff for this example, which also isolates the divergent constant term
\begin{equation}
    \mathcal{A} = \frac{1}{z_{cut}^2} + \frac{2}{z_*^2}\,\left(I_\mathcal{A}^{\lambda}-\frac{1}{2}\right) + \mathcal{O}(z_{cut}^2)\,,
\end{equation}
with the same rescaling of the coordinate $x\!=\!\tfrac{z}{z_*}$ as before in the integral
\begin{equation}
    I_\mathcal{A}^{\lambda} = \int_0^1\mathrm{d} x\,\frac{1}{x^3}\,\left(\frac{N_*}{\sqrt{\Lambda^2x^6-f(z_* x)x^6+N_*^2 f(z_*x)}}-1\right)\,.
\end{equation}
Together with the other perturbative results, the solution of this integral leads to an expression for the area
\begin{align}\label{eq:s107}
    \mathcal{A} =& \frac{1}{z_{cut}^2} - \frac{1}{2c_0^3\ell^2} + (\pi T)^4\ell^2\,\frac{\pi c_0^3}{5\sqrt{3}} + (\pi T)^8\ell^6\,\left(\frac{c_0^6}{12} - \frac{2 c_0^9 \pi^2}{225}\right) \\
    &+\frac{\lambda}{\ell}\,\left(\frac{1}{c_0^3\ell^2} + (\pi T)^4\ell^2\,\frac{2\pi c_0^3}{5\sqrt{3}} + (\pi T)^8\ell^6 \left(\frac{c_0^6}{2} - \frac{4 c_0^9 \pi^2}{75}\right)\right) \nonumber \\
    &+\frac{\lambda^2}{\ell^2}\,\left(-\frac{2}{c_0^3\ell^2} + (\pi T)^4\ell^2\,\frac{2\pi c_0^3}{15\sqrt{3}} + (\pi T)^8\ell^6 \left(\frac{4 c_0^6}{3} - \frac{88 c_0^9 \pi^2}{675}\right)\right) \nonumber \\
    &+ \mathcal{O}(z_{cut}^2) + \mathcal{O}(T^{12}\ell^{10}) + \mathcal{O}(\lambda^3/\ell^3)\,. \nonumber
\end{align}
The second derivative of the area \eqref{eq:s107} with respect to $\pm\lambda$ evaluated at $\lambda\!=\!0$ yields the QNEC quantity $\mathcal{S}''_\pm$ used in the main text \eqref{eq:AdSBB_smallL}.
\begin{equation}
    \frac{1}{2\pi}\,\mathcal{S}'' = -\frac{1}{\pi^2c_0^3\ell^4} + \frac{(\pi T)^4\,c_0^3}{15\sqrt{3}\pi} - (\pi T)^8\ell^4\left(\frac{44c_0^9}{675}-\frac{2c_0^6}{3\pi^2}\right) + \mathcal{O}(T^{12}\ell^8)\,.
\end{equation}
To evaluate EE and QNEC in the large $\ell$ limit for this case, for the backreaction in global AdS or the phase transition requires similar techniques regarding the expansion, but the integrals that need to be solved vary a lot.
The best approach and the optimal way to massage the integrands for \textit{Mathematica} needs to be found for every case separately, which requires a lot of experience and intuition.

\cleardoublepage

\section{Available Resources}
 \label{app:Resources}
 
Some of the \textit{Mathematica} notebooks we used for the numerical and perturbative calculations are available on the websites of my colleagues Christian Ecker (\url{http://christianecker.com}) and my supervisor Daniel Grumiller (\url{http://quark.itp.tuwien.ac.at/~grumil/index.shtml}):
\begin{itemize}
    \item Implementation of the relaxation method (\href{http://christianecker.com/wp-content/uploads/2018/08/AppendixE.nb_.tar.gz}{direct link})
    \item Implementation of the QNEC routine (\href{http://christianecker.com/wp-content/uploads/2018/08/AppendixF.nb_.tar.gz}{direct link})
    \item Implementation of the shooting method (\href{http://christianecker.com/wp-content/uploads/2018/08/AppendixD.nb_.tar.gz}{direct link})
    \item Relaxation method for AdS-Vaidya (\href{http://christianecker.com/wp-content/uploads/2016/02/ShockWavesEE.tar.gz}{direct link})
    \item Shooting method for pure AdS and AdS-Schwarzschild (\href{http://christianecker.com/wp-content/uploads/2018/08/Shooting.nb_.tar.gz}{direct link})
    \item Perturbative calculation for the backreaction in global AdS (\href{http://quark.itp.tuwien.ac.at/~grumil/QNEC_quantum_correct.nb}{direct link})

\end{itemize}
 
\end{appendix}

\cleardoublepage

\addcontentsline{toc}{section}{References}
\bibliography{references}
\bibliographystyle{fullsort}

\end{document}